\documentclass[aps,prd,eqsecnum,11pt,preprintnumbers,nofootinbib]{revtex4}
\usepackage{amssymb,amsmath}
\usepackage{graphicx}
\usepackage{graphics}
\newcommand{\bea}{\begin{eqnarray}}
\newcommand{\eea}{\end{eqnarray}}
\newcommand{\be}{\begin{equation}}
\newcommand{\ee}{\end{equation}}
\newcommand{\bes}{\begin{subequations}}
\newcommand{\ees}{\end{subequations}}
\def\lag{\langle}
\def\rag{\rangle}
\newcommand{\nn}{\nonumber}
\newcommand{\pq}{p\cdot q}

\def\nbox#1#2{\vcenter{\hrule \hbox{\vrule height#2in
\kern#1in \vrule} \hrule}}
\def\sq{\,\raise.5pt\hbox{$\nbox{.09}{.09}$}\,}
\def\sqb{\,\raise.5pt\hbox{$\overline{\nbox{.09}{.09}}$}\,}
\begin{document}
\preprint{LA-UR 10-04531}
\preprint{CERN-PH-TH 2010-158}
\vspace{-7cm}

\title{New Horizons in Gravity: \\ 
The Trace Anomaly, Dark Energy \& Condensate Stars}

\author{\large Emil Mottola}
\affiliation{Theoretical Div., Los Alamos National Laboratory,
Los Alamos, NM 87545 USA\\
and\\
Theoretical Physics Group, PH-TH, CERN
CH-1211, Geneva 23, Switzerland\\
E-mail : emil@lanl.gov, emil.mottola@cern.ch}
\vskip 3mm
\begin{abstract}
General Relativity receives quantum corrections relevant at macroscopic distance scales 
and near event horizons. These arise from the conformal scalar degrees of freedom 
in the extended effective field theory of gravity generated by the trace anomaly of massless 
quantum fields in curved space. The origin of these conformal scalar degrees of 
freedom as massless poles in two-particle intermediate states of anomalous amplitudes 
in flat space is exposed. These are non-local quantum pair correlated states, 
not present in the classical theory. At event horizons the conformal anomaly 
scalar degrees of freedom can have macroscopically large effects on the geometry,  
potentially removing the classical event horizon of black hole and cosmological
spacetimes, replacing them with a quantum boundary layer where the effective 
value of the gravitational vacuum energy density can change. In the effective
theory, the cosmological term becomes a dynamical condensate, whose value depends 
upon boundary conditions near the horizon. In the conformal phase where the anomaly 
induced fluctutations dominate, and the condensate dissolves, the effective cosmological
 ``constant"  is a running coupling which has an infrared stable fixed point at zero. 
By taking a positive value in the interior of a fully collapsed star, the effective 
cosmological term removes any singularity, replacing it with a smooth dark energy interior. 
The resulting {\it gravitational condensate star} configuration resolves all black hole 
paradoxes, and provides a testable alternative to black holes as the final state of 
complete gravitational collapse. The observed dark energy of our universe likewise may 
be a macroscopic finite size effect whose value depends not on microphysics but on the 
cosmological horizon scale. The physical arguments and detailed calculations involving 
the trace anomaly effective action, auxiliary scalar fields and stress tensor in various 
situations and backgrounds supporting this hypothesis are reviewed. Originally delivered 
as a series of lectures at the Krak\'ow School, the paper is pedagogical in style, and wide 
ranging in scope, collecting and presenting a broad spectrum of results on black holes, 
the trace anomaly, and quantum effects in cosmology.
\end{abstract}
\pagestyle{plain}
\maketitle 
\topmargin -1cm
\textheight 9.0in

\section{Introduction: Gravitation and Quantum Theory}
\label{sec:intro}

Although it has been clear for nearly a century that quantum principles govern 
the microscopic domain of atomic, nuclear and particle physics, and certainly 
the Standard Model of Strong \& Electroweak interactions is a fully quantum 
theory of matter, gravitational phenomena are still treated as completely 
classical in Einstein's General Relativity. Perhaps just as significant 
as this formal gap between gravitation and quantum principles, most of our 
intuitions about gravity remain essentially classical, particularly in the 
macroscopic domain. Although it is generally agreed that at the fundamental 
microscopic Planck scale, a theory of gravitational interactions must come 
to terms with the quantum aspects of matter and even of spacetime itself, it 
is usually assumed that quantum effects are negligible on the scale of 
macroscopic phenomena, at astrophysical or cosmological distance scales, 
where classical General Relativity (GR) is presumed to hold full sway. 

In non-gravitational physics quantum effects are present on a wide range 
of scales in a variety of ways, some of them striking, others more subtle 
and less immediately appreciated. Semi-conductors, superfluids, 
superconductors and atomic Bose-Einstein condensates are unmistakable 
macroscopic manifestations of an underlying quantum world. On astrophysical 
scales, the degeneracy pressure of fermions, which at first seemed an 
esoteric feature of quantum statistics is now fully accepted as the basis for 
the stability of such macroscopic objects as white dwarves and neutron stars, 
both of which are ubiquitous throughout the universe. As {\it a posteriori} 
consequences of quantum statistics one may note the periodic table, the 
foundations of chemistry itself and hence of biological processes, which 
being familiar in ordinary experience seem far less exotic than neutron 
stars or superfluidity. However chemical bonding, the structure and function 
of hemoglobin and DNA in the human body, and the overall stability of matter 
itself at ordinary temperatures and densities are every bit a consequence of 
quantum principles as a sample of superfluid $^4$He climbing up the walls 
of its dewar. 

It was not only in the microscopic world of the atom but experiments on 
macroscopic matter and the puzzles they generated for classical mechanics, 
such as the ultraviolet problem of blackbody radiation and the specific heat 
of solids, that led to the development of quantum mechanics \cite{Planck,EincV}. 
Since quantum effects play a role in the properties of bulk matter and macroscopic 
phenomena in most every other area of physics, there is no reason why gravity, 
which couples to the energy content of quantum matter at all scales, should be 
immune from quantum effects on macroscopic scales.

If the effects and predictions of a quantum theory of gravity can be tested only 
at Planck lengths or energies, the quest for such a theory would be 
mostly academic, an exercise better left for future generations possessing 
more complete and accurate information about the ultra microscopic world. 
Thus the important questions at the outset are: Are quantum effects anywhere 
relevant or distinguishable in the macroscopic domain of gravitational 
phenomena? And if so, can one say anything reliable about macroscopic 
quantum effects in gravity, without necessarily possessing a complete,
fundamental and well tested theory valid down to the microscopic Planck 
scale?

There are two problems at the forefront of current research where there are
indications that quantum effects may play a decisive role in gravitational 
physics at macroscopic distance scales. The first of these is that of 
ultimate gravitational collapse, presumed in classical GR to lead a 
singularity of spacetime called a {\it black hole}, which generates a 
number of theoretical paradoxes and challenges for quantum theory.
The second problem of great current interest is the apparent existence 
of {\it cosmological dark energy}, which is causing the expansion of the 
universe to accelerate, and which has the same equation of state 
$p = -\rho$ as that of the quantum vacuum itself.

Both of these problems are at the intersection of gravitation and quantum
theory at {\it macroscopic} length scales. In both cases an approximate or 
{\it effective} theory of quantum effects in gravity far from the Planck scale 
should be the appropriate and only framework necessary. In these notes, such 
an effective field theory (EFT) framework is developed according to the same 
general principles now widely recognized in other areas of physics. The 
essential technical tools involve the use of relativistic quantum field theory 
in curved spacetime, and the key ingredient of our analysis is the {\it conformal 
or trace anomaly} of the stress-energy-momentum tensor $T^a_{\ b}$ of massless 
fields, a quantum effect with low energy implications.

Within this essentially semi-classical EFT framework, the principal qualitative 
result is that Einstein's General Relativity can and does receive quantum 
corrections from the effects of the trace anomaly which are significant and 
in certain circumstances may even dominate at macroscopic distance scales, 
much larger than the Planck scale. No assumptions about the extreme short 
distance degrees of freedom or the precise nature of fundamental interactions 
at that scale will be used or needed. Instead our analysis will rely only upon 
the assumption that the Principle of Equivalence in the form of coordinate 
invariance of the effective action of a metric theory of gravity under smooth 
coordinate transformations applies in the quantum theory at sufficiently low 
energy scales far below the Planck scale. With this moderate theoretical input, 
and without invoking unknown or esoteric physics beyond the Standard Model, we 
shall investigate the macroscopic effects of the quantum conformal anomaly on 
gravitational systems at the astrophysical scale of the event horizon of the 
collapse of massive stars, and on the very largest Hubble scale of the visible 
universe itself.

\section{The Challenge of Black Holes}
\label{sec:BH}

The first problem in which classical General Relativity is challenged by 
quantum theory is in the physics of black holes. Before discussing quantum 
effects, let us review the standard classical theory of the
gravitational collapse of massive stars.

A normal main sequence star sustains itself by nuclear fusion of hydrogen
into helium. The star itself formed when enough predominantly hydrogen gas
collapsed to a high enough density for nuclear fusion reactions to occur. 
The energy generated by the fusion of H into He nuclei generates heat and 
pressure, which supports the star against further gravitational collapse. 
A main sequence star remains in this stable steady state, producing radiant 
energy for typically several billions of years, depending upon its mass. 
Eventually, the hydrogen is exhausted, and the star goes through a sequence 
of less exothermic nuclear reactions, fusing nuclei of heavier and heavier 
elements to extract energy from the difference in rest masses. Since iron 
is the most stable nucleus, this process eventually exhausts all the available 
sources of nuclear fusion energy. At that point the star can no longer sustain 
itself against the force of gravity, and its matter must resume gravitational 
collapse upon itself. 

Classically, nothing can halt this collapse. However, quantum matter obeys
quantum statistics. Because of Fermi-Dirac statistics, if the mass of the star 
is not too great, and it has cooled sufficiently, a new stable configuration, 
a white dwarf star held up by its quantum degeneracy pressure can be formed.
In other cases, the collapse of the outer envelope onto the Fe core produces
a violent explosion, a stellar nova or supernova, in which prodigious amounts
of mass and energy are ejected. This leaves behind a even more compact
object in which the electrons and protons are forced under high pressure
to become neutrons. A neutron star, sustained against further collapse 
by the quantum degeneracy pressure of neutron matter, rotating very rapidly 
at nuclear densities and beaming out radiation guided by its strong
magnetic fields may be observed by astronomers as a pulsar.

If the mass of the stellar remnant core exceeds a certain value, called the
Tolman-Oppenheimer-Volkoff (TOV) limit of $1.5 M_{\odot}$ to $3.2 M_{\odot}$
(depending upon the eq. of state of dense nuclear matter, which is not
very accurately known), not even the neutron degeneracy pressure is enough 
to prevent final and inexorable collapse due to gravity \cite{TOV}. Since 
we have no direct observations of these final stages of complete gravitational 
collapse, it is here that the reliance upon Einstein's theory of General Relativity 
becomes critical, and the discussion takes on a decidely more mathematical flavor.

\subsection{Black Holes in Classical General Relativity}

Just a year after the publication of the field equations of General Relativity 
(GR), K. Schwarzschild found a simple, static, spherically symmetric solution 
of those equations, with the line element \cite{Schw},
\be
ds^2 = -f(r)\, d\tau^2 + \frac{dr^2}{h(r)} + r^2
\left( d\theta^2 + \sin^2\theta\,d\phi^2\right)\,,
\label{sphsta}
\ee
where in this case the two functions of $r$ are equal:
\be
f(r) = h(r) = 1 - \frac{2GM}{c^2 r}\,.
\label{Sch}
\ee
This Schwarzschild solution to the vacuum Einstein's equations, 
with vanishing Ricci tensor $R^a_{\ b} = 0$ and stress tensor $T^a_{\ b} = 0$
for all $r > 0$ describes an isolated, non-rotating object of total mass $M$.
In that sense it is the gravitational equivalent of the Coulomb solution,
\be
\phi = \frac{e}{r}
\label{Coul}
\ee
for the electrostatic potential of an isolated, static charge $e$ in Maxwell's 
theory of electromagnetism. Just as in the Coulomb case, the Schwarzschild 
solution has a singularity at the origin of the spherical coordinates at $r=0$, 
where the gauge invariant field strengths (measured in gravity by the Riemann 
curvature tensor and its contractions) diverge, and there is a delta function 
source.

In classical electromagnetism at the finite scale of the
classical electron radius, $r_{_c} = e^2/mc^2$, where the electrostatic 
self-energy becomes comparable to the rest mass energy, some deviation 
from the simple picture of a structureless point particle is to be expected.
In quantum electromagnetism (QED) the classical linear divergence of 
(\ref{Coul}) is softened somewhat into a logarithmic ultraviolet divergence 
of the self-energy of a charged Dirac particle. This logarithmically divergent
self-energy is absorbed into a renormalization of its total observable mass. 
However already at the larger scale of the electron Compton wavelength
$\hbar/mc > r_{_c}$, the single electron description has to be replaced by the
{\it many body} description of a quantum field theory with vacuum polarization
effects. Hence the pointlike singularity and linear divergence of the classical
Coulomb potential (\ref{Coul}) is not present in the more accurate many body
quantum theory.

Apart from the singularity at $r=0$, analogous to (\ref{Coul}) the Schwarzschld line 
element also possesses  another kind of mathematical singularity at the {\it finite} 
Schwarzschild radius,
\be
r_{_S} \equiv \frac{2GM}{c^2} \simeq 2.953\left(\frac{M}{M_{\odot}}\right)\,{\rm km}\,,
\label{rS}
\ee
where the function $f(r) = h(r)$ vanishes. This macroscopic radius is the location 
of the Schwarzschild {\it event horizon}, the locus of points which defines the 
sphere dividing the exterior region from the interior region. It is the analog
of the classical radius $r_{_c}$ where the Newtonian gravitational self-energy $GM^2/r$
becomes comparable to the total rest mass energy $Mc^2$. Thus $r_{_S}$ is the 
length scale at which some substructure should be expected.

Classical General Relativity does not give much hint of this substructure.
Instead, the change of sign of the functions $f(r),h(r)$ for $r < r_{_S}$ in (\ref{Sch})
indicates that for the interior region $t$ becomes a {\it spacelike} variable, while 
$r$ becomes {\it timelike}. Hence any radiation, even a light ray emanating from a 
point in the interior cannot propagate outward and is drawn inexorably toward the 
singularity at $r=0$, giving rise to the popular name {\it black hole}.

If the Schwarzschild solution (\ref{sphsta})-(\ref{Sch}) for $r< r_{_S}$
is taken seriously, the singularity at $r=0$ is present in Einstein's theory for any 
mass $M>0$, including the certainly macroscopic mass of a collapsed star with 
the mass of the sun, $M_{\odot} \simeq 2 \times 10^{33}$ gm. or even that  of 
supermassive objects with masses $10^6$ to $10^9M_{\odot}$. The collapse 
of such enormous quantities of matter with vastly more degrees of freedom than 
that of a single electron to a single mathematical point at $r=0$ certainly presents 
a challenge to the imagination, and one that it seems Einstein himself sought 
arguments to avoid \cite{Ein}. The situation is scarcely more acceptable if
the singularity is removed only by the intervention of quantum effects at
the extremely tiny Planck length $(G\hbar/c^3)^{\frac{1}{2}} \sim 1.6 \times 10^{-33}$ cm.

A light wave emitted from any $r > r_{_S}$ with local frequency $\omega_{loc}$ 
outward towards infinity is redshifted according to the redshift relation,
\be
\omega_{\infty} = \omega_{loc} \,f^{\frac{1}{2}}= \omega_{loc} 
\sqrt {1 - \frac{r_{_S}}{r}}\,,
\label{redshift}
\ee
showing that a light wave emitted at the horizon becomes redshifted to {\it zero}
frequency and cannot propagate outward at all. Conversely and equivalently,
a light wave with the finite frequency $\omega =\omega_{\infty}$ far from the black hole
is {\it blueshifted} to an {\it infinite} local frequency at the horizon. This gravitational 
redshift/blueshift is purely a kinematic consequence of the classical time dilation 
effect of a gravitational field, which has been tested in a number of experiments 
\cite{MTW,WeinG,Will}. The event horizon is therefore a kind of {\it critical surface} for
the propagation of light rays, and hence all other matter interactions.

Unlike the central singularity at $r=0$, the scalar invariant quantities that can
be constructed from the contractions of the Riemann curvature tensor remain 
finite as $r \rightarrow r_{_s}$. Thus the fully contracted quadratic Riemann
invariant 
\be
R^{abcd}R_{abcd} = \frac{12r_{_S}^2}{r^6}\,,
\label{RieS}
\ee
which diverges at the origin remains finite at $r= r_{_S}$. Moreover, although 
the time for an infalling particle to reach the horizon is infinite for any 
observer remaining fixed outside the horizon, the {\it proper time} measured 
by the particle itself during its infall remains finite as $r \rightarrow r_{_s}$ \cite{MTW,WeinG}. 
Thus despite the singularity of the Schwarzschild coordinates at $r= r_{_S}$,
physics must continue onto smaller values of $r$ in the interior region. Since 
the line element (\ref{sphsta}) is again non-singular for $0 < r < r_{_S}$,
and in the absence of clear evidence to the contrary, the most straightforward 
possibility would seem to be to assume that this non-singular vacuum interior 
(up to the origin or at least some extreme microscopic scale much less than $r_{_S}$) 
can be matched smoothly to the non-singular exterior Schwarzschild solution.

\begin{figure}[htp]
\includegraphics[width=12cm, viewport=-100 0 854 569,clip]{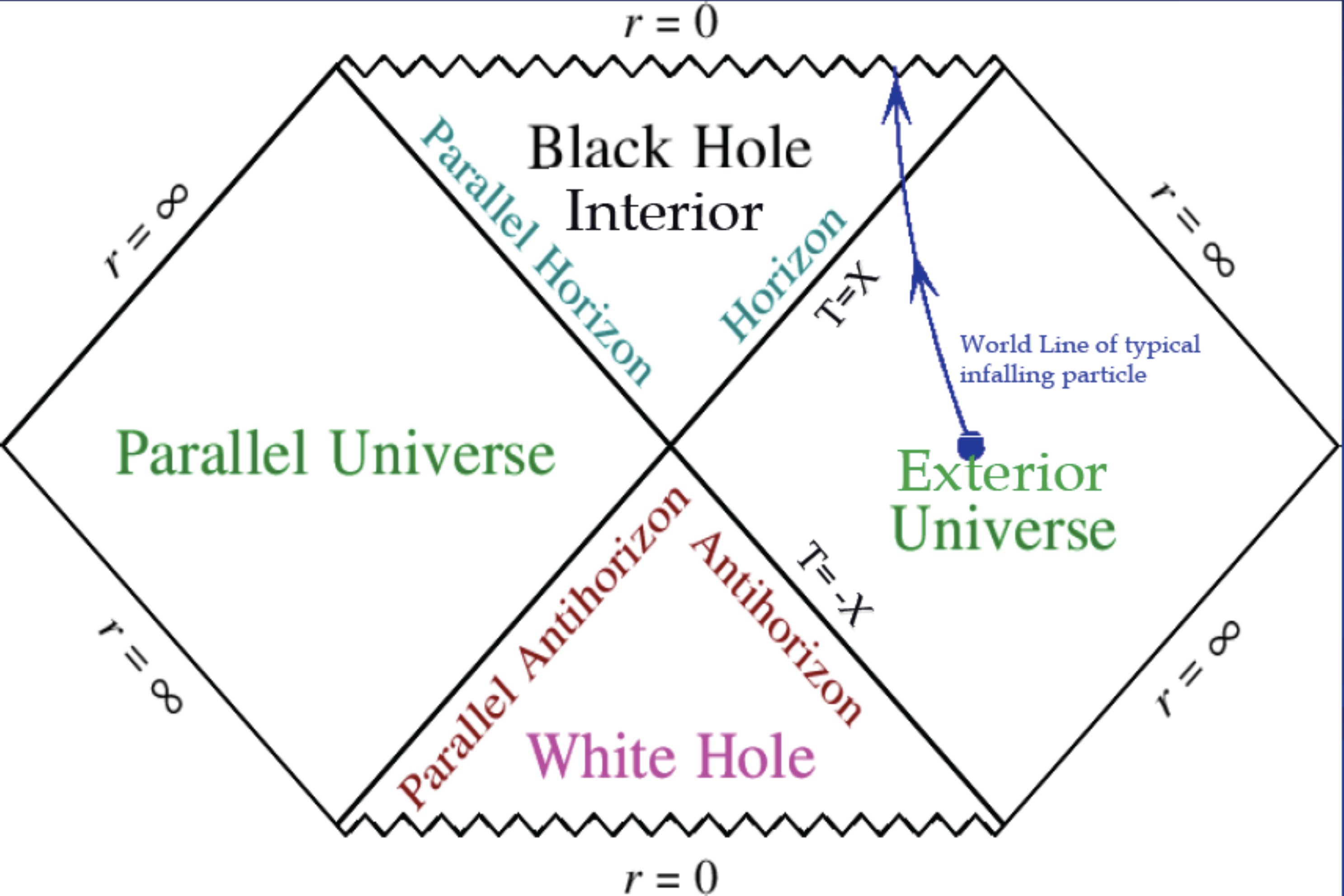}
\caption{The Carter-Penrose conformal diagram of the maximal Kruskal analytic extension
of the Schwarzschild geometry. Radial light rays are represented in this diagram
as $45 ^{\circ}$ lines. The angular coordinates $\theta,\phi$ are suppressed.}
\label{Fig:SchwKruskal}
\end{figure}

This matching was achieved by the coordinate transformations and analytic 
continuation of the Schwarzschild solution found by Kruskal and Szekeres 
\cite{KrusSz}. The Kruskal maximal analytic extension of the Schwarzschild 
geometry is pictured in the Carter-Penrose conformal diagram, 
Fig. \ref{Fig:SchwKruskal}.

The analytic extension of the Schwarzschild geometry relies on finding
a judicious change of the time and radial coordinates $(t,r)$ of (\ref{sphsta})
to new ones $(T,X)$, which are regular on the horizon, and therefore can be 
used to describe the local geometry there without singularities. Explicitly, for 
$r > r_{_S}$ this coordinate transformation to Kruskal-Szekeres coordinates 
$(T,X)$ is given by
\bes
\bea
&& T = \left(\frac{r}{r_{_S}} - 1\right)^{\frac{1}{2}}\, e^{r/2r_{_S}}\, 
\sinh \left(\frac{ct}{2r_{_S}}\right)\,,\\
&& X = \left(\frac{r}{r_{_S}} - 1\right)^{\frac{1}{2}}\, e^{r/2r_{_S}}\, 
\cosh \left(\frac{ct}{2r_{_S}}\right)\,.
\eea
\label{KStrans}\ees
The inverse transformation is
\bes
\bea
&& t = \frac{2 r_{_S}}{c} \tanh^{-1} \left(\frac{T}{X}\right) = \frac{r_{_S}}{c} \, 
\ln \left( \frac{X+T}{X-T}\right)\,\\
&&\hspace{1cm}\left(\frac{r}{r_{_S}} - 1\right)\, e^{r/r_{_S}}= X^2 - T^2\,.
\label{rimp}\eea\label{rTX}\ees
In the new $(T,X)$ coordinates, the Schwarzschild line element (\ref{Sch})
becomes
\be
ds^2 = \frac{4 r_{_S}^3}{r}\, e^{r/2r_{_S}}\, (-dT^2 + dX^2) + r^2 d\Omega^2\,,
\label{SKSz}
\ee
where $r$ is to be regarded as the function of $(T,X)$ given implicitly by (\ref{rimp}),
and $d\Omega^2 = d\theta^2 + \sin^2\theta\, d\phi^2$ is the usual spherical
line element on ${\mathbb S}^2$.

In these Kruskal-Szekeres coordinates, the event horizon at $r= r_{_S}$ is seen actually
to be comprised of two distinct null surfaces, the future and past horizons for $T = \pm X$, 
respectively. The entire exterior region $r > r_{_S}$ is mapped into the region $X > |T| \ge 0$ 
of the $(X,T)$ plane. The $X=+T$ boundary of this region gives the coordinates of a particle 
trajectory infalling into the black hole, a depicted in Fig. \ref{Fig:SchwKruskal}, whereas the 
$X=-T$ boundary corresponds to the time reversed coordinates of a particle trajectory 
outgoing from a {\it white hole}. The time reversed case must be present mathematically 
because of the second order nature of Einstein's eqs., and the static solution (\ref{sphsta}) 
which admits the time reversal symmetry $t \rightarrow -t$ and $T \rightarrow -T$. Whether 
or not this time reversed white hole case corresponds to any real macroscopic body is 
of course another question. 

Since the Schwarzschild line element in Kruskal-Szekeres coordinates (\ref{SKSz}) is
completely regular at $T= \pm X$, one can equally well consider the extension of
the coordinates to the interior regions, $T > X> 0$ and $T < -X < 0$, into the black
hole and white hole interior regions at the top and bottom of Fig. \ref{Fig:SchwKruskal},
at least as far and until the true curvature singularity at $r=0$ is reached. Inspection 
of (\ref{rTX}) shows that this implies an analytic continuation of the original 
Schwarzschild $(t,r)$ coordinates around a logarithmic branch cut to {\it complex} 
values. The relation of the $(T,X)$ coordinates to the original $(t,r)$ coordinates
is singular at $r=r_{_S}$, although the $(T,X)$ coordinates themselves are regular 
at $T = \pm X$. Similarly, by a further complex analytic continuation, one can 
continue to the parallel exterior region on the left of Fig. \ref{Fig:SchwKruskal} with 
$X < -|T| \le 0$. Thus, by the relatively simple but {\it singular} change of coordinates 
(\ref{KStrans})-(\ref{rTX}), we seem to have reached the conclusion that the simplest static 
spherically symmetric Schwarzschild solution to the vacuum Einstein's equations 
predicts the existence not only of a true singularity at $r=0$ but also of an entirely 
separate and macroscopically large, asymptotically flat region in addition 
to the original one. 

Of course, one is free to assert that the white hole interior region and the parallel 
asymptotically flat universe do not exist, and excise them in a gravitational collapse 
from realistic initial conditions, replacing the excised region of Fig. \ref{Fig:SchwKruskal} 
with the non-singular interior of a collapsing matter distribution. However, this is
a matter of initial conditions, and nothing in the equations of General Relativity 
themselves forces us to do this. Dirac expressed skepticism of the interior 
Schwarzschild solution on physical grounds \cite{Dir}. 

The apparently unphysical features of the Schwarzschild solution appear as soon 
as we admit complex analytic continuation of singular coordinate transformations. 
Based on the Principle of Equivalence between gravitational and inertial mass, Einstein's 
theory possesses general coordinate invariance under all {\it regular} and {\it real} 
transformations of coordinates. It is the appending to classical General Relativity 
of the much stronger mathematical {\it hypothesis} of {\it complex} analytic 
continuation through {\it singular} coordinate transformations that leads to the 
global aspects of the Schwarzschild solution which may be unrealized in Nature.

The point which is often left unstated is that the mathematical procedure of analytic 
continuation through the null hypersurface of an event horizon actually involves a physical 
{\it assumption}, namely that the stress-energy tensor $T^a_{\ b}$ is vanishing there. 
Even in the purely classical theory of General Relativity, the hyperbolic character of 
Einstein's equations allows generically for stress-energy sources and hence metric 
discontinuities on the horizon which would violate this assumption. Additional 
{\it physical} information is necessary to determine what happens as the event 
horizon is approached, and the correct matching of interior to exterior geometry. 
What actually happens at the horizon is a matter of this correct physics, which may 
or may not be consistent with complex analytic continuation of coordinates 
(\ref{KStrans})-(\ref{rTX}).

The static Schwarzschild solution of an isolated uncharged mass was generalized 
to include electric charge by Reissner \& Nordstr\o m \cite{RN}, and more interestingly for 
astrophysically realistic collapsed stars, to include rotation and angular momentum
by Kerr \cite{Kerr}. The complete analytic extensions of the Reissner-Nordstr\o m and 
Kerr solutions were found as well \cite{BoyLinCar}. The global properties of these 
analytic extensions are more complicated and arguably even more unphysical than in the 
Schwarzschild case. For slowly rotating black holes with angular momentum $J < GM^2/c$, 
there are an {\it infinite} number of black hole interior and asymptotically flat exterior 
regions, and {\it closed timelike curves} in the interior region(s), which violate causality 
on macroscopic distance scales \cite{HawEll}. Again these apprently unphysical features 
appear in GR only if the mathematical hypothesis of complex analytic 
extension and continuation through real coordinate singularities are assumed.
This analytic continuation is generally invalid if there are stress-tensor sources 
encountered at or before the breakdown of coordinates.

The Schwarzschild white hole and the analytic extension through the horizons 
also raises questions about macroscopic time reversibility. Once classical
particles fall through the future event horizon, there is no way to retrieve them
without violating causality, and something irreversible would seem to have occurred.
This is somewhat troubling from the point of view of thermodynamics, since
if matter disappears completely from view when it falls into a black hole, it carries
any entropy it has in its internal states with it, and the entropy of the visible 
universe would apparently have decreased, violating the second law of thermodynamics,
\be
\Delta S \ge 0\,,
\label{SecLaw}
\ee
which states that the total entropy of an isolated system must be a non-decreasing
function of time in any spontaneous process.

At the same time the infall of matter into a black hole certainly increases its total 
energy. In the general black hole solution characterized by mass $M$, angular 
momentum $J$, and electric charge $Q$, one can define a quantity called the 
{\it irreducible mass} $M_{irr}$ by the relation,
\be
M^2 = \left(M_{irr} + \frac{Q^2}{4G M_{irr}}\right)^2 + \frac{c^2 J^2}{4 G^2 M_{irr}^2}\,,
\label{MQJ}
\ee
or
\be
M^2_{irr} = \frac{M^2}{2\ } - \frac{Q^2}{4G\ } + \frac{1}{2G}\,
\left[G^2M^4 - GM^2Q^2 - c^2J^2\right]^{\frac{1}{2}}\,,
\ee
and show that in any {\it classical} process the irreducible mass can never decrease \cite{Mirr}:
\be
\Delta M^2_{irr} \ge 0\,.
\label{Mirrinc}
\ee
Since one can also show that the irreducible mass is related to the geometrical area
$A$ of the event horizon of the general Reissner-Nordstr\o m-Kerr-Newman black hole via
\be
A = \frac{16 \pi G^2}{c^4} M^2_{irr} \,,
\label{areah}
\ee
this theorem is equivalent to the statement that {\it the horizon area is a 
non-decreasing function of time in any classical process} \cite{Mirr,HawA}.

Simply by taking the differential form of (\ref{MQJ}) one obtains \cite{BCH}
\be
dE = dMc^2=  \frac{c^2}{8\pi G}\,\kappa \,dA + \Omega\,  dJ + \Phi\,dQ
\label{diffSm}
\ee
which is just the differential form of Smarr's formula for a Kerr-Newman rotating,
electrically charged black hole, in which
\bes
\bea
&&\kappa = \frac{1}{M}\left[\frac{c^4}{4G} - \frac{4\pi^2 G}{c^4A^2} 
\left(Q^4 + 4c^2J^2\right)\right],\\
&&\Omega = \frac{4\pi J}{MA}\,,\\
&&\Phi = \frac{Q}{M} \left[\frac{c^2}{2G} + \frac{2\pi Q^2}{Ac^2}\right]\,,\\
&&A = \frac{4\pi G}{c^4}\,\left[2GM^2 -Q^2 + 2\sqrt{G^2M^4 - GM^2Q^2  - c^2J^2}\right],
\eea
\label{kOP}\ees
are the horizon surface gravity, angular velocity, electrostatic potential and area 
respectively \cite{BCH,Smarr}. All dimensionful constants have been retained
to emphasize that (\ref{diffSm})-(\ref{kOP}) are formulae derived from {\it classical} 
GR in which no $\hbar$ whatsoever appears. Notice also that the coefficient of $dA$ in 
(\ref{diffSm}), $c^2\kappa/8\pi G$ has both the form and dimensions of a {\it surface tension}.

The classical conservation of energy is expressed by the first law of black hole mechanics 
(\ref{diffSm}). The classical area theorem (\ref{Mirrinc}) naturally evokes a connection 
to entropy and the second law of thermodynamics (\ref{SecLaw}). {\it If} horizon 
area (or more generally any monotonic function of it) could somehow be identified with entropy, 
and this entropy gain is greater than the entropy lost by matter or radiation falling into 
the hole, then the second law (\ref{SecLaw}) would remain valid for the total or 
generalized entropy of matter plus black hole horizon area. The simplest
possibility would seem to be if entropy is just proportional to area.

Motivated by these considerations, and as suggested by a series of thought
experiments, Bekenstein proposed that the area of the horizon (\ref{areah}) should 
be proportional to the entropy of a black hole \cite{Bek}. Since $A$ does not have 
the units of entropy, it is necessary to divide the area by another quantity with units 
of length squared before multiplying by Boltzmann's constant $k_{_B}$, to obtain an 
entropy. However, classical General Relativity (without a cosmological term) contains 
no such quantity, $G/c^2$ being simply a conversion factor between mass and 
distance. Hence Bekenstein found it necessary for purely dimensional reasons to 
introduce Planck's constant $\hbar$ into the discussion. Then there is a standard 
unit of length, namely the Planck length,
\be
L_{Pl} = \sqrt{\frac{\hbar G }{c^3}} = 1.616 \times 10^{-33} \, {\rm cm.}
\label{LPl}
\ee
Bekenstein proposed that the entropy of a black hole should be
\be
S_{_{BH}} = \gamma k_{_B} \frac{A}{L_{Pl}^2}\,,
\label{SBHg}
\ee
with $\gamma$ a constant of order unity \cite{Bek}. He showed that {\it if} such 
an entropy were assigned to a black hole, so that it is added to the entropy of
matter, $S_{tot} = S_m + S_{_{BH}} $,  then this total generalized entropy would
plausibly always increase. In fact, this is not difficult at all, and the generalized
second law $\Delta S_{tot} \ge 0$ is usually satisfied by a very wide margin, simply
because the Planck length is so tiny, and the macroscopic area of a black hole 
measured in Planck units is so enormous. Hence even the small increase of mass 
and area caused by dropping into the black hole a modest amount of matter 
and concomitant loss of matter entropy $\Delta S_m < 0$ is easily overwhelmed 
by a great increase in $S_{_{BH}}$, $\Delta S_{_{BH}} \gg |\Delta S_m|$, 
guaranteeing that the generalized total entropy increases: $\Delta S_{tot} > 0$.

Since $\hbar$ has entered the assignment of entropy to a black hole horizon,
the discussion can no longer be continued in purely classical terms, and
we must discuss quantum effects in black hole geometries next. It is worth
remarking that whereas Planck's constant enters the thermodynamics
of macroscopic quantum systems, such as the formulae for black body 
radiation, by normalizing the volume of the integral over {\it phase space},
no such interpretation is available for (\ref{SBHg}). Instead $\hbar$ has been
used to form a new quantity $L_{Pl}$ with units of length, which we would
ordinarily associate with the microscopic length scale at which strong 
quantum corrections to Einstein's theory should become important. Why 
such a microscopic, fundamentally quantum length scale should be needed 
to determine the bulk thermodynamic entropy of a macroscopic object on 
the scale of kilometers where classical GR applies, and no large quantum
corrections to GR are expected near the horizon, is far from clear.

\subsection{Quantum Black Holes and Their Paradoxes}

Since classically all matter is irretrievably drawn into a black hole, the idea 
that black holes can instead radiate energy seems quite counterintuitive. More
remarkable still is Hawking's argument that this radiation would necessarily
be {\it thermal} radiation, with a temperature \cite{HawkT}
\be
T_{_H} = \frac{\hbar\kappa}{2\pi c k_{_B}}\  \,\stackrel{J=Q=0}{=}\ \,
 \frac{\hbar c^3}{8 \pi G k_{_B} M}\,,
\label{TH}
\ee
where the first equality is general, and the second equality applies only 
for a Schwarzschild black hole with $J=Q=0$. With the temperature inversely 
proportional to its mass assigned to a black hole by this formula, if 
we assume that the first law of thermodynamics in the form
\be
dE = dM c^2 = T_{_H} \, dS_{_{BH}} + \Omega\,  dJ + \Phi\,dQ\,.
\label{dETdS}
\ee
applies to black holes, then the coefficient $\gamma$ in (\ref{SBHg}) is 
fixed to be $1/4$. This formula is simple and appealing, and has been generally 
accepted since soon after Hawking's paper first appeared. However, 
simultaneously and from the very beginning, a number of problems with 
this thermodynamic interpretation made their appearance as well. 

The first curious feature of (\ref{dETdS}) is that $\hbar$ cancels out between
$T_{_H}$ and $dS_{_{BH}}$. Of course, this is a necessary consequence of
the fact that (\ref{dETdS}) is {\it identical} to the classical Smarr formula
(\ref{diffSm}) in which $\hbar$ does not appear at all. The identification of 
$S_{_{BH}}$ with the entropy of a black hole is founded on the purely classical 
dynamics of Christodoulou's area law (\ref{Mirrinc}), in which quantum mechanics 
played no part whatsoever. Clearly, multiplying and dividing by $\hbar$ does not 
necessarily make a classical relation a valid one in the quantum theory. On the 
other hand, the Hawking temperature (\ref{TH}) is a quantum spontaneous emission 
effect, analogous to the Schwinger pair creation effect in a  strong electric 
field \cite{Schwinger} which vanishes in the classical limit $\hbar \rightarrow 0$. 
Temperature, usually accepted as a classical concept has no apparent meaning for 
a black hole in the strictly classical limit, unless it is identically zero. 
As a consequence, if the identification of the classical area rescaled by 
$k_{_B}L_{Pl}^{-2}$ with entropy and the thermodynamic interpretation of 
(\ref{dETdS}) is to be generally valid in the quantum theory, then the 
classical limit $\hbar \rightarrow 0$ (with $M$ fixed) which yields an 
arbitrarily low Hawking temperature, assigns to the black hole an 
{\it arbitrarily large} entropy, completely unlike the zero temperature 
limit of any other cold quantum system. 

Closely related to this paradoxical result is the fact, pointed out by Hawking 
himself \cite{HawkcV}, that a temperature inversely proportional to the
$M = E/c^2$ implies that the heat capacity of a Schwarzschild  black hole,
\be
\frac{dE}{dT_{_H}} = - \frac{8 \pi G k_{_B} M^2}{\hbar c} = - \frac{Mc^2}{T_{_H}} < 0
\label{dEdT}
\ee
is {\it negative}. In statistical mechanics the heat capacity of any system (at constant 
volume) is related to the energy fluctuations about its mean value $\langle E \rangle$ by
\be
c_{_V} = \left(\frac{d\langle E\rangle}{dT}\right)_{_V} = 
\frac{1}{k_{_B}T^2} \,\Big\langle (E- \langle E\rangle)^2\Big\rangle > 0\,.
\label{cV}
\ee
If pressure or some other thermodynamic variable is held fixed there is an analogous
formula. Hence on general grounds of quantum statistical mechanics, the heat capacity 
of any system in (stable) equilibrium must be {\it positive}. The positivity of the 
statistical average in (\ref{cV}) requires only the existence of a well defined 
{\it stable} ground state upon which the thermal equilibrium ensemble is defined, 
but is otherwise independent of the details of the system or its interactions.

In fact, it is easy to see that a black hole in thermal equilibrium with a heat bath 
of radiation at its own Hawking temperature $T= T_{_H}$ cannot be stable \cite{HawkcV}. 
For if by a small thermal fluctuation it should absorb slightly more radiation in a 
short time interval than it emits, its mass would increase, $\Delta M > 0$, and hence 
from (\ref{TH}) its temperature would decrease, $\Delta T < 0$, so that it would now 
be cooler than its surroundings and be favored to absorb more energy from the heat
bath than it emits in the next time step, decreasing its temperature further
and driving it further from equilibrium. In this way a runaway process of the
black hole growing to absorb all of the surrounding radiation in the heat bath
would ensue. Likewise, if the original fluctuation has $\Delta M< 0$, the temperature
of the black hole would increase, $\Delta T > 0$, so that it would now be hotter
than its surroundings and favored to emit more energy than it absorbs from the
heat bath in the next time step, increasing its temperature further. Then a 
runaway process toward hotter and hotter evaporation of all its mass to its
surroundings would take place. In either case, the initial equilibrium is clearly
{\it unstable}, and hence cannot be a candidate for the quantum ground state 
for the system. This is the physical reason why the positivity property of (\ref{cV})
is violated by the Hawking temperature (\ref{TH}). The instability has also been
found from the negative eigenvalue of the fluctuation spectrum of a black
hole in a box of (large enough) finite volume \cite{BHnegmode}.

The time scale for this unstable runaway process to grow exponentially is the 
time scale for fluctuations away from the mean value of the Hawking flux, {\it not} 
the much longer time scale associated with the lifetime of the hole under continuous 
emission of that flux. This time scale for thermal fluctuations is easily estimated. 
It is the typical time between emissions of a single quantum with typical energy 
(at infinity) of $k_{_B}T$, of a source whose energy emission per unit area per 
unit time is of order $(k_{_B}T_{_H})^4/\hbar^3 c^2$. Multiplying by the area of 
the hole $A \sim (GM)^2/c^4$, and dividing by the typical energy $k_{_B}T$, we find 
the average number of quanta emitted per unit time. The inverse of this, namely
\be
\Delta t \sim \frac{1}{A} \frac{ (\hbar c)^3}{(k_{_B} T_{_H})^3} \sim \frac{r_{_S}}{c}
\sim 10^{-5}\left(\frac{M}{M_{\odot}}\right)\, {\rm sec}
\label{deltat}
\ee
is the typical time interval (as measured by a distant observer) between 
successive emissions of individual Hawking quanta (again as observed far from 
the black hole). This time scale is quite short: $10\ \mu$sec for a solar 
mass black hole. Any tendency for the system to become unstable would be 
expected to show up on this short a time scale, governing the fluctuations
in the mean flux, which is of order of the collapse time itself and before 
a steady state flux could even be established. With the existence of a stable 
equilibrium in doubt, one may well question whether macroscopic equilibrium 
thermodynamic concepts such as temperature or entropy are applicable to black 
holes at all. 

Another rather peculiar feature of the formula (\ref{SBHg}) for the entropy is
that it is {\it non-extensive}, growing not like the volume of the system but its
area. It is non-extensive in a second important respect. The fixing of $\gamma = 1/4$
by (\ref{TH}) and (\ref{dETdS}) is independent of the number or kind of
particle species. Normally, we would expect the entropy of a system to grow 
linearly with the number of distinct particle species it contains. For example, 
if the number of light neutrino species in the universe were doubled, 
the entropy in the primordial plasma would be doubled as well, because 
the available states of one species are independent of and orthogonal to the 
states of a second distinct species, and must be counted separately. The 
formula (\ref{SBHg}) with only fundamental constants and the pure number 
coefficient $\gamma = 1/4$ does not seem to allow for this. `t Hooft found a
way to compensate the number of species factor in the horizon ``atmosphere,"
but this configuration with matter or radiation densely concentrated near
$r=r=r_{_S}$ is no longer then a Schwarzschild black hole \cite{tHatmos}.
It also remains singular at the origin.

Finally, let us state the obvious: a solution of any set of classical field
equations is simply one particular configuration in the space of field configurations.
As such, one would not usually associate any entropy with it. Matter sources to 
the field equations which have internal degrees of freedom may carry entropy of course, 
but a {\it vacuum} solution to the equations such as the Schwarzschild solution 
(\ref{sphsta})-(\ref{Sch}), with at most a singular point source at the origin would 
not ordinarily be expected to carry any entropy whatsoever. What ``entropy" should 
one associate with the structureless classical Coulomb field (\ref{Coul})? 

In this connection it may be worth pointing out that although the Coulomb 
field does not have an event horizon, there would still be an ``information paradox" 
if we allowed charged matter to be attracted into (or emitted from) the Coulomb
field singularity in (\ref{Coul}) at $r=0$ and disappear from (or appear in) the 
visible universe. Such disappearance or appearance processes would clearly violate 
unitarity as well. In quantum theory we exclude such a possibility by restricting 
the Hilbert space of states to those with wave functions that are {\it normalizable} 
at the origin, so that no energy or momentum flux can either vanish or appear at 
the Coulomb singularity. Note that we impose this boundary condition at $r=0$ without 
any detailed knowledge of the extreme short distance structure of a charged particle
in QED, confident that whatever it is, unitarity must be respected. In the case of 
black hole radiance by contrast, the positive Hawking energy flux at infinity 
must be balanced by a compensating flux down the hole and eventually into the 
singularity at $r=0$, which makes clear why problems with unitarity and loss 
of information must result from such a flux. Boundary conditions on the horizon 
have consequences for the behavior of the fluctuations at both the singularity 
and at infinity. As even the example of scattering off the Coulomb potential 
shows, these boundary conditions require physical input. Unphysical 
boundary conditions can easily lead to unphysical behavior.

Returning to the entropy (\ref{SBHg}), it is instructive to evaluate $S_{_{BH}}$ 
for typical astrophysical black holes. Taking again as our unit of mass the mass 
of the sun, $M_{\odot} \simeq 2 \times 10^{33}$ gm., we have
\be
S_{_{BH}} \simeq 1.050 \times 10^{77}\, k_{_B}\, \left(\frac{M}{M_{\odot}}\right)^2\,.
\label{SBH}
\ee
This is truly an enormous entropy. For comparison, we may estimate
the entropy of the sun as it is, a hydrogen burning main sequence star, whose
entropy is given to good accuracy by the entropy of a non-relativistic perfect 
fluid. This is of the order $N k_{_B}$ where $N$ is the number of nucleons 
in the sun $N \sim M_{\odot}/m_{_N} \sim 10^{57}$, times a logarithmic function 
of the density and temperature profile which may be estimated to be of the order 
of $20$ for the sun. Hence the entropy of the sun is roughly
\be
S_{\odot} \sim 2 \times 10^{58} \, k_{_B}\,,
\label{Ssun}
\ee
or nearly $19$ orders of magnitude smaller than (\ref{SBH}).  

A simple scaling argument that the entropy of any gravitationally
bound object with the mass of the sun cannot be much more than (\ref{Ssun}) 
can be made as follows. The entropy of a relativistic gas at temperature $T$ 
in equilibrium in a box of volume $V$ is of order $VT^3$. The total energy is of 
order $VT^4$. Eliminating $T$ from these relations gives 
$S\sim V^{\frac{1}{4}} E^{\frac{3}{4}}$. For a relativistic bound system the 
energy $E \sim M c^2$ while the volume is of order $r_{_S}^3 \sim (GM/c^2)^3$. 
Hence $S \sim (\sqrt{G} M)^{\frac{3}{2}}$. Keeping track of $k_{_B}$, $\hbar$ 
and $c$ in this estimate gives
\be
S \sim k_{_B} \left(\frac{M}{M_{Pl}}\right)^{\frac{3}{2}} \sim 10^{57}\, k_{_B}  
\left(\frac{M}{M_{\odot}}\right)^{\frac{3}{2}}\,,
\label{relstar}
\ee
where $M_{Pl} = \sqrt{\hbar c/G} = 2.176 \times 10^{-5}$ gm. If there are
$\nu$ species of relativistic particles in the object then this estimate should
be multiplied by $\nu^{\frac{1}{4}}$. This estimate applies to the entropy
of relativistic radiation within the body, and is lower than (\ref{Ssun}) because
the radiation pressure in the sun is small compared to the non-relativistic
fluid pressure. However, the entropy from the relativistic radiation pressure 
(\ref{relstar}) grows with the $3/2$ power of the mass, whereas the non-relativistic 
fluid entropy (\ref{Ssun}) grows only linearly with $M$. For stars with masses 
greater than about $50 M_{\odot}$ which are hot enough for their pressure to be 
dominated by the photons' $T^4$ radiation pressure, (\ref{relstar}) indeed gives 
the correct order of magnitude estimate of such a star's entropy at a few times 
$10^{59} k_{_B}$ \cite{ZelNov}. In order to have such an entropy, the temperature 
of the star must be of order $E/S \sim 10$ MeV or $10^{11}\,^\circ$K, while the 
Hawking temperature (\ref{TH}) for the same $50 M_{\odot}$ black hole is a very, 
very cold $10^{-9}\,^\circ$K. It is difficult to see how the entropy of the black 
hole could be a factor of $10^{20}$ larger while its temperature is a factor of 
$10^{20}$ {\it lower} than a relativistic star of the same mass.

The point is that even for the extreme relativistic fluid the entropy (\ref{relstar}) 
for a gravitationally bound system in thermal equilibrium (in which entropy is 
always maximized) grows only like the $3/2$ power of the mass, and hence 
will always be much less than (\ref{SBH}), proportional to $M^2$ for very 
massive objects. Moreover the discrepant factor between (\ref{SBH}) and 
(\ref{relstar}) is of order $(M/M_{Pl})^{\frac{1}{2}} \simeq 10^{19} (M/M_{\odot})^{\frac{1}{2}}$, 
no matter what the non-black hole progenitor of the black hole is. Since the 
formula (\ref{SBH}) makes no reference to how the black hole was formed, 
and a black hole may always be theoretically idealized as forming from an 
adiabatic collapse process, which keeps the entropy constant, (\ref{SBH}) 
states that this entropy must suddenly jump by a factor of order $10^{19}$
for a solar mass black hole at the instant the horizon forms. When Boltzmann's 
formula,
\be
S = k_{_B} \ln W(E)
\ee
is recalled, relating the entropy to the total number of microstates in the system 
$W(E)$ at the fixed energy $E$, we see that the number of such microstates 
of a black hole satisfying (\ref{SBH}) must jump by $\exp(10^{19})$ at that instant 
at which the event horizon is reached, a truly staggering proposition.

This tremendous mismatch between the number of microstates of a black hole inferred
from $S_{_{BH}}$ and that of any conceivable physical non-black hole progenitor is one 
form of the {\it information paradox}. Another form of the paradox is that since black holes are 
supposed to radiate thermally at temperature $T_{_H}$ up until their very last stages, 
when their mass falls to a value of order $M_{Pl}$, there would seem to be no way to 
recover all the information apparently lost in the black hole formation and evaporation 
process. This difficulty with $S_{_{BH}}$ is so severe that it led Hawking to speculate that 
perhaps even the quantum mechanical unitary law of evolution of pure states into pure 
states would have to be violated by black hole physics \cite{Hawkunit}. Although this 
speculation has currently fallen into disfavor \cite{Hawnew}, it is still far from clear what 
the missing microstates of an uncharged black hole are, and how exactly unitarity can 
be preserved in the Hawking evaporation process if (\ref{TH}) and (\ref{SBH})
are correct.

For all of these reasons the thermodynamic interpretation of (\ref{dETdS}) remains 
problematic in quantum theory. On the other hand, if the $\hbar$ is cancelled and one 
simply returns to the differential Smarr relation (\ref{diffSm}), derived from classical GR, 
these difficulties immediately vanish. One would only be left to explain the relationship 
of surface gravity $\kappa$ to surface tension. In Refs. \cite{gstar,PNAS} and Sec. \ref{sec:grava} 
a possible resolution is proposed in which area is not entropy at all but indeed the area of a 
physical surface and the surface gravity can be related to the surface tension of this 
surface.

Although the thermodynamic interpretation of (\ref{dETdS}) and (\ref{SBH}), 
leads to myriad difficulties, it has been essentially universally assumed, 
and great ingenuity has been devoted to postulating the new physics of some kind 
which would be needed to account for the vast number of microstates in the interior 
of a black hole required by (\ref{SBH}). It is not possible to do justice to the 
various approaches in detail here. Curiously the multitude of approaches all seem 
to give the same answer, despite the fact that the states they are counting are
very different \cite{Car}. Of course, any theory that reduces to Einstein's theory in
an appropriate limit will conserve energy and obey the first law of thermodynamics. 
If the effective action of the theory reduces to the Einstein-Hilbert action
then the logarithm of its generating functional $Z$ should produce the entropy (\ref{SBH}),
suggesting that these formal results are actually a feature of the classical theory, 
embodied in the Smarr formula (\ref{diffSm}), and independent of the model used 
or microstates counted.

The lines of research involving exotic internal constituents to obtain $S_{_{BH}}$ 
are all the more remarkable when one recalls where we began, with the classical GR 
expectation that all one has to do is change coordinates as in (\ref{KStrans}) to see
that ``nothing happens" at the event horizon to a particle falling through.  {\it If} 
the horizon is really just a harmless coordinate singularity--the very assumption 
underlying the arguments leading to the Hawking temperature, and hence the entropy 
(\ref{SBH}), how can the semi-classical assumption of no energy or stresses, and 
analyticity and regularity at the geometry at the horizon with no substructure in the
interior then lead to the diametrically opposite conclusion of exotic new physics, 
with $\exp (10^{19})$ additional microstates, and perhaps the radical alteration 
of classical spacetime itself the instant the horizon is reached?

Counting the microstates hiding in electrically charged black holes in string theory 
or other models also leave unanswered the question of how a presumably well-defined 
quantum theory with a stable ground state (which always has a positive heat capacity)
could ever yield the negative heat capacity (\ref{dEdT}) of the original, uncharged 
Schwarzschild black hole in $3+1$ dimensions.

Another line of thought has attempted to identify the entropy (\ref{SBH}) with
quantum entanglement entropy \cite{BKLS,FroFur,IsrBHT}. This is the entropy that results 
when a quantum system is divided into two spatial partitions and one sums over the 
microstates of one of the partitions, forming a mixed state density matrix from
a pure state wave functional even at zero temperature. This has the attractive 
feature that on general grounds it is proportional to the area of the surface 
at which the two partitions are in contact. It has the unattractive feature that 
the coefficient of the area law in quantum field theory is {\it infinite.} The reason 
for this ultraviolet divergence is the same as the reason for the area law itself, 
namely that the largest contribution to the entanglement entropy comes from 
the ultraviolet components of the wave functional within a vanishingly thin layer 
near the surface. If this divergence is cut off at a length scale of the order of 
$L_{Pl}$, a large but finite entropy of order of (\ref{SBHg}) is obtained \cite{FroFur}. 

Counting these states at very high frequencies as contributing to the 
entropy is sensible only if those states are {\it occupied}. In quantum statistical
mechanics the unoccupied states at arbitrarily high energies, no matter how many of 
them, do not contribute to the thermodynamic entropy of the system. In the black hole 
case, the standard classical assumption of the horizon as a harmless coordinate 
singularity and the Hawking-Unruh state corresponding to this classical assumption 
(discussed in more detail in the next section) treats these high frequency states 
as unoccupied {\it vacuum} states with respect to a locally regular coordinate 
freely falling system at the horizon, such as (\ref{KStrans}). Hence it is difficult 
to see why they should be counted as contributing to the entropy. If on the contrary one 
treats these modes as occupied with respect to the singular static frame (\ref{sphsta}), 
then it is difficult to see why their mean energy-momentum or fluctuations in
$\lag T^a_{\ b}\rag$ should be negligible near the horizon. Since the Hawking thermal 
flux originates as radiation closer and closer to $r_{_S}$ with {\it arbitrarily} high 
frequencies at late times, if these states are occupied it is also difficult to see why 
{\it any} cutoff at the Planck scale or otherwise should be imposed to compute the 
entanglement entropy.

The attempt to count microstates near the horizon to account for the
black hole entropy associated with the Hawking effect brings us
face to face with questions about the structure and meaning of the ``vacuum"
itself at {\it trans-Planckian} frequencies. One way or another some
physical input is needed to determine the precise boundary conditions 
on the near horizon modes upon which the entire set of results and physical
consequences for macroscopic black hole physics hinge. At the horizon
the classical supposition that nothing happens at a coordinate singularity 
is in tension with the behavior and assumptions of quantum field theory 
(in a {\it fixed} background) at very high frequencies. This tension is the source 
of the paradoxes, since it is the classical supposition that leads to (\ref{SBHg}) 
and (\ref{TH}) which seem themselves to lead to the {\it opposite} conclusion that
either quantum states at arbitrarily short distances near the horizon are playing
an important physical role (unlike in flat space), or entirely new physics
and degrees of freedom must be invoked to explain black hole entropy,
undermining the semi-classical assumption of mild behavior at the horizon.
Once arbitrarily high frequency modes near the horizon are admitted
into the discussion, one should reconsider whether it is reasonable
to treat gravity classically with the background geometry known and
completely determined and ask whether the near-horizon behavior is 
indeed mild, or whether there might be large quantum backreaction effects 
on the local geometry of spacetime in its vicinity. Notice also that this 
trans-Planckian problem for ultrashort distance modes arises near a black hole 
horizon despite the fact that the horizon radius (\ref{rS}) itself is quite 
macroscopic and very large compared to the Planck scale.

The trans-Planckian problem and the divergence of the entanglement entropy as 
the black hole horizon is approached is also reminiscent of the ultraviolet 
catastrophe of the energy density of classical thermal radiation. 
The cancellation of $\hbar$ from $dE$ in (\ref{dETdS}) is similar to
the cancellation in the energy density of modes of the radiation
field in thermodynamic equilibrium, $\hbar \omega\, n_{_{BE}}(\omega)\omega^2
d \omega \rightarrow k_{_B}T\,\omega^2 d\omega$ in the Rayleigh-Jeans 
limit of very low frequencies, $\hbar \omega \ll k_{_B} T$ where the Bose-Einstein 
distribution $n_{_{BE}}(\omega) = [\exp(\hbar\omega/k_{_B}T) - 1]^{-1}
\rightarrow k_{_B}T/\hbar\omega$. If this low energy relation from classical 
Maxwell theory is improperly extended into the quantum high frequency regime, 
$\hbar \omega \gg k_{_B} T$ it leads to a divergent integral over $\omega$ and 
hence an infinite energy density of the radiation field at any finite temperature. 
This ultraviolet catastrophe and the low temperature thermodynamics of solids
led Planck and Einstein to take the first steps towards a quantum theory of radiation 
and bulk matter \cite{Planck,EincV}. The analogous high frequency divergence of the 
entanglement entropy near a black hole horizon suggests that it results from a similar 
improper extension and misinterpretation of the {\it classical} formula (\ref{diffSm}) 
extended into the high frequency, low temperature regime where quantum effects 
become important.

\subsection{Quantum Fields in Schwarzschild Spacetime}

The preceding discussion indicates that quantum effects, particularly at short distances
need to be treated very carefully when black hole horizons are involved. Given the high 
stakes of the possibility of fundamental revision of the laws of physics and/or vast
numbers of new degrees of freedom and the role of ultrahigh frequency trans-Planckian 
modes to account for the Hawking effect and black hole entropy, it would seem reasonable 
to return to first principles, and re-examine carefully the strictly classical view 
of the event horizon as a harmless kinematic singularity, when $\hbar \neq 0$ and 
the quantum fluctuations of matter are taken into account.

Consider the basic set up of a quantum theory of a scalar field in fixed Schwarschild 
spacetime. Although the locally high energy matter self-interactions and gravitational
self-interactions themselves are almost certainly important near the event horizon, 
we ignore them here in order to simplify the discussion. Then the generally covariant 
free Klein-Gordon eq.,
\be
(-\sq + \mu^2) \Phi = - \frac{1}{\sqrt{-g}}\partial_a \left(\sqrt{-g} g^{ab} \partial_b \Phi\right)
 + \mu^2 \Phi = 0
 \label{KGeq}
\ee
for a scalar field of mass $\mu$ in the static, spherically symmetric Schwarzschild 
geometry (\ref{Sch}) is separable, with eigenfunctions of the form 
\be
\varphi_{\omega \ell m}(t, r, \theta,\phi)= \frac{e^{-i\omega t}}{\sqrt{2 \omega}}
 \, \frac{f_{\omega \ell}(r)}{r} \, Y_{\ell m}(\theta, \phi)\,.
 \label{sepvar}
\ee
Here $Y_{\ell m}$ is a spherical harmonic, and the radial function $f_{\omega \ell}$ 
satisfies the ordinary differential eq.,
\be
\left[- \frac{d^2\ }{dr^{*\,2}} + V_{\ell}\right] \,f_{\omega \ell} = \omega^2\, f_{\omega \ell}\,,
\label{feq}
\ee
in terms of the Regge-Wheeler (``tortoise") radial coordinate,
\be
r^* = r + r_{_S} \, \ln \left( \frac{r}{r_{_S}} - 1\right)\,,
\label{rRW}
\ee
with the potential,
\be
V_{\ell} = \left( 1 - \frac{r_{_S}}{r}\right)\left[ \frac{\ell(\ell + 1)}{r^2} + 
\frac{r_{_S}}{r^3} + \mu^2\right]\,,
\label{Vell}
\ee
which may be viewed as an implicit function of $r^*$ through the relation (\ref{rRW}). 
Note that as $r$ ranges from $r_{_S}$ to $\infty$, $r^*$ ranges over the entire real line 
from $-\infty$ to $+\infty$, and that the potential $V_{\ell}$ vanishes at the lower limit, 
but is otherwise everywhere positive. As a corollary note from (\ref{Vell}) also that 
at the horizon $r=r_{_S}$, the scalar field mass $\mu$ drops out. Since we are interested 
in the near horizon behavior we may concentrate on the massless case and set $\mu =0$.

Eq. (\ref{feq}) defines a standard one dimensional scattering problem, with two 
linearly independent scattering solutions that for $\mu = 0$ have the asymptotic forms, 
$e^{\pm i \omega r^*}$ as $r \rightarrow r_{_S}, r^* \rightarrow - \infty$, and 
as $r, r^* \rightarrow + \infty$. Accordingly, we may define the two fundamental 
linearly independent scattering solutions of (\ref{feq}) $f_{\omega \ell}^{L,R}$ 
by their asymptotic behaviors as \cite{deW}
\bes
\bea
&& f_{\omega \ell}^L  \rightarrow \left\{
\begin{array}{lr} B_{\ell}(\omega) e^{-i\omega r^*} \,, & \quad r^* \rightarrow - \infty\\
e^{-i\omega r^*} + A_{\ell}^R(\omega) e^{i\omega r^*} \,, & \quad r^* \rightarrow + \infty\\
\end{array}\right. \\
&& f_{\omega \ell}^R  \rightarrow \left\{ 
\begin{array}{lr} e^{i\omega r^*} + A_{\ell}^L(\omega) e^{-i\omega r^*} \,,& \quad r^* \rightarrow - \infty\\
B_{\ell}(\omega) e^{i\omega r^*} \,,& \quad  r^* \rightarrow + \infty \end{array} \right.
\eea
\label{scatsoln}\ees
Because of the constancy of the Wronskian associated with eq. (\ref{feq}), the 
reflection and transmission coefficients of (\ref{scatsoln}) obey
\bes
\bea
\vert A_{\ell}^L(\omega)\vert^2 + \vert B_{\ell}(\omega)\vert^2 &=& 
 \vert A_{\ell}^R(\omega)\vert^2 + \vert B_{\ell}(\omega)\vert^2 =1\,,\\
A_{\ell}^{L\, *}(\omega)B_{\ell}(\omega) &=& - A_{\ell}^R(\omega)B^*_{\ell}(\omega)\,,\\
\vert A_{\ell}^L(\omega)\vert  &=& \vert A_{\ell}^R(\omega)\vert\,.
\eea 
\label{ALRB}\ees
Because these two scattering solutions are linearly independent, independent creation
and destruction operators must be introduced for them in the canonical quantization of
the Heisenberg field operator,
\be
\Phi (t, r, \theta, \phi) = \int_0^{\infty} \frac{d\omega}{2 \pi} \sum_{\ell m} 
\sum_{I = L,R} \left(\varphi_{\omega \ell m}^I \, a_{\omega \ell m}^I +  
\varphi_{\omega \ell m}^{I\, *} \, a_{\omega \ell m}^{I\, \dagger}\right)\,.
\label{freqsum}
\ee
The independent canonical commutation relations,
\be
\left[ a_{\omega \ell m}^I, a_{\omega' \ell' m'}^{J\, \dagger}\right]_- = 2 \pi\hbar\, \delta (\omega-\omega')
\delta_{\ell \ell'}\delta_{m m'} \delta^{IJ}\,,
\label{cancom}
\ee 
with each of $I,J$ taking the values $L, R$ enforce the canonical equal time commutation relation,
\be
\left[\Phi(t, r,\theta,\phi), \frac{\partial \Phi}{\partial t}(t, r',\theta',\phi')\right]_-= 
i \hbar\, \frac{\delta (r^* - r^{*\prime})\,\delta(\theta - \theta')\,\delta(\phi-\phi')}{r^2\,\sin\theta}
\label{ETCR}
\ee
on the field, provided the normalization condition
\be
\int_{-\infty}^{\infty}dr^* f^{I*}_{\omega \ell} \ f^J_{\omega' \ell} = 2 \pi\, \delta^{IJ}
\delta (\omega-\omega')
\ee
is satisfied.

From (\ref{scatsoln}) and (\ref{ALRB}) the non-vanishing of the reflection coefficient 
$A^{L,R}_{\ell}$ implies that outgoing spherical waves at the black hole 
horizon are a linear superposition of outgoing and ingoing spherical waves at 
infinity, and similarly for ingoing spherical waves. Notice that this differs from
flat space in spherical coordinates, both in the presence of a scattering potential
and the existence of {\it two} linearly independent regular solutions (\ref{scatsoln})
of the radial wave equation (\ref{feq}). In flat space, $r_{_S} = 0$ and $r^* = r$ so
the corresponding wave equation has a singular point at the origin $r=0$ within
the finite range of the radial variable. This forces one to accept only the solutions 
of (\ref{feq}) which are regular at the origin, namely $r j_{\ell}(kr)$ (with 
$k = \sqrt{\omega^2 - \mu^2}$) and exclude the irregular solutions whose derivatives 
diverge there. The mass $\mu$ does not drop out and there remains a gap between
the positive and negative energy solutions in flat space. 

In contrast, in the Schwarzschild case the change of variables (\ref{feq}) 
and wave eq. (\ref{rRW}) shows that the equation and both its solutions are regular at the 
horizon $r^* \rightarrow -\infty$. The origin $r =0$ is not present within the 
range $-\infty \le r^* \le +\infty$ at all. Hence no particular linear combination of
(\ref{scatsoln}) is preferred {\it a priori}, and we need to retain both solutions.
The frequency integral in (\ref{freqsum}) also extends down to $\omega =0$.
At $\omega =0$ the differential eq. (\ref{feq}) admits solutions which behave 
linearly in $r^*$, hence logarithmically as $\ln( r/r_{_S} - 1)$ near the horizon. 
Whereas modes behaving this way over an infinite domain are excluded by 
initial conditions with compact support, the horizon is a {\it finite} distance
away from any point of fixed $r > r_{_S}$ in the physical Schwarzschild
line element (\ref{sphsta}), and hence these modes are no longer excluded 
{\it a priori}. In several important respects, the radial wave eq. (\ref{feq}) and Hilbert
space spanned by its solutions are discontinuously different in the flat and 
Schwarzschild cases.

Since the static Schwarzschild geometry (\ref{Sch}) approaches ordinary flat space
as $r \rightarrow \infty$ one natural definition of the ``vacuum" would seem to be
the state annihilated by all of the $ a_{\omega \ell m}^I$, {\it viz.,}
\be
a_{\omega \ell m}^L \vert B\rangle = a_{\omega \ell m}^R \vert B\rangle = 0\,.
\label{vacdef}
\ee
This state and its Green's functions were studied in detail by Boulware \cite{Boul} 
and is denoted here by $\vert B \rangle$. If one calculates the expectation value 
of the stress-energy tensor of the massless (conformally coupled) scalar field,
\be
T^a_{\ \ b} = \frac{2}{3}( \nabla^a\Phi)( \nabla_b\Phi)- \frac{1}{6}\,\delta^a_{\ b}\,
(\nabla \Phi)^2 - \frac {1}{3}\, \Phi\, \nabla^a\nabla_b \Phi\,,
\label{Tconf0}
\ee
in the Boulware state, one finds the usual (quartic) divergence of the vacuum energy,
obtained also in flat space, which must be removed, and a finite remainder, which 
vanishes as $r \rightarrow \infty$ (as $r_{_S}^2/r^6$) just as one would expect for 
the Minkowski vacuum far from the black hole. However the renormalized
expectation value of (\ref{Tconf0}) has the property that \cite{ChrFul}
\be
\langle B \vert T^a_{\ b} \vert B \rangle_{_{R}} \rightarrow - \frac{\pi^2}{90} 
\frac{\hbar c}{(4 \pi r_{_S})^4}
\, \left(1 - \frac{r_{_S}}{r}\right)^{-2}\, {\rm diag} \, (-3, 1, 1,1)\,,
\label{TBoul}
\ee
as $r \rightarrow r_{_S}$, where it {\it diverges}. Thus in this Boulware state the 
apparent coordinate singularity of the Schwarzschild horizon is now the locus of {\it arbitarily high}
energy densities. Clearly, if such a state were realized in practice, its stress-energy 
would act as a physical source for the semi-classical Einstein's equations,
\be 
R^a_{\ b} - \frac{R}{2}\, \delta^a_{\ b} + \Lambda\, \delta^a_{\ b} = 8\pi G
\lag T^a_{\ b}\rag_{_R}\,.
\label{scE}
\ee
(with $\Lambda = 0$ here) and necessarily influence the background spacetime 
(\ref{sphsta}) which assumed $T^a_{\ \,b} = 0$. Such large stresses as present in 
(\ref{TBoul}) would cause the solution of (\ref{scE}) to deviate markedly from the 
classical Schwarzschild geometry (\ref{Sch}) near the horizon, and require 
re-evaluation of the entire starting point of the discussion, and certainly the analytic 
continuation (\ref{KStrans}).

The stress-energy in (\ref{TBoul}) is the {\it negative} of that of a scalar field in a thermal 
state at the local blueshifted Tolman-Hawking temperature \cite{TolT},
\be
T_{loc} = T_{_H}\, \left(1 - \frac{r_{_S}}{r}\right)^{-\frac{1}{2}}\,.
\ee
Mathematically, the stress-energy is proportional to an integral over frequencies whose
finite part is proportional to $T_{loc}^4$, and the behavior (\ref{TBoul}) as 
$r \rightarrow r_{_S}$ is obtained. This contribution to the frequency integral is 
dominated by frequencies $\omega \sim c/r_{_S}$, defined by (\ref{sepvar}) with respect 
to the time at infinity, which is the only fixed scale entering the scattering potential 
$V_{\ell}$ in (\ref{Vell}) if $\mu = 0$. (Since near the horizon the potential $V_{\ell}$ 
vanishes and the mass $\mu$ drops out of the leading behavior of $\langle T^a_{\ b}\rangle_{_R}$ 
as $r \rightarrow r_{_S}$, all fields behave essentially as massless fields there in any case). 
The {\it local} frequency of these finite $\omega$ modes becomes arbitrarily large, 
even exceeding the Planck scale on the horizon, which is what leads to the 
divergence in (\ref{TBoul}).

It is clear that the trans-Planckian issue arises because of the infinite
blue shift of frequencies at the event horizon, a necessary consequence
of the gravitational redshift of waves followed backwards to their origin at the
horizon, expressed in the relation (\ref{redshift}). Classically, this
infinite blueshift presents no particular problem, since the energy of
classical waves can be made arbitrarily small, no matter how high their
frequency, simply by making their amplitude small enough. As soon
as $\hbar \neq 0$ (no matter how small), the situation is quite different,
as the amplitude of quantized wave modes is bounded from below by
the Heisenberg uncertainty relation, encoded in the commutation rules
(\ref{cancom})-(\ref{ETCR}). The local energy of the wave mode with
local frequency $\omega_{local}$ is
\be
E_{local} = \hbar \omega_{local} = \hbar \omega \, 
\left(1 - \frac{r_{_S}}{r}\right)^{-\frac{1}{2}} \sim \frac{\hbar c}{r_{_S}}
\, \left(1 - \frac{r_{_S}}{r}\right)^{-\frac{1}{2}}
\label{Eloc}
\ee
which also diverges on the horizon. Since energy-momentum couples
universally to gravity, the very large local vacuum zero point energy 
can affect the geometry there. Let us emphasize that this large effect 
derives from the choice of state in (\ref{vacdef}), and cannot be removed 
by a coordinate transformation, once the state has been specified. 
In the Boulware state the finite vacuum polarization effects and their 
backreaction on the geometry near the horizon are very large in any 
coordinates. 

The relation (\ref{Eloc}) shows that the limits $\hbar \rightarrow 0$ and
$r \rightarrow r_{_S}$ {\it do not commute}. If $\hbar \rightarrow 0$, $E_{local} 
\rightarrow 0$, for all $r > r_{_S}$, and one might entertain the logical possibility 
of analytically continuing the exterior Schwarzschild geometry into the interior 
region, by extending the notion of general coordinate invariance for real, 
differentiable coordinate transformations, $x^{\mu} \rightarrow x^{\prime \mu}(x)$
to complex meromorphic transformations, and get around coordinate singularities 
on the real axis. However, the behavior of the quantum vacuum zero point energy 
near the horizon depends on arbitrarily high local frequencies and is not smooth.
In the Boulware state $\vert B\rangle$ it diverges, (\ref{TBoul}). Hence analytic
continuation around the coordinate singularity there may not be physically justified
in the quantum theory, and certainly not if the matter field is in this state.

Hawking and Unruh argued for a different state in the gravitational collapse problem, 
different from the Boulware state and one which is no longer time symmetric \cite{HawkT,Unr}. 
In that state, denoted by $\vert U \rangle$, only the L ingoing modes are taken to be in 
the vacuum in the first half of (\ref{vacdef}), but the R outgoing modes in the final state 
are thermally populated at infinity with $T= T_{_H}$. The additional finite thermal flux 
has a stress-energy tensor that just cancels the diverging negative energy density 
(\ref{TBoul}) of the Boulware state near the future horizon, $T = +X$ (in any proper 
set of regular coordinates there), and gives a positive flux of Hawking radiation to 
infinity. Indeed, the Unruh state $\vert U \rangle$ is constructed by the requirement 
that its ``vacuum" modes are analytic in the Kruskal null coordinate $U= T - X$ across 
the future horizon. The Hawking flux in the Unruh state may be thought of as bringing the 
quantum expectation value up to its vacuum value at the future horizon where the 
Unruh state is locally similar to the flat space vacuum. This adjustment necessarily 
produces a non-vacuum state of flux of real quanta at infinity. 

In this way, the Hawking-Unruh state maintains the regularity (and smallness) of the 
stress-energy tensor on the future horizon, consistent with the assumption of negligible 
local backreaction of the radiation on the spacetime geometry itself. To obtain this result, 
however, one must use the same set of modes (\ref{scatsoln}) in either case, and follow 
them to arbitrarily large local frequencies near the horizon, with a specific boundary 
condition of analyticity in $U = T- X$ there. This produces an outward Hawking
flux by a compensating negative energy flux through the future horizon and into
the coordinate singularity at $r=0$. In other words, one must assume that ordinary 
quantum field theory and the wave equation (\ref{KGeq}) holds on a {\it fixed} classical 
background geometry with {\it arbitrary accuracy} all the way down distances of the 
order than and even {\it arbitrarily smaller} than the Planck scale $L_{Pl}$, at which
short distances one would normally expect the semi-classical theory to break down 
completely. 

Whereas for example in the Coulomb scattering problem one assumes regular wave functions, 
unitarity and {\it no flux} into or out of the Coulomb singularity (and a formally 
divergent stress-energy density there), in the Schwarzschild case the
Hawking-Unruh state assumes a regular stress-tensor on the future horizon which 
necessarily requires a negative energy flux through it and down to the singularity, 
violating unitarity. Conversely, the Boulware state has no energy flux through the 
horizon but a diverging stress-energy there (\ref{TBoul}). Finally the 
Hartle-Hawking-Israel state is a thermal one for both the L and R modes, and therefore 
has no flux into or out of the singularity, and a regular stress tensor on the horizon, 
but it is a state thermally populated with both ingoing and outgoing quanta even at infinity 
\cite{HarHaw,Isr76,Ein79}. As we already have seen, such a state is thermodynamically unstable. 
Unlike in flat space, there is {\it no} choice of boundary conditions which satisfies 
all three criteria of i) regularity on the future horizon, ii) zero flux there 
(and hence zero flux into the future singularity at $r=0$), and iii) vacuum-like at infinity. 
So an inescapable conclusion is that at least one of these three criteria must be abandoned,
but pure mathematics cannot tell us which.

Despite the apparently thermal expectation values of the two states $\vert B\rangle$ 
and $\vert U\rag$, each are {\it pure} states related to each other by a Bogoliubov 
transformation. In the case of $\vert U \rag$ the apparently thermal emission is
consistent with a pure state because precise quantum mechanical phase correlations 
are set up and maintained between the modes outgoing to infinity and those
infalling into the future singularity. The pure state becomes a mixed thermal state 
if and only if one sums over the modes localized behind the horizon as unobservable
\cite{GibPer,Ful,GibEin79}. Of course, such an averaging procedure entails giving up any hope 
of keeping track of the correlations that might exist between the radiated quanta at 
different times. It is also somewhat paradoxical that although information seems to 
be ``lost" by the pair creation process in which one member of the pair falls into the 
black hole, the mass of the hole and hence the Bekenstein-Hawking entropy (\ref{SBH}) 
is {\it decreased} by the thermal emission. This is very different from a normal thermal 
emission process from a star such as the sun for example. In thermal equilibrium the star's 
radiant energy is supplied by its nuclear interactions in its core, and simply passed outward
at a steady rate. Neither the temperature profile nor the total entropy of the sun changes 
in this steady state process, and the change is its mass is completely negligible. 

Later authors have shown that the stress-tensor for a thermally populated 
state at an arbitrary temperature $T \neq T_{_H}$ behaves like \cite{AndHis}
\be
\langle T^a_{\ b} \rangle_{_{R}} \rightarrow \frac{\pi^2}{90}\,\frac{k_{_B}^4}{(\hbar c)^3}
\, (T^4 - T_{_H}^4)\, \left(1 - \frac{r_{_S}}{r}\right)^{-2}\, {\rm diag} \, (-3, 1, 1,1)\,,
\label{TTemp}
\ee
near the horizon. This divergence and its disappearance if the temperature equals
the Hawking temperature can be understood geometrically. From the Kruskal coordinate
transformation (\ref{KStrans}), we observe that if the Schwarzschild static time coordinate
$t$ is continued to imaginary values, then the resulting Euclidean signature Riemannian
manifold has a {\it conical singularity} at $r=r_{_S}$ unless the Euclidean time
variable is periodic with periodicity an integer multiple of $4 \pi r_{_S}/c \equiv \beta_{_H}$.
Likewise in order to avoid a singularity in the Green's functions and stress-energy
tensor of a quantum field on this background, they too must have this Euclidean
periodicity. This is nothing other than the Kubo-Martin-Schwinger (KMS) Euclidean
periodicity condition \cite{KMS} for the thermal state of a quantum field theory at 
temperature $\hbar/k_{_B} \beta_{_H} = T_{_H}$. In fact this is one way in which
the Hawking temperature was intuitively arrived at in ref. \cite{HarHaw,GibPer}.
If $T \neq T_{_H}$, the Euclidean periodicity of the thermal state does not match 
$2 \pi \hbar/k_{_B}T_{_H}$, and the conical singularity at $r = r_{_S}$ leads 
to the divergence in (\ref{TTemp}).

Although the divergence in the renormalized expectation value is cancelled 
if the modes are populated with a thermal distribution at a temperature 
precisely equal to the Hawking temperature, {\it c.f.} (\ref{TTemp}), our discussion of 
fluctuations in the previous section leads us to expect that even a slight variation 
of the temperature away from the mean value of $\lag T^a_{\ b}\rag$ will 
produce very large fluctuations in the energy density near the horizon. Fluctuations 
are intrinsic to both thermal and quantum theory, and require calculating 
$\lag T^a_{\ b}(x) T^c_{\ d}(x')\rag$, for a linear response treatment of their effect 
on the spacetime for their quantitative analysis \cite{EMfluc}. One would expect 
that the natural time scale for the instability at the horizon to develop 
in such an analysis is the only dynamical timescale available, given by (\ref{deltat}) 
and therefore very rapid. 

The minimal conclusion from these considerations is that the macroscopic quantum 
physics of black holes is quite delicately dependent upon what one assumes for
the population of trans-Planckian frequency modes in the near vicinity of 
the black hole horizon. Depending upon how they are treated, by boundary
conditions, either these ultrahigh frequencies are responsible for the thermal 
evaporation, or they cause the stress-tensor to diverge and can produce significant 
backreaction. The results depend radically on the choice of state, and the
correct physics can be determined only if fluctuations about typical states are 
studied in a systematic way in the collapse problem. To date this investigation 
has not been carried out. In any case it is clear that one cannot maintain 
vacuum boundary conditions at {\it both} the horizon and large distances 
from the black hole. Thus the obstruction to a global vacuum in Schwarzschild 
spacetime has a topological character, related to the possible appearance of
a conical singularity on the horizon and involving singular coordinate 
transformations there. 

Examples of singular coordinate transformations, conical-like singularities and
global obstructions associated with genuine physical effects are known in other areas of 
quantum physics. In classical electromagnetism the gauge potential $A_{\mu}$ is 
unmeasurable locally and may be gauged away. Only the field strength tensor 
$F_{\mu\nu} = \partial_{\mu} A_{\nu} - \partial_{\nu} A_{\mu}$ and functions of it are 
gauge invariant and locally measurable. However, the circulation
integral $\oint_{\cal C} A_{\mu} dx^{\mu}$ is also gauge invariant,
and quantum mechanically therefore so is the phase factor
\be
W({\cal C}) = \exp \left(\frac{i q}{\hbar c}\oint_{\cal C} A_{\mu} dx^{\mu}\right)
\label{Wloop}
\ee
around any closed loop $\cal C$. It measures
the magnetic flux threading the loop. If $W({\cal C}) \neq 1$ , then the gauge
potential cannot be set equal to zero everywhere along $\cal C$ by a {\it regular}
gauge transformation, even if the local electromagnetic field evaluated at all points along 
$\cal C$ vanishes. The singular gauge transformation which is necessary to
set $A_{\mu}=0$ physically corresponds to the creation or destruction of a magnetic
vortex in a superconductor, which would be pointlike and have infinite energy
(were it not for its normal, non-superconducting core of finite radius). The
requirement that the complex valued electron pair density $\lag \Psi\Psi\rag$ be 
single valued around any closed loop leads to flux quantization of magnetic flux 
in superconductors \cite{Lon}, with $q=2e$ for the Cooper pair \cite{BCS,AGD}.

A second example of the physical relevance of the non-local phase factor (\ref{Wloop})
is the Aharonov-Bohm effect \cite{AhaBoh}, which shows that the interference pattern of
electron waves passing around a solenoidal magnetic field confined to a certain
region of space is affected by the presence of the field in the interior enclosed
by the interfering trajectories, even if the electrons' classical trajectories are
confined to the region where the {\it local} field strength tensor $F_{\mu\nu}$ 
{\it vanishes identically}. The non-local gauge invariant phase factor (\ref{Wloop}) has 
physical consequences for interference of electron waves that do not depend upon 
the strength of the local field along the classical electron trajectory.

Both of these physical effects of gauge potentials which can be gauged away locally
but not globally are expressed mathematically by the statement that QED is
properly defined as a theory of a $U(1)$ {\it fiber bundle} over spacetime.
Depending upon the topology of the base manifold of spacetime (for example 
whether or not it is ``punctured" by excising a region where the magnetic field 
of the Aharonov-Bohm solenoid is non-zero), the topology of this fiber bundle 
may be non-trivial and non-local gauge invariant quantities such (\ref{Wloop}) 
can carry information about physical processes.

The global quantum effect of blueshift near a black hole horizon has a
topological aspect which is similar. Although the contractions of the local 
Riemann curvature tensor remain finite in the classical Schwarzschild geometry
(\ref{RieS}), this static geometry has a timelike Killing field with components,
\be
K^a = (1, 0, 0, 0)
\label{Kill}
\ee
in the static coordinates $(t, r, \theta, \phi)$ of (\ref{Sch}). The norm of this
Killing vector is
\be
(-K^aK_a)^{\frac{1}{2}} = \sqrt {-g_{tt}} = f^{\frac{1}{2}}(r)\,,
\label{Knorm}
\ee
exactly the gravitational redshift (blueshift) factor appearing in (\ref{redshift}) or
(\ref{Eloc}), and to the inverse fourth power in (\ref{TBoul}). The quantum state 
of the system, specified in Fourier space by (\ref{vacdef}) retains this global 
information about the infinite blueshift at the horizon relative to the standard of 
time in the asymptotically flat region, $r \rightarrow \infty$, because of the
existence of the global timelike Killing field (\ref{Kill}). This generator of
time translation symmetry has been used in defining the Boulware state 
(\ref{vacdef}) to distinguish positive from negative frequencies, and hence 
distinguish particle-waves from antiparticle-waves in the quantum theory. 
The norm (\ref{Knorm}) is a completely coordinate invariant (albeit non-local) 
scalar quantity, not directly related to the local curvature. Hence there is no 
reason of coordinate invariance that precludes it from having physical effects, 
and in particular, large physical effects at the horizon in a state like $\vert B\rangle$.

The horizon where the norm (\ref{Knorm}) vanishes has topological significance. On 
the Euclidean section $t \rightarrow it$ with $it \rightarrow it + \beta_{_H}$ 
periodically identified as suggested by Hawking and the Unruh state boundary conditions,
the Euler characteristic, defined in terms of the Riemann dual tensor $ ^*\!R_{abcd} 
= \epsilon_{abef}R^{ef}_{\ \ cd}/2$, is
\bea
\chi_{_E} &=& \frac{1}{32\pi^2}\, \int d^4 x \sqrt{g}\  ^*\!R_{abcd} \, ^*\!R^{abcd}\nn\\
&=& \frac{1}{32\pi^2}\, \int d^4 x \sqrt{g}\, \left[R_{abcd}R^{abcd} -4R_{ab}R^{ab} + R^2 \right]\nn\\
&=& \frac{1}{32\pi^2}\,4\pi\beta_{_H} \, \int_{r_{_S}}^\infty \, \frac{12 r_{_S}^2}{r^6}\, r^2 dr
= 2\,,
\label{Eulch}
\eea
where we have used $\beta_{_H} = \hbar/ k_{_B} T_{_H} = 4\pi r_{_S}$ (with $c=1$), (\ref{RieS}) 
and the vacuum Einstein eqs. $R^a_{\ b} = R = 0$ for the Schwarzschild solution. 
The Euclidean Schwarzschild manifold with period $\beta_{_H}$ and $r \ge r_{_S}$ 
has the topology $\mathbb{R}^2 \times \mathbb{S}^2$, unlike flat $\mathbb{R}^4$. This 
is reflected in the Euler characteristic (\ref{Eulch}), and the doubling of regular 
solutions to the radial eq. (\ref{feq}). General theorems in differential geometry 
relate the number of fixed points of a Killing field where the norm (\ref{Knorm}) 
vanishes to the Euler characteristic of the manifold \cite{Bott}, so that $\chi_{_E} = 2$ 
is associated with the vanishing of (\ref{Knorm}) at $r=r_{_S}$. A periodicity condition 
on the orbits of the Killing field (\ref{Kill}), particularly in the complexified domain, 
in order to eliminate the conical singularity at $r_{_S}$ when the period of $it$ is 
different than $\beta_{_H}$ is completely non-trivial at the quantum level and an 
{\it a priori} unwarranted assumption. It is analogous to assuming the triviality of the 
$U(1)$ bundle and of the phase factor (\ref{Wloop}), which would miss genuine physical 
effects such as Abrikosov vortices and the quantization of circulation and magnetic
flux in superfluids and superconductors \cite{AGD}, and the Bohm-Aharanov effect in QED \cite{AhaBoh}.

In the gravitational case the possible non-triviality of the $GL(4)$ bundle of
General Relativity is encoded in the the fact that the Euler density in (\ref{Eulch})
can be expressed as the total divergence of a frame dependent topological
current \cite{Chern}, dual to an anti-symmetric $3$-form gauge potential, 
\be
^*\!R_{abcd} \, ^*\!R^{abcd} = \nabla_a\Omega^a = \nabla_a (\varepsilon^{abcd}A_{bcd})\,,
\label{topcurr}
\ee
and the surface integral of this gauge potential over a closed bounding $3$-surface,
\be
\oint_{\Sigma} \varepsilon^{abcd} A_{abc}\,d\Sigma_d
\label{tform}
\ee
is coordinate (gauge) invariant, under the gauge transformation,
\be
A_{abc} \rightarrow A_{abc} + \nabla_{[a}\theta_{bc]}\,.
\ee
Thus if $\Sigma$ is the $\mathbb{S}^2 \otimes \mathbb{R}$ tube at fixed $r$ from $t_1$ to $t_2$,
the integral (\ref{tform}) measures the topological Gauss-Bonnet-Chern charge residing
in the interior of the tube, much as the circulation integral in the exponent of 
(\ref{Wloop}) measures the magnetic flux from a superconducting vortex or an 
Aharonov-Bohn solenoid threading its interior.

Concerning the Riemann tensor itself, we note that in the general static metric
of (\ref{sphsta}), the tensor component
\be
R^{tr}_{\ \ tr} = \frac{h}{4} \left( \frac{f^{\prime 2}}{f^2} - \frac{2f''}{f} - 
\frac{h'}{h}\frac{f'}{f}\right)
\label{Riem}
\ee
(where primes denote differentiation with respect to $r$) becomes $-f''/2$ and hence 
remains finite when the two functions $f=h$ are equal. However, Einstein's eqs. in the 
static geometry,
\bes\bea
&&-G_{\ t}^t = \frac{1}{r^2} \frac{d}{dr}\left[r\left(1 - h\right)\right]
= -8\pi G\, T_{\ t}^t = 8\pi G \rho \label{Einsa}\,,\\
&& G_{\ r}^r = \frac{h} {r f} \frac{d f}{dr}  + \frac{1}{r^2} \left(h -1\right) =
8\pi G \,T_{\ r}^r = 8\pi G p\,,
\label{Einsb}\eea\label{Einsab}\ees
give
\be
\frac{d}{dr} \left(\frac{h}{f}\right) = - 8 \pi G (\rho + p) \frac{r}{f}\,.
\ee
Hence if the non-vacuum stress-energy has $\rho + p > 0$ in the region where $f$
or $h$ vanishes, in general $h \neq f$ and the cancellation of the singularities at
$h=f=0$, special for static vacuum solutions to Einstein's eqs., will not occur.
Any perturbation of the vacuum Schwarzschild spacetime with $\rho + p \neq 0$ 
in a static frame in the vicinity of the horizon has the potential to produce Riemann 
tensor perturbations, $\delta R^{tr}_{\ \ tr} \sim (r_{_S}^2 f)^{-1}$, which are generically 
large at the horizon, where $f \rightarrow 0$, and thus will produce a large change in
the local geometry there. Further, the eq. of stress-energy conservation in a static, 
spherically symmetric spacetime,
\be
\nabla_aT^a_{\ \,r}= \frac{dp}{dr} + \frac{(\rho + p)}{2f} \frac{df}{dr} 
+ \frac{2\,(p - p_{\perp})}{r} = 0\,,
\label{cons}
\ee
(where $p_{\perp} = T^\theta_{\ \theta} = T^{\phi}_{\ \phi}$ is the transverse pressure
while $p = T^r_{\ r}$ is the radial pressure). This shows that any matter with the effective 
eq. of state $p = p_{\perp} = w \rho$ must have a stress-energy which behaves like 
$f^{-(1+ w)/2w}$, which diverges on the horizon if $w > 0$. Note that this is consistent 
with (\ref{TBoul}) for $w = \frac{1}{3}$ producing a conical singularity.

From these various considerations we conclude that the cancellation of divergences 
on a black hole event horizon is an extremely delicate matter, and there is no reason
to expect it to be generic in quantum theory. This does not
violate the Principle of Equivalence, if that principle is understood to be embodied
in the requirement that physics should respect general coordinate invariance 
under all {\it real, non-singular} coordinate transformations. Singular coordinate 
transformations are another matter. 

Mathematically, gauging away coordinate singularities on the horizon amounts 
to an additional assumption about the triviality of the $GL(4)$ 
frame bundle of General Relativity, which may not be warranted.
Experimentally well-established applications of quantum theory already teach us 
by a number of examples in other fields that such {\it improper} ``gauge" 
transformations generally contain new physical effects. Physically these effects 
are associated with the quantum wavelike nature of matter which cannot be 
idealized as arbitrarily small, isolated pointlike particles, particularly
in gravity where such an extreme local limit must produce infinite energies. 
Instead, matter fields obey wave equations such as (\ref{KGeq}) which require 
boundary conditions for their full solution. In quantum theory the dependence 
of local physics on these boundary conditions, through specification of the 
quantum state of the system as in (\ref{vacdef}) does not violate the Principle 
of Equivalence. It is only our uncritical and unexamined classical notions of 
strict and absolute locality that are violated in known quantum phenomena 
such as entanglement and macroscopic coherence. Expectation values of the 
stress-energy tensor $T^a_{\ b}$ can perfectly well depend on non-local 
invariants such as (\ref{Knorm}) and (\ref{tform}), analogous to (\ref{Wloop}) 
in QED. A new set of calculation tools is needed in order to determine these 
quantum effects in a systematic way, and bring gravity into 
accordance with general quantum principles on macroscopic scales.
This is what we seek to provide in the succeeding sections.

\section{The Challenge of Cosmological Dark Energy}
\label{sec:darkenergy}

The second challenge for macroscopic quantum effects in gravity arise on the
cosmological scale of the Hubble expansion itself, and in particular
upon the discovery of the acceleration of the expansion rate of the
universe.

\subsection{The Cosmological Constant and Energy of the Vacuum}

In classical General Relativity, the requirement that the field equations 
involve no more than two derivatives of the metric tensor allows for the 
possible addition of a constant term, the cosmological term $\Lambda$, 
to Einstein's equations,
\be
R^a_{\ b} - \frac{R}{2} \,\delta^a_{\ b} + \Lambda \,\delta^a_{\ b} 
= \frac{8\pi G}{c^4}\, T^a_{\ b}.
\label{Ein}
\ee
If transposed to the right side of this relation, the $\Lambda$ term corresponds 
to a constant energy density $\rho_{\Lambda} = c^4\Lambda /8\pi G$ and isotropic 
pressure $p_{\Lambda} = - c^4\Lambda/8\pi G$ permeating all of space uniformly, 
and independently of any localized matter sources. Hence, even if the
matter $T^a_{\ b} = 0$, a cosmological term causes spacetime to become 
curved with a radius of curvature of order $\vert\Lambda\vert^{-\frac{1}{2}}$.

With $\Lambda = 0$ there is no fixed length scale in the vacuum Einstein 
equations, $G/c^4$ being simply a conversion factor between the units of 
energy and those of length. Hence in purely classical physics there is no 
natural fundamental length scale to which $\Lambda$ can be compared,
and  $\Lambda$ may take on any value whatsoever with no 
difficulty (and with no explanation) in classical General Relativity. 

As soon as we allow $\hbar \neq 0$, there is a quantity with the dimensions 
of length that can be formed from  $\hbar, G$, and $c$, namely the Planck length
(\ref{LPl}). Hence in quantum theory the quantity
\be
\lambda \equiv \Lambda L_{pl}^2 = \frac{\hbar G \Lambda}{c^3}
\label{Lnum}
\ee
becomes a dimensionless pure number, whose value one might expect 
a theory of gravity incorporating quantum effects to address. Notice
that like the effects we have been considering in black hole physics
this quantity requires {\it both} $\hbar$ and $G$ to be different than
zero.

Some eighty years ago W. Pauli was apparently the first to consider the
question of the effects of quantum vacuum fluctuations on the the curvature 
of space \cite{Pau}. Pauli recognized that the sum of zero point energies of 
the two transverse electromagnetic field modes {\it in vacuo},
\be
\rho_{\Lambda} = 2 \int^{L_{min}^{-1}}\frac{d^3 {\bf k}}{(2\pi)^3} 
\frac{\hbar \omega_{\bf k}}{2}
 = \frac{1}{8\pi^2} \frac{\hbar c}{L_{min}^{\ \ 4}} = -p_{\Lambda}
\label{zeropt}
\ee
contribute to the stress-energy tensor of Einstein's theory as would an 
effective cosmological term $\Lambda > 0$. Since the integral (\ref{zeropt}) 
is quartically divergent, an ultraviolet cutoff $L_{min}^{-1}$ of (\ref{zeropt}) 
at large $\bf k$ is needed. Taking this short distance cutoff $L_{min}$ to be of 
the order of the classical electron radius $e^2/mc^2$, Pauli concluded that if 
his estimate were correct, Einstein's theory with this large a $\Lambda$ would 
lead to a universe so curved that its total size ``could not even reach to 
the moon" \cite{Pau}. If instead of the classical electron radius, the apparently natural 
but much shorter length scale of $L_{min} \sim L_{pl}$ is used to cut off the 
frequency sum in (\ref{zeropt}), then the estimate for the cosmological term in 
Einstein's equations becomes vastly larger, and the entire universe would be 
limited in size to the microscopic scale of $L_{pl}$ itself, in even 
more striking disagreement with observation.

Clearly Pauli's estimate of the contribution of short distance modes of
the electromagnetic field to the curvature of space, by using (\ref{zeropt}) 
as a source for Einstein's eqs. (\ref{Ein}) is wrong. The question is why. 
Here the Casimir effect may have something to teach us \cite{Cas}. The vacuum zero 
point fluctuations being considered in (\ref{zeropt}) are the same ones that contribute 
to the Casimir effect, but this estimate of the scale of vacuum zero point energy, 
quartically dependent on a short distance cutoff $L_{min}$, is certainly {\it not} 
relevant for the effect observed in the laboratory. In calculations of the Casimir 
force between conductors, one subtracts the zero point energy of the electromagnetic 
field in an infinitely extended vacuum (with the conductors absent) from the 
modified zero point energies in the presence of the conductors. It is this 
{\it subtracted} zero point energy of the electromagnetic vacuum, depending upon 
the {\it boundary conditions} imposed by the conducting surfaces, which leads 
to experimentally well verified results for the force between the conductors. In 
this renormalization procedure the ultraviolet cutoff $L_{min}^{-1}$ drops out, 
and the distance scale of quantum fluctuations that determine the magnitude of the 
Casimir effect is not the microscopic classical electron radius, as in Pauli's original 
estimate, nor much less the even more microscopic Planck length $L_{pl}$, but 
rather the relatively {\it macroscopic} distance $d$ between the conducting 
boundary surfaces. The resulting subtracted energy density of the vacuum between
the conductors is \cite{Cas}
\be
\rho_v = -\frac{\pi^2}{720}\, \frac{\hbar c}{d^4} \,.
\label{Casimir}
\ee
This energy density is of the opposite sign as (\ref{zeropt}), leading 
to an attractive force per unit area between the plates of 
$0.013$ dyne/cm$^2$ $(\mu m/d)^4$, a value which is both independent of the 
ultraviolet cutoff $L_{min}^{-1}$, and the microscopic details of the atomic 
constituents of the conductors. This is a clear indication, confirmed 
by experiment, that the {\it measurable} effects associated with 
vacuum fluctuations are {\it infrared} phenomena, dependent upon 
macroscopic boundary conditions, which have little or nothing to do 
with the extreme ultraviolet modes in (\ref{zeropt}).

Actually, the original Casimir calculation of the force between exactly
parallel flat plates of infinite conductivity hides some important
features of the general case. As soon as the conducting plates
have any finite curvature, the local stress-energy tensor {\it diverges}
on the boundary. A general classification of these divergences has been
given in Ref. \cite{CanDeu}. This shows that in the presence of
curved boundaries, there is residual sensitivity to ultraviolet
effects, much as in the divergence of the stress-energy tensor
(\ref{TBoul}) in the Boulware state in the Schwarzschild geometry.
By now the meaning of these divergences and the correct physical
method of handling them have been understood \cite{BorMohMil}. The
divergences are artifacts of the theoretical over-idealization of
the conductors as perfect at arbitrarily high frequencies. 
Any real metal has a finite conductivity which leads to
a finite skin depth and vanishing reflection coefficient at 
arbitrarily high frequencies. When the effects of finite conductivity
of real metals such as gold used in the experiments are taken into
account, all local stresses and energy densities are finite, and
the theoretical predictions in accord with experiment \cite{BorMohMil}.
Thus although the Casimir effect itself and in its original form (\ref{Casimir})
is a macroscopic effect of vacuum zero point energy, the details
do retain some sensitivity to the physics of the surface, which is
short distance compared to the macroscopic separation $d$ (although
of course completely unrelated to an ultrashort distance cutoff on
the scale of the electron radius or $L_{Pl}$).

By the Equivalence Principle, local ultrashort distance behavior in a mildly
curved spacetime is essentially equivalent to that in flat spacetime.
Hence on physical grounds we should not expect the extreme ultraviolet 
cutoff dependence of (\ref{zeropt}) to affect the universe in the large
any more than it affects the force between metallic conductors in the 
laboratory, although any possible surface boundary effects will have to be 
treated carefully.

In the case of the Casimir effect a constant zero point energy of the
vacuum, no matter how large, does not affect the force between the plates.
In the case of cosmology it is usually taken for granted that any effects of
boundary conditions can be neglected.  It is not obvious then what should 
play the role of the conducting plates in determining the magnitude of 
$\rho_v$ in the universe, and the magnitude of any effect of quantum zero 
point energy on the curvature of space has remained unclear from Pauli's 
original estimate down to the present. In recent years this has evolved from 
a question of fundamental importance in theoretical physics to a central one 
of observational cosmology as well. Observations of type Ia supernovae at 
moderately large redshifts ($z\sim 0.5$ to $1$) have led to the conclusion 
that the Hubble expansion of the universe is {\it accelerating} \cite{SNI}. 
According to Einstein's equations this acceleration is possible if and only 
if the energy density and pressure of the dominant component of the universe 
satisfy the inequality, 
\be
\rho + 3 p \equiv \rho\  (1 + 3 w) < 0\,.
\label{accond}
\ee
A vacuum energy with $\rho > 0$ and $w\equiv p_v/\rho_v = -1$ leads to an 
accelerated expansion, a kind of ``repulsive" gravity in which the relativistic
effects of a negative pressure can overcome a positive energy density in
(\ref{accond}). Taken at face value, the observations imply that some $74\%$ of 
the energy in the universe is of this hitherto undetected $w=-1$ dark 
variety. This leads to a non-zero inferred cosmological term in Einstein's 
equations of 
\be
\Lambda_{\rm meas} \simeq (0.74)\, \frac{3 H_0^2}{c^2} 
\simeq 1.4 \times 10^{-56}\ {\rm cm}^{-2}
\simeq  3.6 \times 10^{-122}\ \frac{c^3}{\hbar G}\,.
\label{cosmeas}
\ee
Here $H_0$ is the present value of the Hubble parameter, approximately 
$73$ km/sec/Mpc  $\simeq 2.4 \times 10^{-18}\, {\rm sec}^{-1}$. The last 
number in (\ref{cosmeas}) expresses the value of the cosmological dark 
energy inferred from the SN Ia data in terms of Planck units, 
$L_{\rm pl}^{-2} = \frac{c^3}{\hbar G}$, {\it i.e.} the dimensionless
number in (\ref{Lnum}) has the value 
\be
\lambda \simeq 3.6 \times 10^{-122}\,.
\label{lmeas}
\ee 
Explaining the value of this smallest number in all of physics is the
basic form of the {\it cosmological constant problem.}

\subsection{Classical de Sitter Spacetime}

Just as in our discussion of black hole physics in Sec. \ref{sec:BH}, we 
begin our discussion of cosmological dark energy with a review of the 
simplest classical spacetime that serves as the stage for discussion of 
quantum effects of a cosmological term. The maximally symmetric solution 
to the classical Einstein equations  (\ref{Ein}) with a positive cosmological
term is the one found by W. de Sitter in 1917 \cite{deSs}. In fact, de Sitter 
expressed his solution of the vacuum Einstein eqs. (\ref{Ein}) with $\Lambda > 0$
and $T^a_{\ b} = 0$ in a static, spherically symmetric form analogous to that
of the Schwarzschild solution (\ref{sphsta}). In those coordinates de Sitter's
solution can be presented as
\be
f(r) = h(r) = 1 - H^2 r^2 = 1 - \frac{r^2}{r_{_H}^2}\,,
\label{fhdS}
\ee
instead of (\ref{Sch}). The length scale $r_{_H}$ is related to the cosmological term by
\be
r_{_H} = \frac{1}{H} = \sqrt{\frac{3}{\Lambda}}\,.
\ee
In this spherically symmetric static form, the de Sitter metric has rotational $O(3)$
symmetry about the point $r=0$ and an apparent horizon at $r=r_{_H}$.

Actually the de Sitter solution has a larger $O(4,1)$ symmetry which can be
made manifest in its global analytic extension, analagous to the $T,X$
Kruskal-Szekeres coordinates in the Schwarzschild case. Consider a
{\it five} dimensional Minkowski space with the standard flat metric,
\be
ds^2 = \eta_{_{AB}}\,dX^{\scriptscriptstyle A}\,dX^{\scriptscriptstyle B} =
 -dT^2 + dW^2 + dX^2 + dY^2 + dZ^2\,,
\label{fiveM}
\ee
subject to the condition,
\be
\eta_{_{AB}}\,X^{\scriptscriptstyle A}\, X^{\scriptscriptstyle B} 
= -T^2 + W^2 + X^2 + Y^2 + Z^2 
= r_{_H}^2 = H^{-2}\,.
\label{TWXYZcond}
\ee
with the indices $A, B = 0, 1, 2, 3, 4$ raised and lowered with the five dimensional
Minkowski metric $\eta_{\scriptscriptstyle AB} =$ diag $(-1, 1, 1, 1, 1)$.
Here and henceforward we generally set the speed of light $c=1$, except when
needed for emphasis.

\begin{figure}
\begin{center}
\includegraphics[height=6cm,width=6cm]{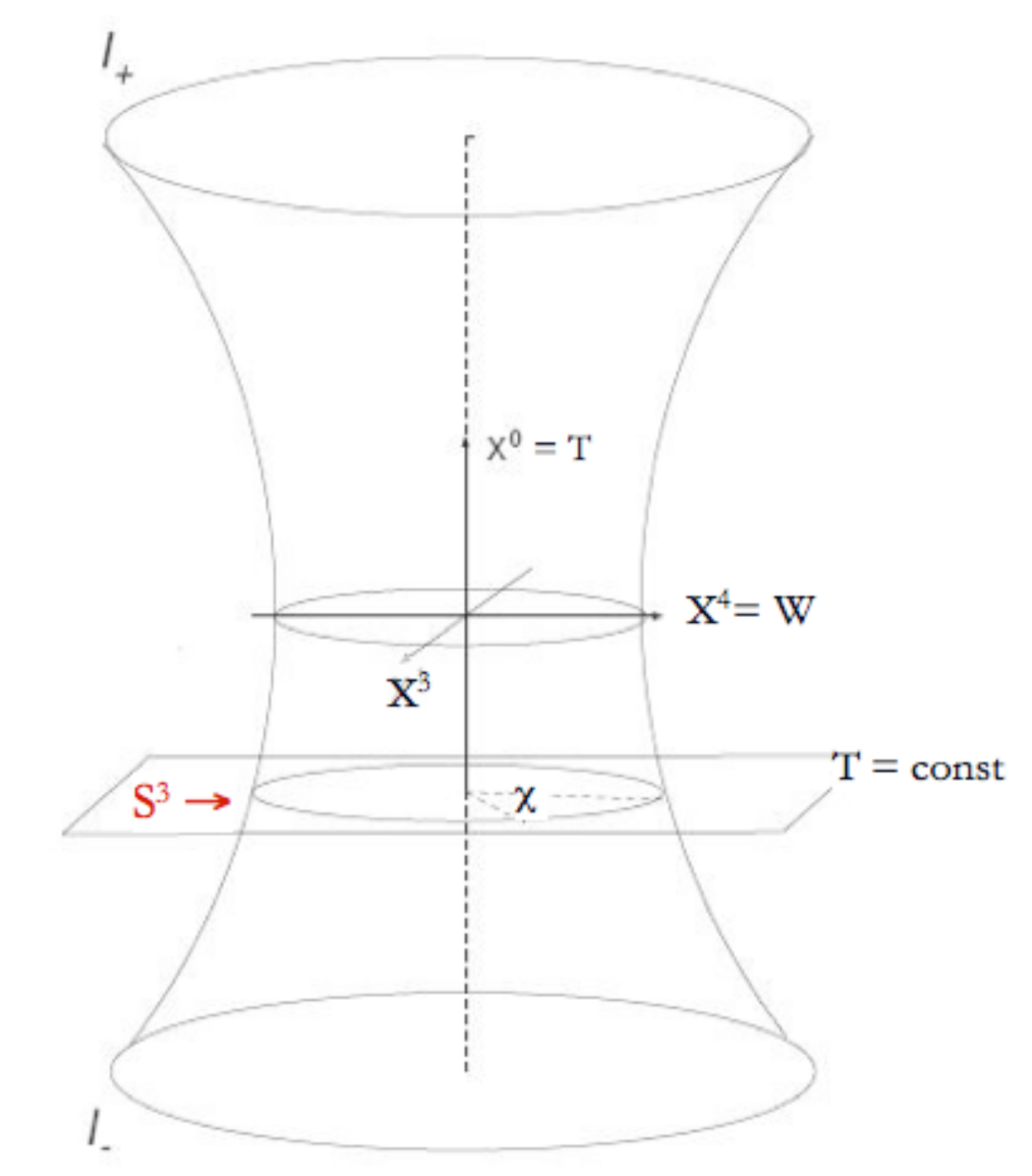}
\caption{The de Sitter manifold represented as a single sheeted hyperboloid of revolution
about the $T$ axis, in which the $X^1$, $X^2$ coordinates are suppressed. The hypersurfaces 
at constant $T$ are three-spheres, $\mathbb{S}^3$. The three-spheres at $T= \pm \infty$ are 
denoted by $I_{\pm}$.}
\label{Fig:deShyper}
\end{center}
\vspace{-7mm}
\end{figure}

If $T$ were a spacelike coordinate rather than timelike so that the four dimensional
manifold defined by (\ref{fiveM}) and (\ref{TWXYZcond}) had a Euclidean signature,
we would clearly be discussing a four-sphere $\mathbb{S}^4$ with an $O(5)$ symmetry 
group. Because of the Lorentzian signature, these relations instead define a single sheeted
hyperbolid of revolution, depicted in Fig. \ref{Fig:deShyper}, with the non-compact 
symmetry group $O(4,1)$,  the maximal possible symmetry for any solution 
of the vacuum Einstein field equations (\ref{Ein}) with a positive cosmological term 
and $T^a_{\ b} = 0$ in four dimensions. 

The curvature tensor of de Sitter space satisfies
\bes
\bea
R^{ab}_{\ \ cd} &=& H^2 \left(\delta^a_{\ c}\,\delta^b_{\ d} - 
\delta^a_{\ d}\,\delta^b_{\ c}\right)\,,\\
R^a_{\ b} &=& 3 H^2\, \delta^a_{\ b}\,,\\
R &=& 12 H^2\,,
\eea
\ees
\noindent
The Lie algebra $so(4,1)$ is generated by the generators of the Lorentz group
in the $4+1$ dimensional flat embedding spacetime (\ref{fiveM}). In the coordinate
(adjoint) representation the $10$ anti-Hermitian generators of this symmetry are
\be
L_{\scriptscriptstyle AB} = X_{\scriptscriptstyle A} \frac{\partial}
{\partial X^{\scriptscriptstyle B}} - X_{\scriptscriptstyle B}
\frac{\partial}{\partial X^{\scriptscriptstyle A}} = - L_{\scriptscriptstyle BA}
\ee
These $10$ generators satisfy
\be
[L_{\scriptscriptstyle AB}, L_{\scriptscriptstyle CD}] = 
-\eta_{\scriptscriptstyle AC} \, L_{\scriptscriptstyle BD} + 
\eta_{\scriptscriptstyle BC} \, L_{\scriptscriptstyle AD} - 
\eta_{\scriptscriptstyle BD} \, L_{\scriptscriptstyle AC}  +
 \eta_{\scriptscriptstyle AD} \, L_{\scriptscriptstyle BC} \,,
\ee
the Lie algebra $so(4,1)$. De Sitter space has $10$ Killing vectors corresponding 
to these $10$ generators.

The hyperbolic coordinates of de Sitter space are defined by
\bes
\bea
T &=& \frac{1}{H}\, \sinh u\,,\\
W &=&  \frac{1}{H}\, \cosh u\,\cos\chi\,,\\
X^i &=& \frac{1}{H}\, \cosh u\,\sin\chi\, \hat n^i\,,\qquad i =1, 2, 3\,.
\eea
\label{hypercoor}
\ees

\noindent
where
\be
\hat n = (\sin\theta\,\cos\phi\,,\sin\theta\,\sin\phi\,,\cos\theta)
\ee
is the unit vector on $\mathbb{S}^2$, and cast the de Sitter line element in the form
\be
ds^2 = \frac{1}{H^2} \left[ -du^2 + \cosh^2 u\, (d\chi^2 + \sin^2\chi\, d\Omega^2)\right]\,.
\label{hypermet}
\ee
The quantity in round parentheses is
\be
d\Omega_3^2 \equiv [d (\sin\chi\, \hat n^i)]^2 + [d (\cos\chi)]^2 
= d\chi^2 + \sin^2\chi\, d\Omega^2\,,
\ee
the standard round metric on $\mathbb{S}^3$. Hence in the geodesically complete
coordinates of (\ref{hypercoor}) the de Sitter line element (\ref{hypermet}) is an
hyperboloid of revolution whose constant $u$ sections are three spheres, 
represented in Fig. \ref{Fig:deShyper}, which are invariant under the $O(4)$ 
subgroup of $O(4,1)$. It is sometimes convenient to define the hyperbolic 
conformal time coordinate $\upsilon$ by $\sec \upsilon = \cosh u$, so that 
(\ref{hypermet}) becomes
\be
ds^2 =  H^{-2}\sec^2\upsilon \left( -d\upsilon^2 + d\Omega_3^2\right)\,,
\ee
conformal to the Einstein static cylinder with $-\pi/2 \le \upsilon \le \pi/2$.

In cosmology it is more common to use instead the Friedmann-Lema\^itre-Robertson-Walker 
(FLRW) line element with flat $\mathbb{R}^3$ spatial sections, {\it viz.}
\bea
ds^2 &=& - d\tau^2 + a^2(\tau)\, d\vec x\cdot d\vec x = 
- d\tau^2 + a^2(\tau)\, (dx^2 + dy^2 + dz^2)\nn\\
&=& - d\tau^2 + a^2(\tau)(d\varrho^2 + \varrho^2 d\Omega^2)\,.
\label{FLRW}
\eea
De Sitter space can be brought in the FLRW form by setting
\bes
\bea
T &=& \frac{1}{2H}\left ( a - \frac{1}{a} \right) + \frac{Ha}{2}\, \varrho^2\,,\\
W &=&  \frac{1}{2H}\left ( a + \frac{1}{a} \right) - \frac{Ha}{2}\, \varrho^2\,,\\
X^i &=& a \,\varrho \,\hat n^i \,,
\eea
\label{RWcoor}
\ees
with
\bes
\bea
a (\tau) &=& e^{H\tau} \label{deSexp}\\
\varrho &=& |\vec x| = \sqrt{x^2 + y^2 + z^2}\,.
\eea\label{rhotau}\ees
\noindent
From (\ref{RWcoor}) and (\ref{rhotau}), $T + W \ge 0$ in these coordinates. 
Hence the flat FLRW coordinates cover only one half of the full de Sitter 
hyperboloid, with the hypersurfaces of constant RW time $\tau$ slicing the 
hyperboloid in Fig. \ref{Fig:deShyper} at a $45^{\circ}$ angle.

The change of time variable to the conformal time coordinate,
\be
\eta = - H^{-1} e^{-H\tau} = -\frac{1}{Ha}\,,\qquad a = - \frac{1}{H\eta}\,,
\ee
is also often used to express the de Sitter line element is the conformally flat form
\be
ds^2 = a^2 \, (- d\eta^2 + d\vec x^2) = \frac{1}{H^2\eta^2}\, (- d\eta^2 + d\vec x^2)\,.
\label{conflat}
\ee
>From (\ref{hypercoor}) and (\ref{RWcoor}), 
\bes
\bea
\cosh u\,\sin\chi = H\varrho\, a &=&  -\frac{\varrho}{\eta}\,,\\
\sinh u + \cosh u\,\cos\chi = a &=&  -\frac{1}{H\eta}\,,
\eea
\ees
\noindent
which gives the direct relation between hyperbolic coordinates and flat FLRW coordinates.

The de Sitter static coordinates $(t, r, \theta, \phi)$ are defined by
\bes
\bea
T &=& \frac{1}{H}\, \sqrt{1 - H^2r^2}\,\sinh(Ht)\,,\\
W &=& \frac{1}{H}\, \sqrt{1 - H^2r^2}\,\cosh(Ht)\,\\
X^i &=& r\, \hat n^i\,.
\eea
\label{staticoor}
\ees

\begin{figure}
\begin{center}
\includegraphics[height=6cm,width=6cm]{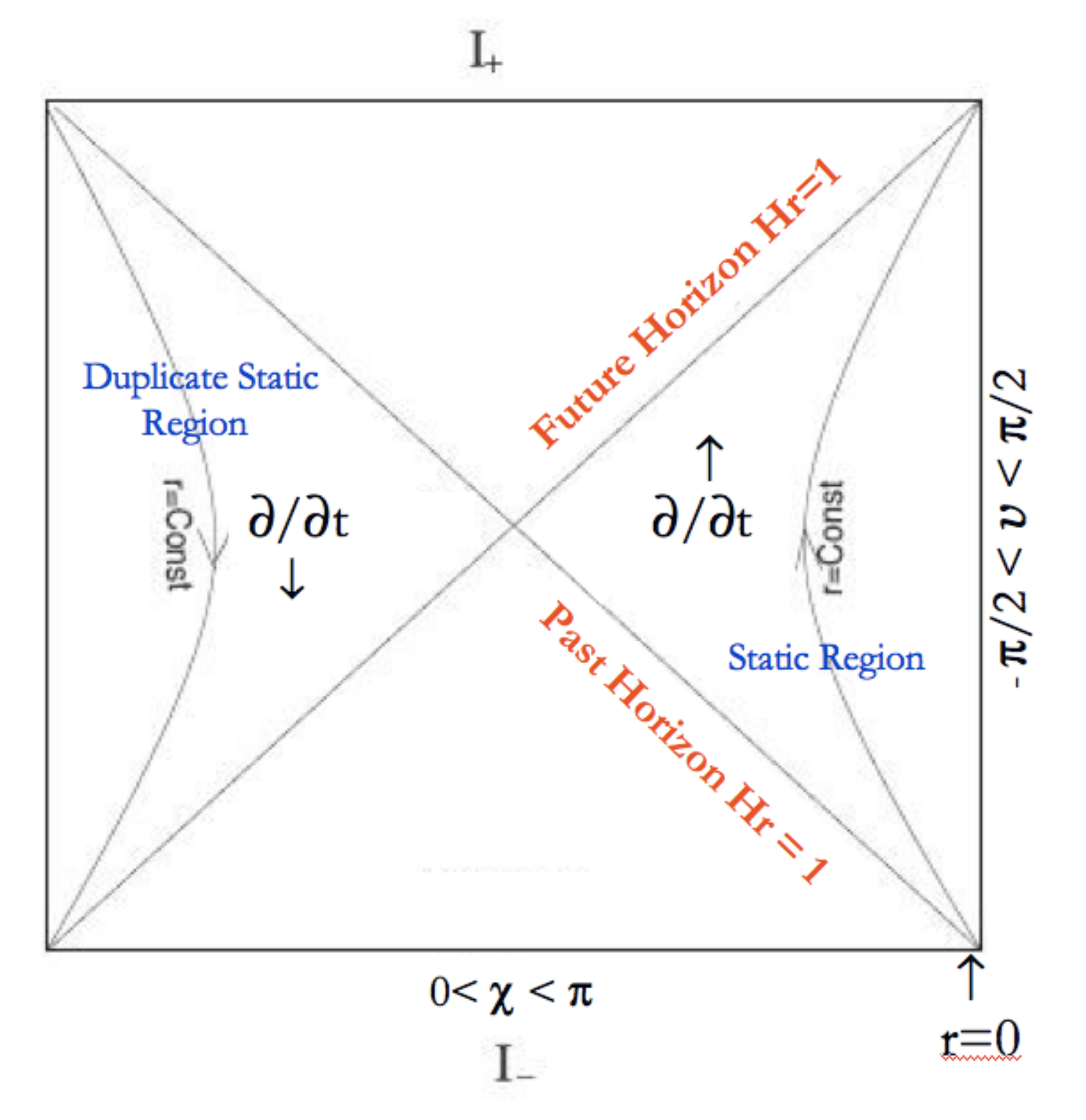}
\caption{The Carter-Penrose conformal diagram for de Sitter space. Future and past infinity 
are at $I_{\pm}$. Only the quarter of the diagram labeled as the static region are covered by
the static coordinates of (\ref{dSstat}). The orbits of the static time Killing field
$\partial/\partial t$ are shown. The angular coordinates $\theta,\phi$ are again suppressed.}
\label{Fig:dSCarPen}
\end{center}
\vspace{-7mm}
\end{figure}

\noindent
They bring the line element (\ref{fiveM}) into the static, spherically symmetric 
form (\ref{sphsta}) with (\ref{fhdS}), {\it i.e.}
\be
ds^2 = - (1-H^2 r^2)\, dt^2 + \frac{dr^2}{(1-H^2r^2)} + r^2 d \Omega^2\,.
\label{dSstat}
\ee
Just as in the Schwarzschild case (\ref{Sch}) these static coordinates 
cover only part of the analytically fully extended de Sitter manifold (\ref{TWXYZcond}).
From (\ref{staticoor}), real static $(t, r)$ coordinates cover only the
quarter of the de Sitter manifold where $W \ge 0$ and both $W \pm T \ge 0$.
This quarter is represented as the rightmost wedge of the Carter-Penrose
conformal diagram of de Sitter space in Fig. \ref{Fig:dSCarPen}.

The Regge-Wheeler radial coordinate $r^*$ can be defined in the static frame by
\bes
\bea
r^*  &=& \frac{1}{2H}\, \ln\left(\frac{1+ Hr}{1 - Hr}\right) = \frac{1}{H} \, \tanh^{-1} (Hr)\,,\\
r &=&  \frac{1}{H} \, \tanh (Hr^*)\,\qquad {\rm so\ that}\\
dr^* &=& \frac{dr}{1-H^2r^2}\,,\qquad \sqrt{1 - H^2r^2}=  {\rm sech} (Hr^*)\,,
\eea
\label{RegWhdS}
\ees
and
\bea
ds^2 &=&{\rm sech}^2 (Hr^*)\, (-dt^2 + dr^{*\,2}) 
+ \frac{1}{H^2} \tanh^2(Hr^*)\, d\Omega^2\nn\\
&=& \frac{1}{H^2}\, {\rm sech}^2 (Hr^*)\,\left[ -H^2dt^2 + H^2dr^{*\,2} + 
\sinh^2(Hr^*) \,d\Omega^2\right].\qquad
\label{optmet}
\eea
Note that the horizon at $r = H^{-1} = r_{_H}$ is mapped to $r^* = \infty$ in these coordinates, and 
that the spatial part of the line element (\ref{optmet}) in square brackets (the ``optical" metric) is
\be
ds^2_{opt} = H^2dr^{*\,2} + \sinh^2(Hr^*) \,d\Omega^2 
=4\,\frac{d\vec {\sf y} \cdot d\vec {\sf y}}{(1 - {\sf y}^2)^2} 
\label{Lob}
\ee
where the second form is obtained by defining
\bes
\bea
\vec {\sf y}&\equiv& {\sf y} \, \hat n\,,\\
{\sf y} &\equiv & \tanh \left(\frac{Hr^*}{2}\right) = \frac{Hr}{1 + \sqrt{1-H^2r^2}} 
\quad {\rm so\  that}\\
r&= &\frac{2}{H}\frac{\sf y}{1+ {\sf y}^2}\,,\qquad r^* = \frac{1}{H} \,
\ln \left(\frac{1 + {\sf y}}{1 - {\sf y}}\right)\,.
\eea
\ees
Eq. (\ref{Lob}) is a standard form of the line element of three-dimensional \L obachewsky 
(hyperbolic or Euclidean anti-deSitter) space ${\mathbb H}^3$. Thus, one expects conformal 
field theory (CFT) behavior at the horizon boundary, ${\sf y}=1$, $r= r_{_H}$ \label{AdSCFT}.
This is reflected also in the wave eq. analogous to (\ref{feq}) in which the 
corresponding potential $V_{\ell}$ of (\ref{VelldS}) vanishes at the horizon, which 
we consider in the next subsection.

Since these various coordinatizations of the de Sitter manifold are quite different
globally, and involve singular coordinate transformations at the horizon, much of 
the discussion of the Schwarzschild geometry have their analogs in de Sitter 
spacetime. The Carter-Penrose conformal diagram for the analytically extended 
de Sitter hyperboloid, Fig. \ref{Fig:dSCarPen} is similar to the corresponding 
diagram in the Schwarzschild case, Fig. \ref{Fig:SchwKruskal}. In each case the 
horizon is bifurcate ({\it i.e.} has two distinct parts) and the region covered by the static 
coordinates (\ref{dSstat}) is duplicated by a second region centered on the antipodal 
point of $\mathbb{S}^3$ where the sense of increasing static time $t$ is reversed. 
Thus the static Killing vector $K^a$ (\ref{Kill}) becomes null on either horizon and 
spacelike in the upper and lower quarter wedges of Fig. \ref{Fig:dSCarPen}.

\subsection{Quantum Effects in de Sitter Spacetime}

Quantum fluctuations and their backreaction effects in de Sitter spacetime 
were considered in \cite{EMfluc,EMdSp,EMSt,EMdSMM,EMtime}. These studies
indicate that fluctuations at the horizon scale $r_{_H}$ are responsible 
for important backreaction effects on the classical de Sitter expansion that
could potentially relax the effective cosmological vacuum energy to zero. 
The authors of Refs. \cite{Wood} have performed a perturbative
analysis of long wavelength gravitational fluctuations in non de Sitter 
invariant initial states up to two-loop order. This work indicates the 
presence of secular terms in the quantum stress tensor of fluctuations 
about de Sitter space, tending to decrease the effective vacuum energy 
density, consistent with the earlier considerations in Refs.
\cite{EMfluc,EMdSp,EMSt,EMdSMM,EMtime}. The authors of Ref. \cite{MAB} 
have studied the stress tensor for long wavelength cosmological 
perturbations in inflationary models as well, and also found a backreaction 
effect of the right sign to slow inflation. See also Refs. \cite{AW}, \cite{IWald},
and a discussion of these various approaches in Ref. \cite{NJP}.

One of the first indications of non-trivial quantum effects in de Sitter spacetime
is particle creation \cite{EMdSp}. A space or time dependent background field generally
creates particles. Schwinger first studied this effect in QED in a series of classic 
papers \cite{Schwinger}. The rate of spontaneous decay of the
electric field into charged particles is
\be
\Gamma = \frac{(eE)^2}{c\hbar^2\pi^2}  \exp \left(-\frac{m^2c^3}{eE\hbar}\right)
\ee
in the limit $eE \ll m^2c^3/\hbar$. From this point of view the exponential de Sitter
expansion (\ref{FLRW})-(\ref{rhotau}) provides a time dependent background field 
which can create particle pairs from the ``vacuum," converting the energy of the classical 
gravitational background into that of particle modes. The rate of this spontaneous 
creation of matter in de Sitter space can be calculated in analogy to the Schwinger effect 
in an electric field, with a similar result for the decay rate per unit volume \cite{EMdSp},
\be
\Gamma = \frac{16 H^4}{\pi^2}\exp\left(- \frac{2 \pi m}{\hbar H}\right)
\label{dSdecay}
\ee
for a scalar massive field with arbitrary curvature coupling ({\it i.e.} $m^2 = \mu^2 + \xi R
= \mu^2 + 12\xi H^2$). 

In the case of a spatially uniform electric field (and no magnetic fields) the Maxwell 
equation,
\be
\frac{\partial \vec E}{\partial t} = -  \langle\vec j\rangle 
\label{Max}
\ee
indicates that the creation of these particle pairs tends (at least initially)
to decrease the strength of the electric field. In cosmology the Friedmann 
equation,
\be
\left(\frac{\dot a}{a}\right)^2 = H^2= \frac{8 \pi G}{3}\,\rho\,,
\ee
together with the equation of covariant energy conservation,
\be
{\dot \rho} + 3\, H\, (\rho + p) = 0, 
\label{ener}
\ee
(with over dots here denoting differentiation with respect to proper time $\tau$)
imply that
\be
{\dot H} = - \frac{4 \pi G}{c^2} \ (\rho + p). 
\label{hd}
\ee
In both cases there is a classical static background that solves the equation 
trivially, namely $H$ or $\vec E$ a constant with zero source terms on the 
right hand side of (\ref{Max}) or (\ref{hd}). In the case of (\ref{hd}) this is de Sitter 
spacetime with $\rho_{\Lambda} + p_{\Lambda} = 0$. Any creation of matter with 
$\rho+ p  > 0$ will tend to decrease the strength of the classical background 
field $H$. The particle creation process has been studied in greater detail 
in the electric field \cite{KlugMot} and de Sitter \cite{HabMolMot} cases, 
and backreaction effects were also taken into account in the QED case the 
large $N$ limit \cite{KlugMot}. In this limit self-interactions between the created charge 
particles are neglected, but such scatterings are essential to the final decay 
of the coherent classical field into particles. In the gravitational case it has 
so far not been feasible to carry out a full dynamical calculation of backreaction
and particle scattering which would establish $\rho + p > 0$ and final
decay of the background value of $H$. Unlike electromagnetism in gravity 
there are massless particles which couple to the background field, and from
(\ref{dSdecay}) these will give the largest effect. However for minimally coupled 
massless scalar particles and for gravitons themselves, the definition of the 
``vacuum" state becomes more subtle, so that computing the infrared effects
of massless particles and their interactions requires a different approach.

One possible way to define a preferred state of a system is by its symmetries.
Since the symmetry group of the maximally analytically extended
de Sitter hyperboloid is $O(4,1)$, a maximally $O(4,1)$ symmetric
state can be defined, for most field theories, including massive
and massless conformally invariant fields, such as the photon.
Since under analytic continuation of $T \rightarrow iT$ the de Sitter
manifold becomes $\mathbb{S}^4$, which is compact, one can define a
state by the requirement of maximal $O(5)$ invariance, regularity
on $\mathbb{S}^4$, whose $n$-point functions are all analytic under the
continuation. This is often referred to as the Bunch-Davies (BD)
state in the literature \cite{BD}. Since the Euclidean $\mathbb{S}^4$  
radius is $r_{_H}$, the Euclidean Green's functions are periodic in imaginary 
time with period $2\pi r_{_H}/c$. This is again the KMS property of a thermal Green's 
function, and hence the BD state defined by this analytic 
continuation from $\mathbb{S}^4$ is a thermal state with temperature \cite{GibPer,GibEin79}, 
\be
T_{_H} = \frac{\hbar H}{2 \pi k_{_B}}\,,
\label{HawkdS}
\ee
the Hawking temperature of de Sitter space. Hence the BD state in de Sitter space 
is not a ``vacuum" state, but is a state thermally populated with quanta at temperature 
$T_{_H}$. In this respect it is rather like the Hartle-Hawking state in the Schwarzschild 
background. In these fully time symmetric states there is no net particle creation
on average and no decay rate. Because of the full $O(4,1)$ symmetry of the BD state, 
the expectation value of the energy-momentum tensor of any matter fields in this state 
must itself be of the form of a quantum correction to the cosmological term with 
$\rho = - p \sim H^4$ constant. The BD state usually assumed, tacitly or explicitly in 
models of inflation. It is interesting that because of infrared divergences the
BD state does not exist for massless, minimally coupled scalar fields and gravitons
themselves \cite{AllFol}, which implies that these fields must be in a state with
less than the full $O(4,1)$ de Sitter symmetry. Once the de Sitter symmetry is
broken, however, it is not clear what the residual symmetry
and final state of the system is, or how to go about determining it.

Even the existence of a maximally symmetric state does not guarantee its 
stability against small fluctuations. For example, it can be shown that
in the uniform constant electric field background there are exactly $10$ 
isometries, the same number as with zero external field \cite{Beers}. Thus 
one can construct mathematically a state which respects all these isometries,
including a discrete symmetry corresponding to time reversal invariance,
and no net pair creation, without necessarily guaranteeing the stability of 
the vacuum against pair creation in an external electric field. A subtle
point, easily overlooked in Schwinger's elegant effective action method,
is that either time asymmetric boundary/initial conditions must be specified
in such a decay problem, or it must be shown that the time symmetric state
is dynamically unstable to quantum fluctuations, spontaneously breaking
the larger isometry group to a smaller subgroup.

One may construct an argument analogous to that in the Schwarzschild case that the 
de Sitter invariant BD should be quantum mechanically unstable, under fluctuations 
in its Hawking temperature. Note that the vacuum energy 
within a spherical volume of radius $r_{_H}$ is
\be
E_{_H} =  \frac{4 \pi r_{_H}^3}{3}\,\rho_{\Lambda} = \frac{c^4}{2G}\ \frac{1}{H}\,,
\label{EdS}
\ee
which is {\it inversely} proportional to (\ref{HawkdS}). Therefore,
\be
\frac{dE_{_H}}{dT_{_H}} = - \frac{E_{_H}}{T_{_H}}  = - \frac{\pi c^3 k_{_B}}{GH^2} <0\,,
\ee
and the heat capacity of the region of de Sitter space within one horizon volume is
apparently negative. This indicates that if the region can exchange energy with its
environment external to the cosmological horizon, the BD thermal state will be 
unstable to such energy exchanges, analogously to the black hole case.
The problem of negative heat capacity is also similar.

It has also been suggested that the Bekenstein-Hawking $1/4$ area formula for the entropy
\be
S_{_H} \stackrel{?}{=} k_{_B} \frac{A_H}{4 L_{pl}^2}  =\frac{\pi c^5 k_{_B}}{\hbar GH^2}
\label{SH}
\ee
be taken over to the de Sitter case \cite{GibHaw}. However what degrees of freedom this 
``entropy" counts are even less clear than in the black hole case. Note also that from (\ref{EdS})
\bea
dE_{_H} &=& d(\rho_{\Lambda} V_{_H}) = d\rho_{\Lambda} \,V_{_H} 
+  \rho_{\Lambda} \, dV_{_H} \nn\\
&=& d\rho_{\Lambda} V_{_H} -p_{\Lambda} dV_{_H} \nn\\
&\neq & T_{_H} dS_{_H} - p_{\Lambda} dV_{_H} \,.
\label{dEdeS}
\eea
In fact $T_{_H} dS_{_H}$ and $d\rho_{\Lambda} V_{_H}$ have opposite signs. This implies that
$S_{_H}$ cannot be interpreted as an entropy and/or additional contributions, such
as surface terms, are missing from (\ref{dEdeS}). 

Like the Schwarzschild case, de Sitter spacetime admits a static Killing field
which is timelike in one region of its maximal analytic extension. If one considers 
any thermal state with a temperature $T$ different from $T_{_H}$, defined with
respect to the static Killing time in one patch of de Sitter space, then
the renormalized stress-energy tensor is \cite{AndHis}
\be
\lag T^a_{\ b}\rag_{_R}  \rightarrow \frac{\pi^2}{90} \,\frac{k_{_B}^4}{(\hbar c)^3}\, 
(T^4 - T_{_H}^4)\left(1 - H^2r^2\right)^{-2}\, {\rm diag} \, (-3, 1, 1,1)\,,
\label{dSthermal}
\ee
as $r \rightarrow r_{_H}$, analogous to (\ref{TTemp}) in the Schwarzschild case.
Thus any finite deviation of the temperature from the Hawking-de Sitter temperature 
(\ref{HawkdS}) in the region interior to the horizon will produce a {\it arbitrarily large} 
stress-energy on the horizon. The previous discussion in Sec. {\ref{sec:BH} about the 
dependence of $\lag T^a_{\ b}\rag_{_R}$ on the gauge invariant but non-local norm of the static 
Killing vector field, the non-commutivity of the limits $\hbar \rightarrow 0$ and 
$r \rightarrow r_{_H}$, and the breakdown of the analytic continuation hypothesis 
through coordinate or conical singularities apply to the de Sitter case as well.

The wave equation of a quantum field propagating in de Sitter space can
also be separated in the static coordinates (\ref{dSstat}).  By following the
analogous steps as used in (\ref{KGeq})-(\ref{Vell}) in the Schwarzschild
case, one arrives at the identical form of the radial scattering eq. (\ref{feq})
in terms of the $r^*$ coordinate (\ref{RegWhdS}),
with the scattering potential in de Sitter space given by
\bea
&&V_{\ell}\big\vert_{dS} = (1-H^2r^2)\left[\frac{\ell(\ell +1)}{r^2} + \mu^2 
+ 2H^2 (6\xi - 1)\right]
\nn\\
&& = H^2 \left[\ell(\ell +1) \,{\rm csch}^2(Hr^*) + \left(\frac{1}{4} - \nu^2\right)
 {\rm sech}^2(Hr^*)\right],
\label{VelldS}
\eea
where $-\xi/2$ is the $R\Phi^2$ coupling in the scalar field Lagrangian, and
\be
\nu \equiv \sqrt{\frac{9}{4} - \frac{\mu^2}{H^2} - 12 \xi}\,.
\label{defnu}
\ee
A significant difference of the effective one dimensional scattering problem of
spherical waves in the de Sitter case is that the coordinate singularity at the origin 
$r=0$ is in the physical range, compared to the Schwarzschild case in which
the $r^*$ coordinate range $(-\infty, \infty)$ covers only the exterior Schwarzschild
geometry. Thus in order to avoid a singularity at the origin we must require that the
scattering solutions to (\ref{feq}) with (\ref{VelldS}) satisfy
\be
f_{\omega\ell}\big\vert_{dS} \sim r^{\ell(\ell +1)}\qquad {\rm as}\qquad r\rightarrow 0\,,
\label{origcond}
\ee
thereby excluding a possible singular $r^{-\ell}$ behavior for $\ell \ge 1$.
This means that in de Sitter space only the particular linear combination 
of ingoing and outgoing solutions of the corresponding (\ref{feq}) vanishing
at the origin should be used in the quantization of the $\Phi$ field and not 
the general linear combination in (\ref{scatsoln})-(\ref{freqsum}) appropriate 
in the Schwarzschild background. As a corollary this implies that there can be 
no analog of the Unruh state in de Sitter spacetime, since (\ref{origcond}) is 
equivalent to the requirement of no net flux into or out of the origin at $r=0$ in 
static de Sitter coordinates (\ref{dSstat}). Otherwise the quantization of a
scalar field in de Sitter space in these coordinates is analogous 
to the Schwarzschild case, and one can again easily find states regular at the
origin which have diverging stress tensors on the horizon, as in (\ref{dSthermal}).
The appearance of such states and large stress tensors would necessarily mean
large quantum backreaction effects in the vicinity of the horizon $r= r_{_H}$,
and the breakdown of $O(4,1)$ de Sitter symmetry down to $O(3)$, with 
perhaps a very different global geometry than the analytic continuation 
implied by extending the coordinates (\ref{hypermet}) globally. As in
the Schwarzschild case, this is ultimately a question of physics, not 
simply mathematical analytic continuation of coordinates. 

The earlier work on the behavior of the graviton propagator in
de Sitter space \cite{AMJMP,AlTur,FIT} also indicates the existence of infrared divergent 
contributions to correlations at large distances. This violates cluster decomposition 
and makes the definition of a graviton scattering matrix in global de Sitter space 
problematic. The existence of long distance correlations and infrared divergences 
is a signal of the breakdown of the global BD state. The relevance of matter 
self-interactions, first studied in Refs. \cite{Mhyr,Ford} has been taken up 
again recently in Refs. \cite{Poly0810}, where the phenomenon of runaway stimulated 
emission in de Sitter space is explored. This adds to the by now substantial 
literature on non-trivial quantum infrared behavior in de Sitter space \cite{NJP}. 
Since infrared effects are the common feature of all these studies, one might 
well suppose that there should be some relevant operator(s) in the low energy 
effective theory of gravity which describes in a general and universal way 
({\it i.e.} independently of specific matter self-interactions) the
quantum effects of matter/radiation fields on the macroscopic scales of both 
the black hole and cosmological horizons, and which points to a possible resolution
of the physics in both cases. This is what a predictive effective field theory
(EFT) approach should provide.

\section{Effective Field Theory: The Role of Anomalies}
\label{sec:EFT}

The important role of quantum anomalies in the low energy effective field theory
approach to the strong interactions is reviewed in this section, and their
relevance to the EFT of gravity discussed. The first and best known example
of this is the axial anomaly in QED and QCD. Since the two-particle correlations
that give rise to non-trivial infrared effects can be seen already in this
flat space case, we digress and review that here before returning to the 
curved spacetime application in gravity.

\subsection{The Axial Anomaly in QED and QCD}
\label{subsec:axial}

Although the triangle axial anomaly in both QED has been known for some time
\cite{Adler,BellJack,Jackcurr}, the general behavior of the amplitude off mass shell, 
its spectral representation, the appearance of a massless pseudoscalar pole in the limit 
of zero fermion mass, and the infrared aspects of the anomaly generally have received 
only limited attention\,\cite{DolZak,Horej}. It is this generally less emphasized infrared 
character of the axial anomaly which will be important for our EFT considerations in gravity, 
so we begin by reviewing this aspect of the axial anomaly in QED in some detail.

The vector and axial currents in QED are defined by\footnote{We use the conventions 
that $\{\gamma^{\mu}, \gamma^{\nu}\} = -2\,g^{\mu\nu} = 2\,{\rm diag}\ (+---)$, 
so that $\gamma^0 = (\gamma^0)^{\dagger}$, and $\gamma^5 \equiv i\gamma^0 \gamma^1 
\gamma^2 \gamma^3 = (\gamma^5)^{\dagger}$ are hermitian, and 
tr$(\gamma^5\gamma^{\mu}\gamma^{\nu}\gamma^{\rho}\gamma^{\sigma})
= -4i \epsilon^{\mu\nu\rho\sigma}$, where $\epsilon^{\mu\nu\rho\sigma}
= - \epsilon_{\mu\nu\rho\sigma}$ is the fully anti-symmetric Levi-Civita tensor, 
with $\epsilon_{0123} =+1$.} 
\bes\bea
J^{\mu}(x) = \bar\psi (x) \gamma^{\mu} \psi (x)\,.\\
J_5^{\mu}(x) = \bar\psi (x) \gamma^{\mu} \gamma^5 \psi (x)\,.
\eea\label{currents}\ees
The Dirac eq.,
\be
-i \gamma^{\mu} ( \partial_{\mu} - ieA_{\mu})\psi + m \psi = 0\,.
\label{Dirac}
\ee
implies that the vector current is conserved,
\be
\partial_{\mu} J^{\mu} = 0\,,
\label{vecons}
\ee
while the axial current apparently obeys 
\be
\partial_{\mu} J_5^{\mu} = 2 i m\, \bar\psi\gamma^5 \psi \qquad {\rm (classically)}.
\label{axclass}
\ee
In the limit of vanishing fermion mass $m\rightarrow 0$, the classical Lagrangian has a
$U_{ch}(1)$ global symmetry under $\psi \rightarrow e^{i\alpha\gamma^5}\psi$, in 
addition to $U(1)$ local gauge invariance, and $J_5^{\mu}$ is the Noether current
corresponding to this chiral symmetry. As is well known, both symmetries cannot be 
maintained simultaneously at the quantum level. Let us denote by $\lag J_5^{\mu}(z) \rag_{_A}$ 
the expectation value of the chiral current in the presence of a background electromagnetic 
potential $A_{\mu}$. Enforcing $U(1)$ gauge invariance (\ref{vecons}) on the full quantum theory 
leads necessarily to a finite axial current anomaly,
\be
\partial_{\mu}\lag J_5^{\mu}\rag_{_A}\Big\vert_{m=0} = \frac{e^2}{16\pi^2} \
\epsilon^{\mu\nu\rho\sigma}F_{\mu\nu}F_{\rho\sigma} = \frac{e^2}{2\pi^2}\,\vec E \cdot \vec B\,,
\label{axanom}
\ee
in an external electromagnetic field.  Varying this expression twice with respect to the
external $A$ field we see that anomaly must appear in the amplitude,
\bea
&&\Gamma^{\mu \alpha\beta}(p,q) \equiv - i \int d^4x\int d^4y\, e^{ip\cdot x+ i q\cdot y}\  
\frac{\delta^2 \lag J_5^{\mu}(0) \rag_{_A}}{\delta A_{\alpha}(x) \delta A_{\beta}(y)} 
\nn\Bigg\vert_{A=0}\\
&&= ie^2 \int d^4 x \int d^4 y\, e^{i p\cdot x + i q \cdot y}\ \lag {\cal T}
J_5^{\mu}(0) J^{\alpha}(x) J^{\beta}(y)\rag\big\vert_{A=0}\,.
\label{GJJJ}
\eea
At the lowest one-loop order it is given by the triangle diagram of Fig. \ref{Fig:tri}, plus
the Bose symmetrized diagram with the photon legs interchanged. 

\begin{figure}
\begin{center}
\includegraphics[width=12cm, viewport=170 575 400 685,clip]{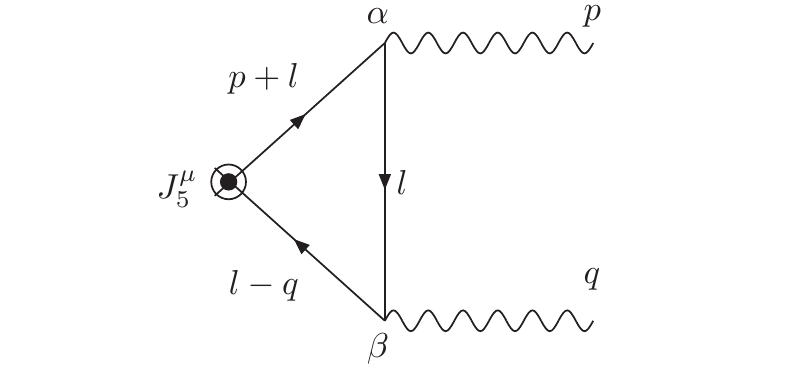}
\caption{The triangle diagram contributing to the axial current anomaly. 
The four-momentum of integration is $l$.}
\label{Fig:tri}
\end{center}
\vspace{-7mm}
\end{figure}

Elementary power counting indicates that the triangle diagram of Fig.  \ref{Fig:tri} is superficially 
linearly divergent. The formal reason why (\ref{vecons}) and (\ref{axclass}) cannot both be 
maintained at the quantum level is that verifying them requires the ability to shift the loop momentum 
integration variable $l$ in the triangle amplitude. Because the diagram is linearly divergent, such 
shifts are inherently ambiguous, and can generate finite extra terms. It turns out that there is no 
choice for removing the ambiguity which satisfies both the vector and chiral Ward identities simultaneously, 
and one is forced to choose between them. Thus although the ambiguity results in a well-defined finite term, 
the axial anomaly has most often been presented as inherently a problem of regularization of an apparently 
ultraviolet linearly divergent loop integral\,\cite{Adler,BellJack,Jackcurr}.

There is an alternative derivation of the axial anomaly that emphasizes instead its infrared character. 
The idea of this approach is to use the tensor structure of the triangle amplitude to extract its 
well-defined ultraviolet {\it finite} parts, which are homogeneous of degree three in the external 
momenta $p$ and $q$. Then the remaining parts of the full amplitude may be determined by the joint 
requirements of Lorentz covariance, Bose symmetry under interchange of the two photon legs, and 
electromagnetic current conservation,
\be
p_\alpha \Gamma^{\mu\alpha\beta}(p,q)=0 = q_\beta\Gamma^{\mu\alpha\beta}(p,q)\,,
\label{chicons}
\ee
at the two vector vertices. By this method the full one-loop triangle contribution to 
$\Gamma^{\mu\alpha\beta}(p,q)$, becomes completely determined in terms of well-defined 
ultraviolet {\it finite} integrals which require no further regularization \cite{Rosenb, Horejb}. 
The divergence of the axial current may then be computed unambiguously, and one obtains
(\ref{axanom}) in the limit of vanishing fermion mass. It is this latter method
which makes clear that the anomaly is a consequence of symmetries of the low energy
theory, no matter how its UV behavior is tamed, provided only that the regularization
respects these symmetries. There is of course no contradiction between these two 
points of view, since it is the same Ward identities which are imposed in either method, 
and in any case in the conformal limit of vanishing fermion mass the infrared and ultraviolet 
behavior of the triangle amplitude are one and the same. 

The details of the calculation by this method may be found in Refs. \cite{Rosenb,Horejb,GiaMot}.
One first uses the Poincar{\'e} invariance of the vacuum to assert that $\Gamma^{\mu\alpha\beta}(p,q)$ 
can be expanded in the set of all three-index tensors constructible from the $p$ and $q$, 
with the correct Lorentz transformation properties. There are exactly eight such tensors, 
two of which are linear in $p$ or $q$, namely $\varepsilon^{\mu\alpha\beta\lambda} p_{\lambda}$ 
and $\varepsilon^{\mu\alpha\beta\lambda} q_{\lambda}$, while the remaining six are 
homogeneous of degree three in the external momenta. However only certain linear
combinations of these eight tensors satisfy (\ref{chicons}). Define first
the two index tensor,
\be
\upsilon^{\alpha\beta}(p,q) \equiv \epsilon^{\alpha\beta \rho\sigma} p_{\rho}q_{\sigma}\,,
\label{Wdef}\ee
which satisfies
\bes\bea
\upsilon^{\alpha\beta}(p,q) &=& \upsilon^{\beta\alpha}(q,p)\,,\\
p_{\alpha} \upsilon^{\alpha\beta}(p,q) = &0& = q_{\beta}\upsilon^{\alpha\beta}(p,q) \,.
\eea\ees
Then the six third rank tensors, $\tau_i^{\mu\alpha\beta}(p,q)$,
$i= 1, \dots, 6$ which satisfy the conditions (\ref{chicons}),
\be
p_\alpha \tau_i^{\mu\alpha\beta}(p,q)=0=\tau_i^{\mu\alpha\beta}(p,q)\,q_\beta = 0 \,,
\qquad i= 1, \dots, 6,
\label{taucons}
\ee
are given in Table \ref{chitens}. 

\begin{table}[htp]
$$
\begin{array}{|@{\hspace{.8cm}}c @{\hspace{.8cm}}|@{\hspace{.8cm}}c @{\hspace{.8cm}}|} 
\hline
i & \tau_i^{\mu\alpha\beta}(p,q) \\ \hline \hline
1 & -\pq \,\epsilon^{\mu\alpha\beta\lambda} p_\lambda  -
p^\beta \,\upsilon^{\mu\alpha}(p,q) \\ \hline
2 & p^2 \epsilon^{\mu\alpha\beta\lambda} q_\lambda \, +
 p^{\alpha} \upsilon^{\mu\beta}(p,q) \\ \hline
3 & p^\mu\,\upsilon^{\alpha\beta}(p,q)\\ \hline \hline
4 & \pq \,\epsilon^{\mu\alpha\beta\lambda} q_\lambda  +
q^{\alpha} \,\upsilon^{\mu\beta}(p,q) \\ \hline
5 & -q^2 \varepsilon^{\mu\alpha\beta\lambda} p_\lambda \,
-q^{\beta} \upsilon^{\mu\alpha}(p,q)\\ \hline
6 & q^\mu\,\upsilon^{\alpha\beta}(p,q) \\ \hline
\end{array}
$$
\caption{The $6$ third rank (pseudo)tensors obeying (\ref{taucons}) \label{chitens}}
\end{table}

Hence  we may express the amplitude (\ref{GJJJ}) satisfying (\ref{chicons}) as a linear combination,
\be
\Gamma^{\mu\alpha\beta}(p,q)= \sum_{i=1}^6 f_i\, \tau_i^{\mu\alpha\beta}(p,q)\,,
\label{tencomp}
\ee
where $f_i = f_i(k^2; p^2,q^2)$ are dimension $-2$ scalar functions of the three invariants,
$p^2, q^2$, and $k^2$. Note from Table \ref{chitens} that the two tensors 
of dimension one which could potentially have logarithmically divergent scalar coefficient 
functions occur only in linear combination with dimension three tensors. Hence the coefficient 
functions $f_i$ of these linear combinations obeying (\ref{chicons}) are all ultraviolet {\it finite}.
These finite contributions can be obtained unambiguously from the imaginary part of
the triangle graph of Fig. \ref{Fig:ImTri}, which are finite {\it a priori} and then determining
the real part from the imaginary part by an {\it unsubtracted, i.e.} UV finite dispersion 
relation. This leads to the finite coefficients given 
in the literature \cite{Rosenb,Horejb,GiaMot}, {\it viz.}
\bes\bea
&& f_1=f_4
=\frac{e^2}{\pi^2}\int_0^1 dx\int_0^{1-x} dy\  \frac{x y}{D}\,\\
&& f_2=\frac{e^2}{\pi^2}\int_0^1 dx\int_0^{1-x} dy\ 
\frac{x(1-x)}{D}\,,\\
&& f_5=\frac{e^2}{\pi^2}\int_0^1 dx\int_0^{1-x} dy\ 
\frac{y(1-y)}{D}\,,\\
&& f_3 = f_6 = 0\,,
\eea\label{chiamps}\ees
where the denominator of the Feynman parameter integral is 
\be
D \equiv p^2 x(1-x) + q^2 y(1-y) + 2\pq \, xy + m^2 = 
(p^2\,x +q^2\,y)(1-x-y)+ xy\,k^2+m^2\,,
\label{denom}
\ee
strictly positive for $m^2 >0$, and spacelike momenta, $k^2, p^2, q^2 > 0$.

Thus, the full amplitude $\Gamma^{\mu\alpha\beta}(p,q)$ satisfying
\vspace{-1mm}
\begin{enumerate}
\item [(i)] Lorentz invariance of the vacuum\,,\vspace{-.3cm}
\item [(ii)] Bose symmetry under interchange of photon lines\,,\vspace{-.3cm}
\item [(iii)] vector current conservation (\ref{chicons})\,,\vspace{-.3cm}
\item [(iv)] unsubtracted dispersion relation of real and imaginary parts\,,\vspace{-1mm}
\end{enumerate}
with the finite imaginary parts determined by the cut triangle diagram of Fig. \ref{Fig:ImTri}, 
is given by (\ref{tencomp}) and (\ref{chiamps}), without any need of regularization of ultraviolet 
divergent loop integrals at any step. 

\begin{figure}
\begin{center}
\includegraphics[width=12cm, viewport=170 575 400 685,clip]{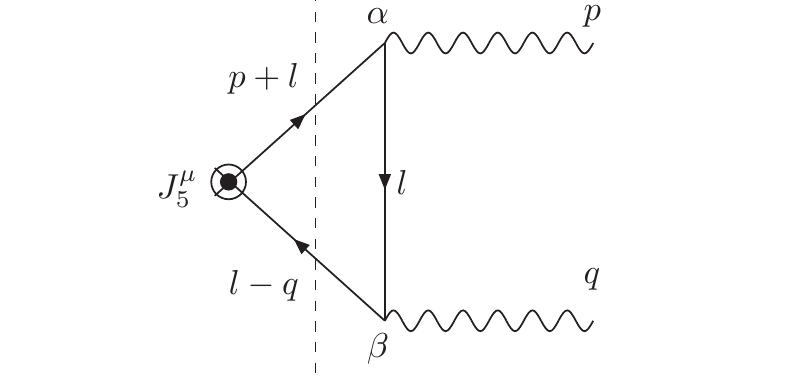}
\caption{The discontinuous or imaginary part of the triangle diagram of Fig. \ref{Fig:tri} 
with respect to $k^2$. The propagators which are cut by the dashed line are replaced by their
corresponding on-shell delta functions.}
\label{Fig:ImTri}
\end{center}
\vspace{-7mm}
\end{figure}

Since this amplitude is fully determined by (i)-(iv), its contraction with $k_{\mu}$, and the
divergence of the axial vector current is determined as well. There is no further freedom
to demand the naive axial Ward identity corresponding to (\ref{axclass}). Instead we find
\be
k_\mu \, \Gamma^{\mu \alpha\beta}(p,q) = {\cal A} \, \upsilon^{\alpha\beta} (p, q)\,,
\ee
where
\bea
&&{\cal A}(k^2; p^2,q^2) = 2\pq\, f_1 + p^2 f_2 + q^2 f_5 \nn\\
&& = \frac{e^2}{\pi^2}\int_0^1 dx\int_0^{1-x} dy\  \frac{D - m^2}{D}\nn\\
&&\qquad = \frac{e^2}{2\pi^2} - \frac{e^2}{\pi^2} \ m^2\,
\int_0^1 dx\int_0^{1-x} dy\  \frac{1}{D}\,.
\label{anomcalc}
\eea
The second term proportional to $m^2$ is what would be expected from the naive axial vector 
divergence (\ref{axclass})\,\cite{Horejb}. The first term in (\ref{anomcalc}) in which the 
denominator $D$ is cancelled in the numerator is
\be
\frac{e^2}{\pi^2} \int_0^1 dx\int_0^{1-x} dy = \frac{e^2}{2\pi^2}\,,
\ee
and which remains finite and non-zero in the limit $m \rightarrow 0$ is the axial anomaly. 

Thus the finite anomalous term is unambiguously determined by our four requirements above, 
and may be clearly identified even for finite $m$, when the chiral symmetry is broken. This
construction of the amplitude from only symmetry principles and its finite parts may be
regarded as a proof that the same finite axial anomaly must arise in {\it any} regularization
of the original triangle amplitude which respects these symmetries and leaves the finite
parts unchanged. Explicit calculations in dimensional regularization and Pauli-Villars 
regularization schemes, which respect these symmetries confirm this \cite{Bertl}.

Now the important consequence of the anomaly in (\ref{anomcalc}) in the divergence
is that when the photons $p^2 = q^2=0$ are on shell, the amplitude $\Gamma^{\mu \alpha\beta}(p,q)$ 
develops a simple pole at in $k^2$ as the electron mass $m \rightarrow 0$. This follows
from the first line of (\ref{anomcalc}) which shows that when $p^2 = q^2 = 0$, $2\pq = k^2$ so 
${\cal A}$ is explicitly proportional to $k^2$. Since $\cal A$ is finite when $m \rightarrow 0$,
the amplitude function $f_1(k^2; 0, 0)$ must develop a pole in $k^2$. Indeed by
explicit calculation from (\ref{chiamps})-(\ref{denom}), we have
\be
f_1(k^2; p^2=0, q^2=0)\big\vert_{m=0} = \frac{e^2}{2\pi^2}\frac{1}{k^2}\,.
\label{axpole}
\ee
The corresponding imaginary part (discontinuity in $k^2$) becomes a $\delta(k^2)$ in the same limit.
Moreover, even for arbitrary $p^2, q^2, m^2 \ge 0$, the spectral function obtained
from this imaginary part obeys an ultraviolet finite sum rule \cite{Horejb,GiaMot}.

The appearance of a massless pseudoscalar pole (\ref{axpole}) in the triangle anomaly amplitude
in the massless fermion limit suggests that this can be described as the propagator of a 
pseudoscalar field which couples to the axial current. Indeed it is not difficult to find 
the field description of the pole. To do so let us note first that the axial current 
expectation value $\lag J_5^{\mu}\rag_{_A}$ can be obtained from an extended action principle 
in which we introduce an axial vector field source, ${\cal B}_{\mu}$ into the Dirac Lagrangian,
\be
i\bar\psi \gamma^{\mu}\left(\stackrel{\leftrightarrow}{\partial}_{\mu} - ie A_{\mu}\right)\psi
-m \bar\psi \rightarrow  i\bar\psi \gamma^{\mu}\left(\stackrel{\leftrightarrow}{\partial}_{\mu} 
- ie A_{\mu} - ig\gamma^5{\cal B}_{\mu}\right)\psi -m \bar\psi\psi
\label{addax}
\ee
so that the variation of the corresponding action with respect to ${\cal B}_{\mu}$ gives
\be
\frac{\delta {\cal S}} {\delta {\cal B}_{\mu}}= g \lag J_5^{\mu}\rag_{_A} \,.
\label{expect}
\ee
Henceforth we shall set the axial vector coupling $g=1$. Next let us decompose the axial vector 
${\cal B}_{\mu}$ into its transverse and longitudinal parts,
\be
{\cal B}_{\mu} = {\cal B}_{\mu}^{\perp} + \partial_{\mu} {\cal B}
\ee
with $\partial^{\mu} {\cal B}_{\mu}^{\perp} =0$ and $\cal B$ a pseudoscalar. Then, by an integration
by parts in the action corresponding to (\ref{addax}), we have
\be
\partial_{\mu} \lag J_5^{\mu}\rag_{_A} = -\frac{\delta {\cal S}} {\delta {\cal B}}\,
\label{varS}.
\ee
Thus the axial anomaly (\ref{axanom}) implies that there is a term in the one-loop effective action 
in a background $A_{\mu}$ and ${\cal B}_{\mu}$ field, linear in $\cal B$ of the form,
\be
{\cal S}_{eff} = -\frac{e^2}{16\pi^2} \,\int d^4x \, 
\epsilon^{\mu\nu\rho\sigma}F_{\mu\nu}F_{\rho\sigma}\,{\cal B}\,,
\label{axnonl}
\ee
or since $\partial^{\lambda} {\cal B}_{\lambda} = \sq {\cal B}$,
\be
{\cal S}_{eff} = -\frac{e^2}{16\pi^2} \,\int d^4 x\int d^4y \
, [\epsilon^{\mu\nu\rho\sigma}F_{\mu\nu}F_{\rho\sigma}]_x
\sq^{-1}_{xy} \,[\partial^{\lambda} {\cal B}_{\lambda}]_y\,,
\label{nonlF}
\ee
where $\sq^{-1}_{xy}$ is the Green's function for the massless scalar wave operator 
$\sq = \partial_{\mu}\partial^{\mu}$. 
Thus from (\ref{expect}), this non-local action gives\,\cite{SmaiSpal}
\be
\lag J_5^{\mu}\rag_{_A} = \frac{e^2}{16\pi^2} \partial^{\mu} \sq^{-1}
\epsilon^{\alpha\beta\rho\sigma}F_{\alpha\beta}F_{\rho\sigma}\,,
\label{nonlpole}
\ee
which exhibts the massless scalar pole in the massless limit of (\ref{axpole}), and which 
agrees with the explicit calculation of the physical $\lag 0\vert J_5^{\mu}\vert p,q\rag$ 
triangle amplitude to two photons for $p^2=q^2 = m^2 = 0$. The existence of this pole or 
$\delta$ function in the two-particle intermediate $e^+e^-$ intermediate state is the 
first indication that a phenomenon similar to Cooper pairing in condensed matter systems 
can exist in relativistic quantum field theory, and this effect is connected to anomalies.

In Ref. \cite{GiaMot} a local action corresponding to (\ref{Sanom}) was found by 
introducing two pseudoscalar fields. The introduction of two fields is necessary
if the only term in the effective action with this massless pole is the term
(\ref{nonlF}) which is off-diagonal in $F \tilde F$ and $\partial \cdot {\cal B}$. If on
the other hand the effective action contains the perfect square,
\be
\int d^4 x \left( \partial^{\lambda}{\cal B}_{\lambda} + 
\sq^{-1} F_{\mu\nu}\tilde F^{\mu\nu}\right)^2
\nn
\ee
then this effective action can be represented by a single pseudoscalar field.
The possible existence of the additional terms in massless QED necessary to complete
the square is currently under investigation,

Since it contains kinetic terms for the additional pseudoscalar field(s), the effective
action of Ref. \cite{GiaMot} describes additional pseudoscalar degree(s) of freedom. 
These degrees of freedom are two-particle $0^-$ correlated $e^+e^-$ states, composite
bilinears of $\bar \psi$ and $\psi$, which appear in anomalous amplitudes as massless
poles. In condensed matter physics, or electrodynamics at finite temperature or 
in polarizable media, where Lorentz invariance is broken, it is a familiar 
circumstance that there are low energy collective modes of the many body theory, 
which are not part of the single particle constituent spectrum. This occurs 
also {\it in vacuo} in the two dimensional massless Schwinger model, whose 
anomaly and longitudinal ``photon" can be described by the introduction of an 
effective scalar field composed of an $e^+ e^-$ pair\,\cite{LSB}. In $3+1$ 
dimensions, relativistic kinematics and symmetries severely limit the possibilities 
for the appearance of such composite massless scalars, with the triangle anomaly 
the only known example\,\cite{ColGro}. The fact that the $e^+e^-$ pair becomes 
collinear in the massless limit shows that this effectively reduces the dimensionality 
back to $1+1$. In the well studied $1+1$ dimensional case, the commutation relations 
of fermion bilinear currents $J^{\mu}$ and $J^{\nu}_5$, which create the composite 
$e^+e^-$ massless state are due to the anomaly\,\cite{AdBertHof}. A similar 
phenomenon occurs in the triangle amplitude in $3+1$ dimensions. The axial 
anomaly thus implies additional long range correlations and collective degrees 
of freedom in the many body quantum theory, similar to Cooper pairs in a superconductor, 
which are not present in the classical or first quantized single particle theory.

In real QED these infrared effects are suppressed by the non-zero physical 
electron mass $ m> 0$, and the additional fact that macroscopic chirality 
violating sources for $J_5^{\mu}$ which would be sensitive to the 
anomaly are difficult to create. In QCD the situation is complicated both
by the strong interactions in the infrared and the spontaneous breaking of
chiral symmetry. The neutral member of the isotriplet of pseudoscalar Goldstone 
bosons in the low energy EFT is the $\pi^0$, whose decay to two photons, 
$\pi^0 \rightarrow 2 \gamma$ is correctly given by the triangle amplitude 
\cite{BarFriGM,Wein}. In fact, it was the experimental agreement between the 
measured decay rate to that predicted by the axial anomaly computed 
in the UV theory of $3$ colors of fractionally charged quarks that gave 
one of the strongest early confirmations of QCD. The fact that this 
amplitude is non-vanishing in the chiral limit, yet cannot be described
by a local operator of dimension $4$ or less in the chiral Lagrangian,
violates naive decoupling and illustrates how an anomaly couples UV 
to low energy physics. It is the fact that the anomaly may be computed 
in the UV theory of QCD but gives rise to a low energy amplitude of
meson decay, $\pi^0 \rightarrow 2 \gamma$ that led to the principle of 
anomaly {\it matching} \cite{tHooft}. 

The apparent massless pseudoscalar anomaly pole of (\ref{axpole}) in the 
{\it isosinglet} channel in the chiral limit of QCD is even more interesting. 
The $0^-$ state described by this pole in the isosinglet channel
mixes with the pseudoscalar axial gluon density ${\cal Q}(x) = 
G^a_{\mu\nu}(x)\tilde G^{a \mu\nu}(x)$,
and gives rise to a non-vanishing susceptibility of axial gluon densities,
\be
\chi(k^2) = \int d^4x\,e^{ik\cdot x}\lag {\cal Q}(x)\, {\cal Q}(0)\rag\,,
\ee
as $k^2 \rightarrow 0$, despite the fact that $\cal Q$ is a total derivative
and therefore one would naively expect $\chi(k^2)$ to be proportional to
$k^2$ and vanish in this limit. The fact that the susceptibility $\chi(0)$ is
non-vanishing is a direct effect of the massless anomaly pole \cite{Ven}.
The degree of freedom this infrared pole represents combines with a
non-dynamical but gauge invariant ${\cal Q}^2$ term in the effective
action of QCD to yield finally one propagating massive isosinglet 
psuedoscalar state which can be identified with the $\eta'$ meson,
solving the $U(1)$ problem in QCD \cite{Ven}. Thus there is no
doubt that the pseudoscalar $0^-$ state which appears in the isosinglet
anomaly channel in perturbation theory is physical and propagating in 
the final S-matrix of the theory, but it becomes massive by a topological 
variant of the Higgs mechanism \cite{ATT}.

The lesson to be taken away from this QCD example is that anomalies are a unique 
window which the low energy EFT provides to short distance physics. Two particle
correlated pair states appear in anomalous amplitudes, which can be described
as propagating massless fields that have consequences for low energy physics.  
The anomalous Ward identities and the long distance effects they generate must 
be taken into account by explicitly adding the IR relevant terms 
they induce in the low energy effective action \cite{BarFriGM,Wein}.

\subsection{The $\lag TJJ\rag$ Triangle Amplitude in QED}
\label{subsec:TJJ}

In this section we consider the amplitude for the trace anomaly in flat space that 
most closely corresponds to the triangle amplitude for the axial current anomaly 
reviewed in the previous section, and give a complete calculation of the full 
$\lag T^{\mu\nu} J^{\alpha} J^{\beta}\rag$ amplitude for all values of the mass and the
off-shell kinematic invariants. Although the tensor structure of this amplitude is more 
involved than the axial vector case, the kinematics is essentially the same, and the 
appearance of the massless pole very much analogous to the axial case.

Classical fields satisfying wave equations with zero mass, which are
invariant under conformal transformations of the spacetime metric, 
$g_{\mu\nu} \rightarrow e^{2\sigma} g_{ab}$ have stress tensors with zero 
classical trace, $T^{\mu}_{\ \mu} = 0$. In quantum theory the stress tensor
$T^{\mu}_{\ \nu}$ becomes an operator with fluctuations about its mean value.
The mean value itself $\langle T^{\mu}_{\ \nu}\rangle$ is formally UV divergent,
due to its zero point fluctuations, as in (\ref{zeropt}), and requires a 
careful renormalization procedure. The result of this renormalization
consistent with covariant conservation in curved spacetime is that
classical conformal invariance cannot be maintained at the quantum level. 
The trace of the stress tensor is generally non-zero when $\hbar \ne 0$, 
in non-trivial background fields, provided one preserves the covariant 
conservation of $T^{\mu}_{\ \nu}$ (a necessary requirement of any theory 
respecting general coordinate invariance and consistent with the 
Equivalence Principle) yields an expectation value of the quantum 
stress tensor with a non-zero trace.  

The fundamental quantity of interest for us now is the expectation value 
of the energy-momentum tensor bilinear in the fermion fields in an external 
electromagnetic potential $A_{\mu}$, 
\be
\lag T^{\mu\nu}\rag_{_A} = \lag T^{\mu\nu}_{free}\rag_{_A} + 
\lag T^{\mu\nu}_{int}\rag_{_A} \,
\label{Texpect}
\ee
where
\bes\bea
&&T^{\mu\nu}_{free} = -i \bar\psi \gamma^{(\mu}\!\!  
\stackrel{\leftrightarrow}{\partial}\!^{\nu)}\psi + g^{\mu\nu}
(i \bar\psi \gamma^{\lambda}\!\!\stackrel{\leftrightarrow}{\partial}\!\!_{\lambda}\psi
- m\bar\psi\psi)\\
&&T^{\mu\nu}_{int} = -\, e J^{(\mu}A^{\nu)} + e g^{\mu\nu}J^{\lambda}A_{\lambda}\,,
\eea\label{EMT}\ees 
are the contributions to the stress tensor of the free and interaction terms of the Dirac 
Lagrangian (\ref{addax}). The notations, $t^{(\mu\nu)} \equiv (t^{\mu\nu} + t^{\nu\mu})/2$ 
and $\stackrel{\leftrightarrow}{\partial}\!\!_{\mu} \equiv 
(\stackrel{\rightarrow}{\partial}\!\!_{\mu} - \stackrel{\leftarrow}{\partial}\!\!_{\mu})/2$,
for symmetrization and ant-symmetrization have been used. 
The expectation value $\lag T^{\mu\nu}\rag_{_A}$ satisfies the partial 
conservation equation,
\be
\partial_{\nu} \lag T^{\mu\nu}\rag_{_A} = eF^{\mu\nu} \lag J_{\nu}\rag_{_A}\,,
\label{conspsi}
\ee
upon formal use of the Dirac eq. of motion (\ref{Dirac}). Just as in the
chiral case, the relation is formal because of the {\it a priori} ill-defined nature 
of the bilinear product of Dirac field operators at the same spacetime point in (\ref{EMT}).
Energy-momentum conservation in full QED ({\it i.e.} when the electromagnetic
field $A_{\mu}$ is also quantized) requires adding to the fermionic $T^{\mu\nu}$ of
(\ref{EMT}) the electromagnetic Maxwell stress tensor,
\be
T^{\mu\nu}_{\ Max} = F^{\mu\lambda}F^{\nu}_{\ \ \lambda} - \frac{1}{4} g^{\mu\nu}
F^{\lambda\rho}F_{\lambda\rho}
\label{MaxEMT}
\ee
which satisfies $\partial_{\nu}T^{\mu\nu}_{\ Max} = - F^{\mu\nu} J_{\nu}$. This
cancels (\ref{conspsi}) at the operator level, so that the full stress tensor of QED 
is conserved upon using Maxwell's eqs., $\partial_{\nu} F^{\mu\nu} = J^{\mu}$. Since 
in our present treatment $A_{\mu}$ is an arbitrary external potential, rather than a 
dynamical field, we consider only the fermionic parts of the stress tensor (\ref{EMT}) 
whose expectation value satisfies (\ref{conspsi}) instead.

At the classical level, {\it i.e.} again formally, upon use of (\ref{Dirac}), the trace of 
the fermionic stress tensor obeys 
\be
T^{\mu\ (cl)}_{\ \mu} \equiv g_{\mu\nu}T^{\mu\nu\, (cl)} = - m \bar\psi\psi \qquad
{\rm (classically)}\,,
\label{trclass}
\ee
analogous to the classical relation for the axial current (\ref{axclass}). From this 
it would appear that $\lag T^{\mu\nu}\rag_{_A}$ will become traceless in the 
massless limit $m \rightarrow 0$, corresponding to the global dilation symmetry 
of the classical theory with zero mass. However, as in the case of the classical 
chiral symmetry, this symmetry under global scale transformations cannot be 
maintained at the quantum level, without violating the conservation law satisfied 
by a related current, in this case the partial conservation law (\ref{conspsi}), 
implied by general coordinate invariance. Requiring that (\ref{conspsi}) {\it is} 
preserved at the quantum level necessarily leads to a well-defined anomaly 
in the trace \cite{ChanEll,AdlColDun,DrumHath},   
\be
\lag T^{\mu}_{\ \mu}\rag_{_A} \big\vert_{m=0} = -\frac{e^2}{24\pi^2}\, F_{\mu\nu}F^{\mu\nu}\,,
\label{tranomF}
\ee
analogous to (\ref{axanom}). It is the infrared consequences of this modified, anomalous
trace identity and the appearance of massless scalar degrees of freedom for vanishing 
electron mass $m= 0$, analogous to those found in the axial case that we will study.

The one-loop triangle amplitude analogous to (\ref{GJJJ}) of
the axial anomaly must satisfy vector current conservation,
\be
p_\alpha \Gamma^{\mu\nu\alpha\beta}(p,q) = q_\beta\Gamma^{\mu\nu\alpha\beta}(p,q)=0  \,.
\label{gauge}
\ee
and the (partial) conservation law of the fermion stress-tensor (\ref{conspsi})
gives the Ward identity,
\be
k_{\nu}\Gamma^{\mu\nu\alpha\beta} (p,q) =
(g^{\mu\alpha} p_{\nu} -\delta^{\alpha}_{\nu} p^{\mu})\Pi^{\beta\nu}(q) +
(g^{\mu\beta} q_{\nu} -\delta^{\beta}_{\nu}q^{\mu})\Pi^{\alpha\nu}(p)\,,
\ee
or since the polarizarion $\Pi^{\mu\nu}(p)$ is also a correlator of conserved
currents,
\be
\Pi^{\alpha\beta} (p) = (p^2 g^{\alpha\beta} - p^{\alpha}p^{\beta}) \Pi (p^2)\,.
\label{trans}
\ee
we obtain
\bea
&&k_\nu\,\Gamma^{\mu\nu\alpha\beta}(p,q)= \left(q^\mu p^\alpha p^\beta
-q^\mu g^{\alpha\beta} p^2 +g^{\mu\beta}q^\alpha p^2
-g^{\mu\beta}p^\alpha p\cdot q \right)   \Pi(p^2) \nn\\
&&\qquad + \left(p^\mu q^\alpha q^\beta - p^\mu g^{\alpha\beta} q^2
+g^{\mu\alpha}p^\beta q^2 - g^{\mu\alpha}q^\beta p\cdot q\right) \Pi(q^2) \,.
\label{TWard}
\eea
These relations are still formal since the one-loop expressions are formally divergent.
\!However, analogously to the axial case, the joint requirements of:\!\!
\vspace{-1mm}
\begin{itemize} 
\item [(i)] Lorentz invariance of the vacuum, \vspace{-.3cm} 
\item [(ii)] Bose symmetry, $\Gamma^{\mu\nu\alpha\beta}(p,q) 
= \Gamma^{\mu\nu\beta\alpha}(q,p)$ \vspace{-.3cm}
\item [(iii)] vector current conservation (\ref{gauge}), \vspace{-.3cm}
\item [(iv)] unsubtracted dispersion relation of real and imaginary parts, and\vspace{-.3cm}
\item [(v)]  energy-momentum tensor conservation (\ref{TWard}),  \vspace{-1mm}
\end{itemize}
are sufficient to determine the full amplitude $\Gamma^{\mu\nu\alpha\beta}(p,q)$
in terms of its explicitly finite pieces, and yield a well-defined finite trace anomaly.
As in the axial anomaly case conisdered previously, this method of constructing the full 
$\Gamma^{\mu\nu\alpha\beta}(p,q)$ may be regarded as a proof that the same finite trace 
anomaly must be obtained in any regularization scheme that respects (i)-(v) above.
It is particularly important to recognize that the last condition (v) is necessary to
obtain a covariantly conserved stress tensor and avoid any gravitational anomaly
({\it i.e.} breaking of general coordinate invariance) from arising at the quantum level. 
If this fifth condition is not applied, and the naive conformal Ward identity arising from
(\ref{trclass}) is used instead, there is a gravitational anomaly, general coordinate
invariance is broken, but scale invariance is preserved and the $\beta$ function of the 
electromagnetic or QCD coupling would vanish \cite{GiaMot,ChanEll}, 
in contradiction with experiment \cite{L3}.

The tensor analysis in this case is somewhat more involved than in the axial case and is
given detail in Ref. \cite{GiaMot}. Lorentz invariance of the vacuum (i) is again assumed 
first by expanding the amplitude in terms of all the possible tensors with four indices 
depending on $p^{\alpha}$, $q^{\beta}$ and the flat spacetime metric 
$g^{\alpha\beta}= \eta^{\alpha\beta}$. Define the two-index tensors,
\bes\bea
&&u^{\alpha\beta}(p,q) \equiv (p\cdot q) g^{\alpha\beta} - q^{\alpha}p^{\beta}\,,\\
&&w^{\alpha\beta}(p,q) \equiv p^2 q^2 g^{\alpha\beta} + (p\cdot q) p^{\alpha}q^{\beta}
- q^2 p^{\alpha}p^{\beta} - p^2 q^{\alpha}q^{\beta}\,,
\eea \label{uwdef}\ees
each of which satisfies the conditions of Bose symmetry,
\bes\bea
&&u^{\alpha\beta}(p,q) = u^{\beta\alpha}(q,p)\,,\\
&&w^{\alpha\beta}(p,q) = w^{\beta\alpha}(q,p)\,,
\eea\ees
and vector current conservation,
\bes\bea
&&p_{\alpha} u^{\alpha\beta}(p,q) = 0 = q_{\beta}u^{\alpha\beta}(p,q)\,,\\
&&p_{\alpha} w^{\alpha\beta}(p,q) = 0 = q_{\beta}w^{\alpha\beta}(p,q)\,.
\eea\ees
Making use of $u^{\alpha\beta}(p,q)$ and $w^{\alpha\beta}(p,q)$, one finds that 
there are exactly $13$ linearly independent four-tensors $t_i^{\mu\nu\alpha\beta}(p,q)$, 
$i=1, \dots, 13$, which satisfy
\be
p_{\alpha} t_i^{\mu\nu\alpha\beta}(p,q) = 0 = q_{\beta} t_i^{\mu\nu\alpha\beta}(p,q) \,,
\qquad i=1, \dots, 13\,.
\label{vcons}
\ee
These $13$ tensors are catalogued in Table \ref{genbasis}. 
\begin{table}
$$
\begin{array}{|c|c|}\hline
i & t_i^{\mu\nu\alpha\beta}(p,q)\\ \hline\hline
1 &
\left(k^2 g^{\mu\nu} - k^{\mu } k^{\nu}\right) u^{\alpha\beta}(p.q)\\ \hline 
2 &
\left(k^2g^{\mu\nu} - k^{\mu} k^{\nu}\right) w^{\alpha\beta}(p.q)  \\ \hline
3 & \left(p^2 g^{\mu\nu} - 4 p^{\mu}  p^{\nu}\right) 
u^{\alpha\beta}(p.q)\\ \hline
4 & \left(p^2 g^{\mu\nu} - 4 p^{\mu} p^{\nu}\right)
w^{\alpha\beta}(p.q)\\ \hline
5 & \left(q^2 g^{\mu\nu} - 4 q^{\mu} q^{\nu}\right) 
u^{\alpha\beta}(p.q)\\ \hline
6 & \left(q^2 g^{\mu\nu} - 4 q^{\mu} q^{\nu}\right) 
w^{\alpha\beta}(p.q) \\ \hline
7 & \left[p\cdot q\, g^{\mu\nu}   
-2 (q^{\mu} p^{\nu} + p^{\mu} q^{\nu})\right] u^{\alpha\beta}(p.q) \\ \hline
8 & \left[p\cdot q\, g^{\mu\nu} 
-2 (q^{\mu} p^{\nu} + p^{\mu} q^{\nu})\right] w^{\alpha\beta}(p.q)\\ \hline
9 & \left(p\cdot q \,p^{\alpha}  - p^2 q^{\alpha}\right) 
\big[p^{\beta} \left(q^{\mu} p^{\nu} + p^{\mu} q^{\nu} \right) - p\cdot q\,
(g^{\beta\nu} p^{\mu} + g^{\beta\mu} p^{\nu})\big]  \\ \hline
10 & \big(p\cdot q \,q^{\beta} - q^2 p^{\beta}\big)\, 
\big[q^{\alpha} \left(q^{\mu} p^{\nu} + p^{\mu} q^{\nu} \right) - p\cdot q\,
(g^{\alpha\nu} q^{\mu} + g^{\alpha\mu} q^{\nu})\big]  \\ \hline
11 & \left(p\cdot q \,p^{\alpha} - p^2 q^{\alpha}\right)
\big[2\, q^{\beta} q^{\mu} q^{\nu} - q^2 (g^{\beta\nu} q^ {\mu} 
+ g^{\beta\mu} q^{\nu})\big]  \\ \hline
12 & \big(p\cdot q \,q^{\beta} - q^2 p^{\beta}\big)\,
\big[2 \, p^{\alpha} p^{\mu} p^{\nu} - p^2 (g^{\alpha\nu} p^ {\mu} 
+ g^{\alpha\mu} p^{\nu})\big] \\ \hline
13 & \big(p^{\mu} q^{\nu} + p^{\nu} q^{\mu}\big)g^{\alpha\beta}
+ p\cdot q\, \big(g^{\alpha\nu} g^{\beta\mu} 
+ g^{\alpha\mu} g^{\beta\nu}\big) - g^{\mu\nu} u^{\alpha\beta} \\
& -\big(g^{\beta\nu} p^{\mu} 
+ g^{\beta\mu} p^{\nu}\big)q^{\alpha} 
- \big (g^{\alpha\nu} q^{\mu} 
+ g^{\alpha\mu} q^{\nu }\big)p^{\beta}  \\ \hline
\end{array}
$$
\caption{The 13 fourth rank tensors satisfying (\ref{vcons}) \label{genbasis}}
\end{table}

Only the first two of the thirteen tensors possess a non-zero trace,
\bes\bea
&& g_{\mu\nu}t_1^{\mu\nu\alpha\beta}(p,q)  = 3k^2\, u^{\alpha\beta}(p,q)\,,\\
&& g_{\mu\nu}t_2^{\mu\nu\alpha\beta}(p,q)  = 3k^2\, w^{\alpha\beta}(p,q)\,,
\eea\ees
while the remaining eleven tensors are traceless,
\be
g_{\mu\nu}t_i^{\mu\nu\alpha\beta}(p,q) = 0\,, \qquad i=3, \dots , 13\,.
\ee
In the limit of zero fermion mass, the entire trace anomaly 
will reside only in the first amplitude function, $F_1 (k^2; p^2,q^2)$.

To proceed, one expands $\Gamma^{\mu\nu\alpha\beta}(p,q)$ in terms of these $13$
tesnors with scalar coefficient functions $F_i$ of the invariants  $k^2, p^2, q^2$
analogous to (\ref{tencomp}), and then fixes as many of the $13$ scalar functions $F_i$ 
as possible by examining the finite terms in the formal expressions for the triangle 
amplitude. This turns out to give enough information to fix $10$ linear combinations 
of the $13$ scalar functions. The information needed to fix the remaining three functions 
comes from our fifth and final requirement on the amplitude namely the Ward identity 
(\ref{TWard}). In this way, after some algebra one finds that all the $F_i$ are completely
determined and hence there is no remaining freedom in the trace part $F_1$.
The result of this calculation is \cite{GiaMot}
\bea
&&3k^2F_1 = \frac{e^2}{2\pi^2} \int_0^1\,dx\int_0^{1-x}\,dy \, 
(1-4xy)\, \frac{(D-m^2)}{D} \nn\\
&& \qquad = \frac{e^2}{6\pi^2} - \frac{e^2\ m^2}{2\pi^2}\int_0^1\,dx\int_0^{1-x}\,dy \, 
\frac{(1-4xy)}{D}\,.
\label{trace1}
\eea
The second term which vanishes in the limit $m\rightarrow 0$ is the non-anomalous part
which one would have expected by naive application of the tree level Ward identity
(\ref{trclass}). The first term where the denominator $D$ appears also in the numerator 
of the integrand and cancels, giving a well-defined contribution to the trace independent
of $m$ is the anomaly.  As in the axial case it again implies the existence of a pole in 
$k^2$ for $F_1$ in the full amplitude in the massless limit (and when the photons are 
on-shell: $p^2 = q^2 = 0$). In the imaginary part of the $\lag TJJ\rag$ triangle amplitude 
a $\delta$ function develops in the corresponding spectral density when 
$p^2 =q^2 =0$ and $m \rightarrow 0^+$. Otherwise there is again an exact ultraviolet
finite sum rule for this spectral density \cite{GiaMot}, showing that the state
persists even away from the conformal on-shell limit.

The kinematics of the state appearing in the imaginary part and spectral function 
in this limit is essentially $1+1$ dimensional, and can be represented as the two-particle 
collinear $e^+e^-$ pair in Fig. \ref{Fig:epair}. This is the only configuration possible 
for one particle with four-momentum $k^{\mu}$ converting to two particles of zero mass, 
$p^2= q^2 = 0$ as $k^2 \rightarrow 0$ as well. Although this special collinear kinematics 
is a set of vanishing measure in the two particle phase space, the $\delta (s)$ in the spectral 
function and finiteness of the anomaly itself shows that this pair state couples 
to on shell photons on the one hand, and gravitational metric perturbations on the other hand, 
with finite amplitude. 

\begin{figure}[htp]
\includegraphics[width=30cm, viewport=40 640 1000 690,clip]{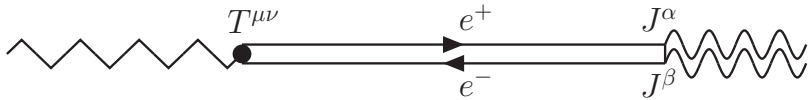}
\caption{The two particle intermediate state of a collinear $e^+e^-$ pair responsible 
for the $1/k^2$ pole and $\delta$-fn. in the $k^2$ discontinuity of the $\lag TJJ\rag$
triangle amplitude.}
\label{Fig:epair}
\end{figure}

The two-particle correlated state of $e^+e^-$ behaves like an effective massless scalar 
exchange, and can be so described in an effective field theory approach. Because
this effective field theory contains massless fields, it can have long distance,
macroscopic effects. Before giving the general form of this effective theory in four
dimensional curved space, let us consider the somewhat simpler case of two dimensions.

\subsection{The Trace Anomaly and Quantum Gravity in Two Dimensions}

In two dimensional curved space the trace anomaly takes the simple form, \cite{BirDav}.
\be
\langle T^a_{\ a} \rangle = \frac{N}{24\pi}\, R\,,\qquad (d=2)
\label{trtwo}
\ee
where $N = N_S + N_F$ is the total number of massless fields, either scalar ($N_S$) 
or (complex) fermionic ($N_F$). The fact that the anomalous trace is independent 
of the quantum state of the matter field(s), and dependent only on the geometry 
through the local Ricci scalar $R$ suggests that it should be regarded as a 
geometric effect. However, no local coordinate invariant action exists whose 
metric variation leads to (\ref{trtwo}). This is important because it shows immediately 
that understanding the anomalous contributions to the stress tensor will bring 
in some non-local physics or boundary conditions on the quantum state
at large distance scales. 

A non-local action corresponding to (\ref{trtwo}) can be found by 
introducing the conformal parameterization of the metric,
\be
g_{ab} = e^{2\sigma}\bar g_{ab}\,,
\label{confdef}
\ee
and noticing that the scalar curvature densities of the two metrics 
$g_{ab}$ and $\bar g_{ab}$ are related by
\be
R \,\sqrt{-g} = \bar R \,\sqrt{-\bar g} - 2 \,\sqrt{-\bar g} \sqb \sigma\,,
\qquad (d=2)
\label{RRbar}
\ee
a linear relation in $\sigma$ in two (and only two) dimensions. Multiplying 
(\ref{trtwo}) by $\sqrt{-g}$, using (\ref{RRbar}) and noting that 
$\sqrt{-g}\langle T^a_{\ a} \rangle$ defines the conformal variation,
$\delta \Gamma^{(2)}/\delta \sigma$ of an effective action $\Gamma^{(2)}$, 
we conclude that the $\sigma$ dependence of $\Gamma^{(2)}$ can be at most 
quadratic in $\sigma$. Hence the Wess-Zumino effective action \cite{WZ} in two 
dimensions, $\Gamma_{WZ}^{(2)}$ is
\be
\Gamma_{WZ}^{(2)} [\bar g ; \sigma ] = \frac{N}{24\pi}  
\int\,d^2x\,\sqrt{-\bar g}
\left( - \sigma \sqb \sigma + \bar R\,\sigma\right)\,.
\label{WZact}
\ee
Mathematically the fact that this action functional of the base metric $\bar g_{ab}$ 
and the Weyl shift parameter $\sigma$ cannot be reduced to a single local
functional of the full metric (\ref{confdef}) means that the local Weyl
group of conformal transformations has a non-trivial cohomology,
and $\Gamma_{WZ}^{(2)}$ is a one-form representative of this cohomology 
\cite{AMMc,MazMot}. This is just a formal mathematical statement of the 
fact that a effective action that incorporates the trace anomaly 
in a covariant EFT consistent with the Equivalence Principle 
must exist but that this $S_{anom}[g]$ is necessarily {\it non-local}.

It is straightforward in fact to find a non-local scalar functional 
$S_{anom}^{(2)}[g]$ such that \cite{Polyanom}
\be
\Gamma_{WZ}^{(2)} [\bar g ; \sigma ] = S_{anom}^{(2)}[g= e^{2\sigma}\bar g]
- S_{anom}^{(2)}[\bar g]\,.
\label{cohom}
\ee
By solving (\ref{RRbar}) formally for $\sigma$, and using the 
fact that $\sqrt{-g} \sq = \sqrt{-\bar g} \sqb$ is conformally invariant 
in two dimensions, we find that $\Gamma_{WZ}^{(2)}$ can be written as 
a Weyl shift (\ref{cohom}) with
\be
S_{anom}^{(2)}[g] = \frac{Q^2}{16\pi} \int\,d^2x\,\sqrt{-g}
\int\,d^2x'\,\sqrt{-g'}\, R(x)\,{\sq}^{-1}(x,x')\,R(x')\,,
\label{acttwo}
\ee
and ${\sq}^{-1}(x,x')$ denoting the Green's function inverse of the scalar 
differential operator $\sq$. The parameter $Q^2$ is $-N/6$ if only matter fields 
in a fixed spacetime metric are considered. It becomes $(25 - N)/6$ if account is 
taken of the contributions of the metric fluctuations themselves in addition to 
those of the $N$ matter fields, thus effectively replacing $N$ by $N-25$ \cite{dress}. 
In the general case, the coefficient $Q^2$ is arbitrary, related to the matter 
central charge, and can be treated as simply an additional free parameter of 
the low energy effective action, to be determined.

The anomalous effective action (\ref{acttwo}) is a scalar under coordinate 
transformations and therefore fully covariant and geometric in character,
as required by the Equivalence Principle. However since it involves the 
Green's function $\sq^{-1}(x,x')$, which requires boundary conditions for 
its unique specification, it is quite non-local, and dependent upon more 
than just the local curvature invariants of spacetime. In this important 
respect it is quite different from the classical terms in the action,
and describes rather different physics. In order to expose that physics 
it is most convenient to recast the non-local and non-single valued 
functional of the metric, $S_{anom}^{(2)}$ into a local form by 
introducing auxiliary fields. In the case of (\ref{acttwo}) a single 
scalar auxiliary field, $\varphi$ satisfying
\be
- \sq \varphi = R
\label{auxeomtwo} 
\ee
is sufficient. Then varying
\be
S_{anom}^{(2)}[g;\varphi]  \equiv \frac{Q^2}{16\pi} \int\,d^2x\,\sqrt{-g}\,
\left(g^{ab}\,\nabla_a \varphi\,\nabla_b \varphi - 2 R\,\varphi\right)
\label{actauxtwo}
\ee
with respect to $\varphi$ gives the eq. of motion (\ref{auxeomtwo})
for the auxiliary field, which when solved formally by $\varphi =-{\sq}^{-1}R$
and substituted back into $S_{anom}^{(2)}[g;\varphi]$ returns the non-local
form of the anomalous action (\ref{acttwo}), up to a surface term. 
The non-local information in addition to the local geometry which was previously 
contained in the specification of the Green's function ${\sq}^{-1}(x,x')$ 
now resides in the local auxiliary field $\varphi (x)$, and the freedom 
to add to it homogeneous solutions of (\ref{auxeomtwo}).

The variation of (\ref{actauxtwo}) with respect to the metric yields 
a stress-energy tensor,
\bea
&&T_{ab}^{(2)}[g; \varphi] \equiv -\frac{2}{\sqrt{-g}} 
\frac{\delta S_{anom}^{(2)} [g; \varphi]}{\delta g^{ab}}  \nn\\
&&\hspace{-1.2cm}= \frac{Q^2}{4\pi}\left[-\nabla_a\nabla_b \varphi
+ g_{ab}\, \sq\varphi - \frac{1}{2}(\nabla_a\varphi)(\nabla_b\varphi)
+ \frac{1}{4} g_{ab}\, (\nabla_c\varphi)(\nabla^c\varphi)\right]\!,
\label{anomTtwo}
\eea
which is covariantly conserved, by use of (\ref{auxeomtwo}) and the 
vanishing of the Einstein tensor, $G_{ab} = R_{ab} - Rg_{ab}/2 = 0$ 
in two (and only two) dimensions. The {\it classical} trace of the 
stress tensor (\ref{anomTtwo}) is
\be
g^{ab}T_{ab}^{(2)}[g; \varphi] = \frac{Q^2}{4\pi} \sq \varphi
=- \frac{Q^2}{4\pi}\,R
\label{trTtwo}
\ee
which reproduces the {\it quantum} trace anomaly in a general classical background
(with $Q^2$ proportional to $\hbar$). Hence (\ref{actauxtwo}) is exactly the 
local auxiliary field form of the effective action which should be added to the 
action for two dimensional gravity to take the trace anomaly of massless quantum 
fields into account.

Since the integral of $R$ is a topological invariant in two dimensions, 
the classical Einstein-Hilbert action contains no propagating degrees of freedom 
whatsoever in $d=2$, and it is $S_{anom}^{(2)}$ which contains the {\it only} kinetic 
terms of the low energy EFT. In the local auxiliary field form (\ref{actauxtwo}), 
it is clear that $S_{anom}$ describes an additional scalar degree of freedom 
$\varphi$, not contained in the classical action $S_{cl}^{(2)}$. Once the anomalous
term is treated in the effective action on a par with the classical terms,
its effects become non-perturbative and do not rely on fluctuations
from a given classical background to remain small.

Extensive study of the stress tensor (\ref{trTtwo}) and its correlators, arising 
from this effective action established that the two dimensional trace anomaly 
gives rise to a modification or gravitational ``dressing" of critical exponents 
in conformal field theories at second order critical points \cite{dress}. 
Since critical exponents in a second order phase transition depend only 
upon fluctuations at the largest allowed infrared scale, this dressing is 
clearly an infrared effect, independent of any ultraviolet cutoff. These 
dressed exponents and shift of the central term from $N-26$ to $N-25$
are evidence of the infrared fluctuations of the additional scalar degree 
of freedom $\varphi$ which are quite absent in the classical action. The scaling 
dimensions of correlation functions so obtained are clearly non-perturbative 
in the sense that they are not obtained by considering perturbatively small 
fluctuations around flat space, or controlled by a uniform expansion in 
$\lambda \ll 1$. The appearance of the gravitational dressing exponents and 
the anomalous effective action (\ref{acttwo}) itself have been confirmed in 
the large volume scaling limit of two dimensional simplicial lattice simulations 
in the dynamical triangulation approach \cite{DT,CatMot}. Hence there can be 
little doubt that the anomalous effective action (\ref{actauxtwo}) correctly 
accounts for the infrared fluctuations of two dimensional geometries.

The importance of this two dimensional example is the lessons it allows
us to draw about the role of the quantum trace anomaly in the low energy
EFT of gravity, and in particular the new dynamics it contains in the
conformal factor of the metric. The effective action generated by the 
anomaly in two dimensions contains a {\it new} scalar degree of freedom, 
relevant for infrared physics, beyond the purely local classical action. 
It is noteworthy that the new scalar degree of freedom in (\ref{auxeomtwo}) 
is massless, and hence fluctuates at all scales, including the very
largest allowed. In two dimensions its propagator ${\sq}^{-1}(x,x')$ 
is logarithmic, and hence is completely unsuppressed at large distances. 
This is the precise analog of the massless pole found in flat space triangle 
amplitudes in four dimensions. In $d=2$ this pole appears already in
two-point amplitudes of current correlators in the Schwinger model \cite{LSB},
and in correlators of the energy momentum tensor of conformal fields \cite{AdBertHof}.
Physically this means that the quantum correlations at large distances 
require additional long wavelength information such as macroscopic 
boundary conditions on the quantum state.

The action (\ref{actauxtwo}) due to the anomaly is exactly the missing 
relevant term in the low energy EFT of two dimensional gravity responsible 
for non-perturbative fluctuations at the largest distance scales. This 
modification of the classical theory is required by general covariance 
and quantum theory, and  essentially unique within the EFT framework. 

\subsection{The General Form of the Trace Anomaly in Four Dimensions}

The line of reasoning in $d=2$ dimensions just sketched to find the 
conformal anomaly and construct the effective action may be followed 
also in four dimensions. In $d=4$ the trace anomaly takes the 
somewhat more complicated form,
\be
\langle T^a_{\ a} \rangle
= b F + b' \left(E - \frac{2}{3}\sq R\right) + b'' \sq R + \sum_i \beta_iH_i\,,
\label{tranom}
\ee
in a general four dimensional curved spacetime, where we employ the notation,
\bes\bea
&&E \equiv ^*\hskip-.2cmR_{abcd}\,^*\hskip-.1cm R^{abcd} = 
R_{abcd}R^{abcd}-4R_{ab}R^{ab} + R^2 
\,,\qquad {\rm and} \label{EFdef}\\
&&F \equiv C_{abcd}C^{abcd} = R_{abcd}R^{abcd}
-2 R_{ab}R^{ab}  + \frac{R^2}{3}\,.
\eea\ees 
with $R_{abcd}$ the Riemann curvature tensor, 
$^*\hskip-.1cmR_{abcd}= \varepsilon_{abef}R^{ef}_{\ \ cd}/2$ its dual, 
and $C_{abcd}$ the Weyl conformal tensor. Note that $E$ is the four 
dimensional Gauss-Bonnet combination whose integral gives the Euler 
number of the manifold, analogous to the Ricci scalar $R$ in $d=2$.
The coefficients $b$, $b'$ and $b''$ are dimensionless parameters multiplied
by $\hbar$. Additional terms denoted by the sum $\sum_i \beta_i H_i$ in 
(\ref{tranom}) may also appear in the general form of the trace anomaly, 
if the massless conformal field in question couples to additional long range 
gauge fields. Thus in the case of massless fermions coupled to a background 
gauge field, the invariant $H =$tr($F_{ab}F^{ab}$) appears in (\ref{tranom}) 
with a coefficient $\beta$ determined by the anomalous dimension of the 
relevant gauge coupling. 

As in $d=2$ the form of (\ref{tranom}) and the coefficients $b$ and $b'$ are 
independent of the state in which the expectation value of the stress tensor 
is computed, nor do they depend on any ultraviolet short distance cutoff. 
Instead their values are determined only by the number of massless fields 
\cite{BirDav,anom}, 
\bes
\bea
b &=& \frac{1}{120 (4 \pi)^2}\, (N_S + 6 N_F + 12 N_V)\,,\\
b'&=& -\frac{1}{360 (4 \pi)^2}\, (N_S + \frac{11}{2} N_F + 62 N_V)\,,
\label{bprime}
\eea\label{bbprime}\ees
with $(N_S, N_F, N_V)$ the number of fields of spin 
$(0, \frac{1}{2}, 1)$ respectively and we have taken $\hbar = 1$.
Notice also that $b >0$ while $b' < 0$ for all fields of lower spin 
for which they have been computed. Hence the trace anomaly can lead 
to stress tensors of either sign. 
The anomaly terms can be utilized to generate an effective positive 
cosmological term if none is present initially. Such anomaly driven 
inflation models \cite{Starob} require curvatures comparable to the 
Planck scale, unless the numbers of fields in (\ref{bbprime}) is 
extremely large. It is clear that conformally flat cosmological models 
of this kind, in which the effects of the anomaly can be reduced to a 
purely local higher derivative stress tensor, are of no relevance to the 
very small cosmological term (\ref{cosmeas}) we observe in the acceleration 
of the Hubble expansion today. Instead it is the essentially {\it non-local} 
effects of the anomaly on the horizon scale, much larger than 
$L_{Pl}$ which should come into play. This requires a covariant action
functional analogous to (\ref{actauxtwo}) for a proper treatment.
This is what we now turn to computing.

Three local fourth order curvature invariants $E, F$ and $\sq R$ appear in 
the trace of the stress tensor (\ref{tranom}), but only the first two 
(the $b$ and $b'$) terms of (\ref{tranom}) cannot be derived from a local 
effective action of the metric alone. If these terms could be derived from 
a local gravitational action we would simply make the necessary finite 
redefinition of the corresponding local counterterms to remove them from 
the trace, in which case the trace would no longer be non-zero or anomalous. 
This redefinition of a local counterterm (namely, the $R^2$ term in the 
effective action) is possible only with respect to the third $b''$ coefficient 
in (\ref{tranom}), which is therefore regularization dependent and not part of 
the true anomaly. Only the non-local effective action corresponding to the 
$b$ and $b'$ terms in (\ref{tranom}) are independent of the UV regulator and 
lead to effects that can extend over arbitrarily large, macroscopic distances. 
The distinction of the two kinds of terms in the effective action is emphasized 
in the cohomological approach to the trace anomaly \cite{MazMot}.

To find the WZ effective action corresponding to the $b$ and $b'$ terms in
(\ref{tranom}), introduce as in two dimensions the conformal parameterization 
(\ref{confdef}), and compute
\bes
\bea
&&\sqrt{-g}\,F = \sqrt{-\bar g}\,\bar F\,\label{Fsig}\\
&&\sqrt{-g}\,\left(E - \frac{2}{3}\sq R\right) = \sqrt{-\bar g}\,
\left(\overline E - \frac{2}{3}\sqb\overline R\right) + 4\,\sqrt{-\bar g}\,
\bar\Delta_4\,\sigma\,,
\label{Esig}
\eea
\label{FEsig}
\ees
\vskip-.3cm
\noindent whose $\sigma$ dependence is no more than linear. The 
fourth order differential operator appearing in this expression is 
\cite{AntMot,MazMot,Rie}
\be
\Delta_4 \equiv \sq^2 + 2 R^{ab}\nabla_a\nabla_b - \frac{2}{3} R \sq + 
\frac{1}{3} (\nabla^a R)\nabla_a \,,
\label{Deldef}
\ee
which is the unique fourth order scalar operator that is conformally covariant, 
{\it viz.}
\be
\sqrt{-g}\, \Delta_4 = \sqrt{-\bar g}\, \bar \Delta_4 \,,
\label{invfour}
\ee
for arbitrary smooth $\sigma(x)$ in four (and only four) dimensions. 
Thus multiplying (\ref{tranom}) by $\sqrt{-g}$ and recognizing that the
result is the $\sigma$ variation of an effective action $\Gamma_{WZ}$, we
find immediately that this quadratic effective action is
\be
\hspace{-1mm}
\Gamma_{WZ}[\bar g;\sigma] = b  \int\,d^4x\,\sqrt{-\bar g}\, \bar F\,\sigma
+ b' \int\,d^4x\,\sqrt{-\bar g}\,\left\{\left(\bar E - \frac{2}{3}
\sqb \bar R\right)\sigma + 2\,\sigma\bar\Delta_4\sigma\right\},
\label{WZfour}
\ee
up to terms independent of $\sigma$. This Wess-Zumino action is a 
one-form representative of the non-trivial cohomology of the local 
Weyl group in four dimensions which now contains two distinct cocycles, 
corresponding to the two independent terms multiplying $b$ and $b'$ \cite{MazMot}. 
By solving (\ref{Esig}) formally for $\sigma$, using (\ref{invfour}), and substituting
the result in (\ref{WZfour}) we obtain
\be
\Gamma_{WZ}[\bar g;\sigma] = S_{anom}[g=e^{2\sigma}\bar g] - S_{anom}[\bar g],
\label{Weylshift}
\ee
with the {\it non-local} anomalous action 
\be
S_{anom}[g] = \frac {1}{2}\int d^4x\sqrt{g}\int d^4x'\sqrt{g'}\,
\left(\frac{E}{2} - \frac{\sq R}{3}\right)_x \, \Delta_4^{-1} (x,x')\left[ bF + b'
\left(\frac {E}{2} - \frac{\sq R}{3}\right)\right]_{x'}\,,
\label{Sanom}
\ee
and $\Delta_4^{-1}(x,x')$ denotes the formal Green's function inverse of the fourth 
order differential operator defined by (\ref{Deldef}). From the foregoing
construction it is clear that if there are additional Weyl invariant terms in the 
anomaly (\ref{tranom}) they should be included in the $S_{anom}$ by making the 
replacement $bF \rightarrow bF + \sum_i\beta_i H_i$ in the last square bracket 
of (\ref{Sanom}). The case of the stress-energy of charged fermions coupled to 
photons in QED of Sec. \ref{subsec:TJJ} produces just such an additional
term in the anomaly action, the effective action for which is the flat space
limit of the general anomaly action.

\subsection{Anomaly Effective Action and Massless Scalar Fields}

As detailed in Ref.\,\cite{MotVau} we may render the non-local anomaly action (\ref{Sanom}) 
into a local form, by the introduction of two scalar auxiliary fields $\varphi$ and $\psi$ 
which satisfy fourth order differential eqs.,
\bes\bea
&& \Delta_4\, \varphi = \frac{1}{2} \left(E - \frac{2}{3} \sq R\right)\,,
\label{auxvarphi}\\
&& \Delta_4\, \psi = \frac{1}{2}C_{\alpha\beta\mu\nu}C^{\alpha\beta\mu\nu} 
+ \frac{c}{2b} F_{\alpha\beta}F^{\alpha\beta} \,,
\label{auxvarpsi}
\eea\label{auxeom}\ees
where we have added the last term in (\ref{auxvarpsi}) to take account of the 
background gauge field. For the case of Dirac fermions, $b = 1/320\pi^2$, 
$b' = - 11/5760\pi^2$, and $c= -e^2/24\pi^2$. The local effective action 
corresponding to (\ref{Sanom}) in a general curved space is given by
\be
S_{anom} = b' S^{(E)}_{anom} + b S^{(F)}_{anom} + \frac{c}{2} \int\,d^4x\,\sqrt{-g}\ 
F_{\alpha\beta}F^{\alpha\beta} \varphi\,,
\label{allanom}
\ee
where
\bea
&& \hspace{-5mm}S^{(E)}_{anom}\! \!\equiv \!\frac{1}{2}\! \!\int \!d^4x\sqrt{-g}\!\left\{\!
-\!\left(\sq \varphi\right)^2\!\! +\! 2\!\left(\!R^{\mu\nu}\!\! - \!\frac{R}{3} g^{\mu\nu}\!\right)\!\!
(\nabla_{\mu} \varphi)\!(\nabla_{\nu} \varphi)\! + \!\left(\!E - \frac{2}{3}\! \sq\! R\!\right)
\!\varphi\!\right\}\nn\\
&& S^{(F)}_{anom} \equiv \,\int\,d^4x\,\sqrt{-g}\ \left\{ -\left(\sq \varphi\right)
\left(\sq \psi\right) + 2\left(R^{\mu\nu} - \frac{R}{3}g^{\mu\nu}\right)(\nabla_{\mu} \varphi)
(\nabla_{\nu} \psi)\right.\nn\\
&& \qquad\qquad\qquad + \left.\frac{1}{2} C_{\alpha\beta\mu\nu}C^{\alpha\beta\mu\nu}
\varphi + \frac{1}{2} \left(E - \frac{2}{3} \sq R\right) \psi \right\}\,.
\label{SEF}\eea
The free variation of the local action (\ref{allanom})-(\ref{SEF}) with respect to $\varphi$ 
and $\psi$ yields the eqs. of motion (\ref{auxeom}). Each of these terms when 
varied with respect to the background metric gives a stress-energy tensor in terms 
of the auxiliary fields satisfying eqs. (\ref{auxeom}). 

If we are interested in only the first variation of the action with respect to $g_{\mu\nu}$, 
around flat spacetime, in order to compare to our calculation of the $\lag TJJ\rag$
amplitude in Sec. \ref{subsec:TJJ}, we may drop all terms in (\ref{allanom}) which are 
second order or higher in the metric deviations from flat space. Since $\delta R$ is 
first order in variation around flat space, we may assume from (\ref{auxvarphi}) that 
$\varphi$ is as well. Then the entire $b'S^{(E)}$ contribution to (\ref{allanom}) is 
at least second order in this variation from flat space and cannot contribute to 
$\lag TJJ\rag$. From (\ref{auxvarpsi}) the field $\psi$ has a contribution from 
$F_{\mu\nu}F^{\mu\nu}$ even in flat space, and a potential second order pole, 
$\sq^{-2} \rightarrow k^{-4}$ in its stress tensor. However, retaining only terms 
in (\ref{allanom}) and (\ref{SEF}) which contribute at first order in variation of 
the metric from flat space, we find that the second order pole is partially
cancelled, and we obtain the simpler first order pole form,
\be
S_{anom}[g,A]  \rightarrow  -\frac{c}{6}\int d^4x\sqrt{-g}\int d^4x'\sqrt{-g'}\, R_x
\, \sq^{-1}_{x,x'}\, [F_{\alpha\beta}F^{\alpha\beta}]_{x'}\,,
\label{SSimple}
 \ee
which is valid to first order in metric variations around flat space, or 
\be
S_{anom} [g,A;\varphi,\psi'] =  \int\,d^4x\,\sqrt{-g} 
\left[ -\psi'\sq\,\varphi - \frac{R}{3}\, \psi'  + \frac{c}{2} F_{\mu\nu}F^{\mu\nu} \varphi\right]\,,
\label{effact}
\ee
its local equivalent, where
\bes\bea
&&\psi' \equiv  b \sq\, \psi\,, \label{tpsidef}\\
&&\sq\,\psi' =  \frac{c}{2}\, F_{\mu\nu}F^{\mu\nu} \label{tpsieom}\,,\\
&&\sq\, \varphi = -\frac{R}{3}\,.
\label{phieom}\eea  
\ees
Then after variation we may set $\varphi = 0$ in flat space, and the only terms which remain 
in the stress tensor derived from (\ref{allanom}) are those linear in $\psi'$, {\it viz.}
\be
T^{\mu\nu}[\psi'(z)]=  \frac{2}{\sqrt{-g}}
\frac{\delta S_{anom}}{\delta g_{\mu\nu}(z)}\Bigg\vert_{flat, \varphi=0} = 
\frac{2}{3}\, (g^{\mu\nu}\, \sq - \partial^{\mu}\partial^{\nu})\psi' (z)\,,
\label{anomten}
\ee
which is independent of $b$ and $b'$, and contain only second order differential
operators, after the definition (\ref{tpsidef}). Solving (\ref{tpsieom}) formally 
for $\psi'$ and substituting in (\ref{anomten}), we find
\be
T^{\mu\nu}_{anom}(x) =
\frac{c}{3}  \left(g^{\mu\nu}\sq - \partial^{\mu}\partial^{\nu}\right) \int\,d^4x'\,
 \sq_{x,x'}^{-1}[F_{\alpha\beta}F^{\alpha\beta}]_{x'}\,,
\label{TanomF}
\ee
a result that may be derived directly from (\ref{SSimple}) as well.

By varying (\ref{TanomF}) again with respect to the background gauge potentials,
and Fourier transforming, we obtain
\bea
&&\Gamma_{anom}^{\mu\nu\alpha\beta}(p,q) = \int\,d^4x\,\int\,d^4y\, 
e^{ip\cdot x + i q\cdot y}\,\frac{\delta^2 T^{\mu\nu}_{anom}(0)}
{\delta A_{\alpha}(x) A_{\beta}(y)} \nn\\
&& \qquad = \frac{e^2}{18\pi^2} \frac{1}{k^2} \left(g^{\mu\nu}k^2 - k^{\mu}k^{\nu}\right)
u^{\alpha\beta}(p,q)\,,
\label{Gamanom}
\eea
which gives the full trace for massless fermions,
\be
g_{\mu\nu}T^{\mu\nu}_{anom} = c F_{\alpha\beta}F^{\alpha\beta} = 
-\frac{e^2}{24\pi^2} F_{\alpha\beta}F^{\alpha\beta}\,,
\ee
in agreement with (\ref{tranom}). The {\it tree} amplitude of the effective action (\ref{effact})
which reproduces the pole in the trace part of the $\lag TJJ\rag$ triangle amplitude 
computed in Sec. \ref{subsec:TJJ} is illustrated in Fig. \ref{Fig:psiphi}. 

\begin{figure}[hb]
\includegraphics[height=5cm, viewport=-20 10 200 100,clip]{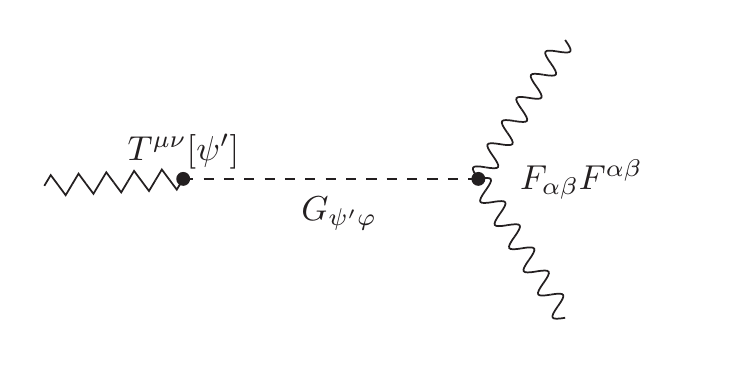}
\caption{Tree Diagram of the Effective Action (\ref{effact}), which reproduces the trace 
of the triangle anomaly. The dashed line denotes the propagator $G_{\psi'\varphi} = \sq^{-1}$ 
of the scalar intermediate state, while the jagged line denotes the 
gravitational metric field variation $h_{\mu\nu} = \delta g_{\mu\nu}$.
Compare to Fig. \ref{Fig:epair}}
\label{Fig:psiphi}
\end{figure}

The massless degrees of freedom $\varphi$ and $\psi'$ are a necessary consequence 
of the trace anomaly, required by imposition of all the other symmetries. In this case these 
are scalar rather than pseudoscalar degrees of freedom. An important physical difference 
with the axial case is that the introduction in QED of a chiral current $J_5^{\mu}$
and axial vector source ${\cal B}_{\mu}$ corresponding to it appear rather artificial, 
and difficult to realize in Nature, whereas the trace of the stress tensor obtained by a 
conformal variation of the effective action is simply a particular metric variation already 
present in the QED Lagrangian in curved space, required by general coordinate 
invariance and the Equivalence Principle, without any additional couplings or 
extraneous fields. Since the stress-energy tensor couples to the universal force of 
gravity, we should expect that physical processes can excite the scalar $\varphi$ 
and $\psi'$ scalar degrees of freedom required by the trace anomaly with a 
gravitational coupling strength, which can produce effects of arbitrarily long
range. 

\subsection{The Scalar Anomaly Pole and Gravitational Scattering Amplitudes} 

In order to verify the existence of the massless scalar pole in a physical process, 
consider the simple tree diagram of gravitational exchange between an arbitrary 
conserved stress-energy source $T^{\prime\,\mu\nu}$ and photons illustrated in 
Fig. \ref{Fig:GravScat}. 
\vspace{-3mm}
\begin{figure}[htp]
\includegraphics[width=28cm, viewport=10 560 800 710,clip]{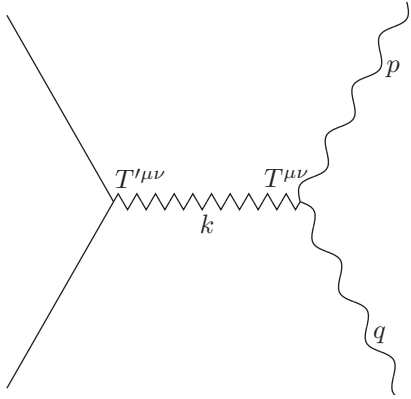}
\caption{Tree Level Gravitational Scattering Amplitude}
\label{Fig:GravScat}
\end{figure}

This process is described by the scattering amplitude\,\cite{Feyn},
\be
{\cal M} = 8\pi G \int d^4x'\int d^4x\, \left[ T^{\prime\,\mu\nu}(x')\, 
\left(\frac{\hspace{.05cm}1}{\sq}\right)_{\hspace{-.1cm}x',x}T_{\mu\nu}(x)
-\frac{1}{2}\,T^{\prime\,\mu}_{\ \ \mu}(x')\,
\left(\frac{\hspace{.05cm}1}{\sq}\right)_{\hspace{-.1cm}x',x}
T^{\nu}_{\ \nu}(x)\right].
\label{ScatM}
\ee
The relative factor of $-\frac{1}{2}$ between the two terms is dictated by the 
requirement that there be no scalar or ghost state exchanged between the two 
sources, and is exactly the prediction of General Relativity, linearized about flat 
space. That only a spin-$2$ propagating degree of freedom is exchanged between 
the two sources in Fig. \ref{Fig:GravScat} can be verified by introducing the following 
$3+1$ decomposition for each of the conserved stress tensors,
\bes\bea
&& T^{00} = T_{00}\,,\\
&& T^{0i} = -V^{\perp\,i} - \partial^i \frac {1\hspace{.1cm}}{\nabla^2}\, \dot T_{00}\,,\\
&& T^{ij} = T^{\perp\,ij} + \partial^i \frac {1\hspace{.1cm}}{\nabla^2}\, \dot V^{\perp\,j}  
+ \partial^j \frac {1\hspace{.1cm}}{\nabla^2}\, \dot V^{\perp\,i} 
+ \frac{1}{2} \left(g^{ij} - \partial^i  \frac {1\hspace{.1cm}}{\nabla^2}\,\partial^j\right) 
(T^{\mu}_{\ \mu} + T_{00}) \nn\\
&& \qquad\qquad - \frac{1}{2} \left(g^{ij} 
- 3\,\partial^i  \frac {1\hspace{.1cm}}{\nabla^2}\,\partial^j\right)
\frac {1\hspace{1mm}}{\nabla^2}\,\ddot T_{00}\,,
\eea\label{decomp}\ees
where $\partial_iV^{\perp\,i} = 0$, $\partial_iT^{\perp\,ij} = T^{\perp\,i}_{\ i} = 0$, and 
$\nabla^{-2}$ denotes the static Green's function of the Laplacian operator, 
$\nabla^2 = \partial^i\partial_i$ in flat space. This parameterization assumes only 
the conservation of the stress-tensor source(s), {\it i.e.} $\partial_{\mu} T^{\mu\nu} = 0$, 
so that there remain six independent components of $T^{\mu\nu}$ which must be 
specified, and we have chosen these six to be $T_{00}, V^{\perp\,i}, T^{\perp\,ij}$
and the total trace $T^{\mu}_{\ \mu}$, which is a spacetime scalar. Substituting the 
decomposition (\ref{decomp}) into (\ref{ScatM}) gives
\bea
&&{\cal M} = 8\pi G \int d^4x'\int d^4x\, \left[ T^{\prime\perp}_{ij} \left(\frac{1}{\sq}\right)_{\hspace{-.1cm}x',x}\hspace{-.2cm}T^{\perp}_{ij}
-2\, V^{\prime\perp}_i  \left(\frac {1\hspace{.1cm}}{\nabla^2}\right)_{\hspace{-.1cm}x',x}
\hspace{-.2cm}V^{\perp}_i\right. \nn\\
&& \hspace{-1cm}\left. + \frac{3}{2}\, T'_{00}\frac {1\hspace{.2cm}}{(\nabla^2)^2}_{\hspace{-.1cm}x',x}
\hspace{-.2cm}\sq \,T_{00} + \frac{1}{2}\,T'_{00} \left(\frac {1\hspace{.1cm}}{\nabla^2}\right)_{\hspace{-.1cm}x',x}\hspace{-.2cm}\,T^{\mu}_{\ \mu}
+ \frac{1}{2}\,T^{\prime\mu}_{\ \ \mu}\,\left(\frac {1\hspace{.1cm}}{\nabla^2}\right)_{\hspace{-.1cm}x',x}\hspace{-.2cm}T_{00}\right]\,,
\label{Scatpos}
\eea
which becomes the five terms
\be
{\cal M} \rightarrow -8\pi G \left[ T^{\prime\perp}_{ij} \frac{1}{k^2}\,T^{\perp}_{ij}\,
-2\, V^{\prime\perp}_i\, \frac{1}{\vec k^2}\,V^{\perp}_i
+  \frac{3}{2}\, T'_{00}\,\frac{k^2}{(\vec k^2)^2}\, \,T_{00}
+ \frac{1}{2}T'_{00}\, \frac{1}{\vec k^2}\,T^{\mu}_{\ \mu}  + \frac{1}{2}\,T^{\prime\mu}_{\ \ \mu}\, 
\frac{1}{\vec k^2}\,T_{00}\right]
\label{Scatdecom}
\ee
in momentum space. These expressions show that only the spatially transverse 
and tracefree components of the stress tensor, $T^{\perp}_{ij}$ exchange a physical 
propagating helicity $\pm 2$ graviton in the intermediate state, characterized by a 
Feynman (or for classical interactions, a retarded) massless propagator 
$-\sq^{-1} \rightarrow k^{-2}$ pole in the first term of (\ref{Scatpos}) or (\ref{Scatdecom}). 
All the other four terms in either expression contain only an instantaneous Coulomb-like 
interaction $-\nabla^{-2} \rightarrow \vec k^{-2}$ or $\nabla^{-4} \rightarrow \vec k^{-4}$ 
between the sources, in which no propagating physical particle appears in the 
intermediate state of the cut diagram. This is the gravitational analog of the decomposition,
\bes\bea
&& J^0 = \rho\,,\\
&& J^i = J^{\perp\,i} - \partial^i \frac {1\hspace{.1cm}}{\nabla^2}\, \dot \rho\,,
\eea\ees
of the conserved electromagnetic current and corresponding tree level scattering amplitude, 
\be
\int\! d^4x'\!\int\!\! d^4x\, J^{\prime\mu}(x')\! \left(\frac{1}{\sq}\right)_{\hspace{-.1cm}x',x} 
\hspace{-.3cm}J_{\mu}(x) 
\rightarrow -J^{\prime\mu} \frac{1}{k^2} J_{\mu} = -J^{\prime\perp}_i\, \frac{1}{k^2}\,J^{\perp}_i 
+ \rho'\, \frac{1}{\vec k^2}\,\rho \,,
\label{Coulint}
\ee
which shows that only a helicity $\pm 1$ photon is exchanged between the transverse components of the current, the last term in (\ref{Coulint}) being the instantaneous Coulomb interaction between the charge densities.

We now replace one of the stress tensor sources by the matrix element of the
one-loop anomalous amplitude, considering first the trace term with the anomaly pole in $F_1$.
This corresponds to the diagram in Fig. \ref{Fig:GravPhoTri}.
We find for this term,
\bes\bea
&& \lag 0\vert T_{00} \vert p,q\rag_{_1}= - \vec k^2 F_1(k^2)\, u^{\alpha\beta}(p,q)
\tilde A_{\alpha}(p) \tilde A_{\beta}(q)\,\\
&& \lag 0\vert T^{\mu}_{\ \mu}  \vert p,q\rag_{_1} = 3k^2 F_1(k^2)\, 
u^{\alpha\beta}(p,q)\tilde A_{\alpha}(p) \tilde A_{\beta}(q)\,\\
&& \lag 0\vert V^{\perp}_i \vert p,q\rag_{_1} = \lag 0\vert T^{\perp}_{ij} \vert p,q\rag_{_1} = 0\,.
\eea\ees
Hence the scattering amplitude (\ref{Scatdecom}) becomes simply,
\be
{\cal M}_1 = 4\pi G \, T^{\prime\mu}_{\ \ \mu}\,F_1(k^2)\,u^{\alpha\beta}(p,q)
\tilde A_{\alpha}(p) \tilde A_{\beta}(q)
=  \frac{4\pi G}{3} \, T^{\prime\mu}_{\ \ \mu}\,\frac{1}{k^2}\,
\lag 0\vert T^{\nu}_{\ \nu}  \vert p,q\rag_{_1}\,
\label{scat1}
\ee
where (\ref{trace1}) has been used for $m=0$. Thus for massless fermions the pole
in the anomaly amplitude becomes a scalar pole in the gravitational scattering amplitude, 
appearing in the intermediate state as a massless scalar exchange between the traces of the 
energy-momentum tensors on each side. The standard gravitational interaction with the 
source has produced an effective interaction between the scalar auxiliary field $\psi'$ 
and the trace $T^{\prime\mu}_{\ \ \mu}$ with a well defined gravitational coupling. 
Thus we may equally well represent the scattering as Fig. \ref{Fig:GravPhoTri} involving 
the fermion triangle, or as the tree level diagram Fig. \ref{Fig:GravScal} of the effective theory, 
with a massless scalar exchange. 

\begin{figure}[htp]
\includegraphics[width=9.6cm, viewport=-70 10 220 160,clip]{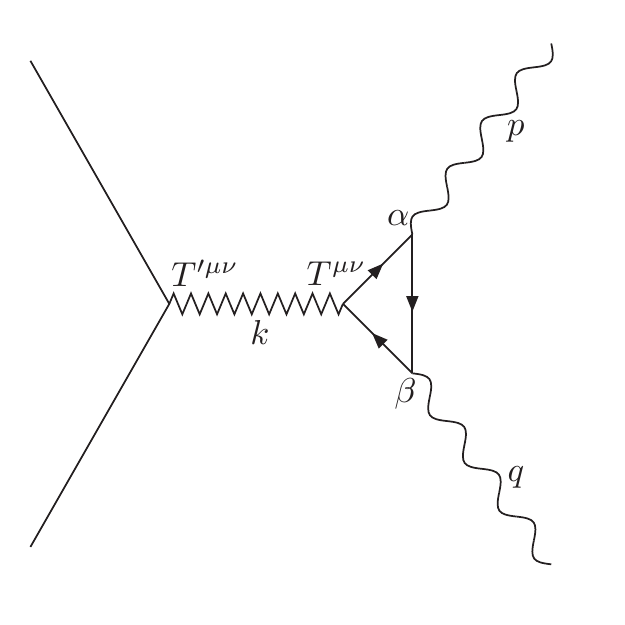}
\caption{Gravitational Scattering of photons from the source $T^{\prime\mu\nu}$ via the triangle amplitude}
\label{Fig:GravPhoTri}
\end{figure}

\begin{figure}[htp]
\includegraphics[width=8.8cm,viewport=-60 10 220 150,clip]{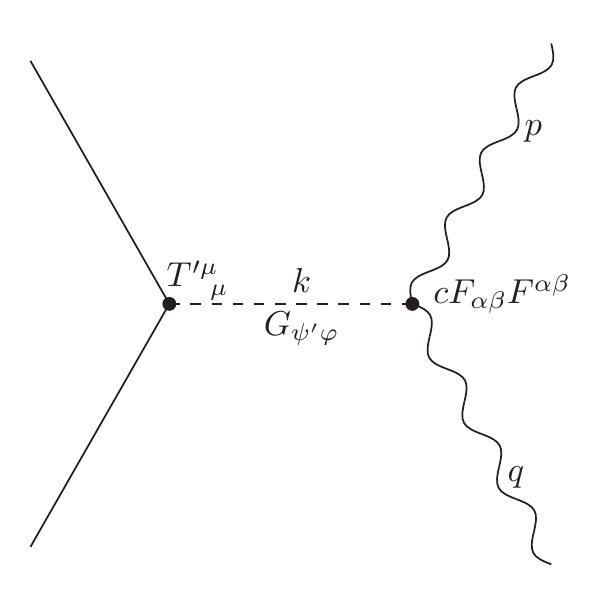}
\caption{Gravitational scattering of photons from the trace of a source $T^{\prime\mu}_{\ \ \mu}$
via massless scalar exchange in the effective theory of (\ref{effactT}).}
\label{Fig:GravScal}
\end{figure}

This tree diagram is generated by the effective action in flat space modified from (\ref{effact}) to
\be
S_{eff} [g,A;\varphi,\psi'] =  \int\,d^4x\,\sqrt{-g} 
\left[ -\psi'\sq\,\varphi + \frac{8\pi G}{3}\, T^{\prime\mu}_{\ \ \mu}\, \psi'
+ \frac{c}{2}\, F_{\alpha\beta}F^{\alpha\beta} \varphi\right],
\label{effactT}
\ee
to include the coupling to the trace of the energy-momentum tensor of any matter 
$T^{\prime\mu}_{\ \ \mu}$ source. Correspondingly the eq. (\ref{phieom}) for $\varphi$ becomes
\be
\sq \varphi = \frac{8\pi G}{3} \,T^{\prime\mu}_{\ \ \mu}\,,
\label{newphieom}
\ee
instead of (\ref{phieom}).
The eq. of motion for $\psi'$ remains (\ref{tpsieom}). We note that if the 
source $T^{\prime\mu\nu}$ generates the curvature $R$ by Einstein's eqs., then
$R = -8\pi G\,T^{\prime\mu}_{\ \ \mu}$, so that (\ref{effactT}) and (\ref{newphieom}) 
are equivalent to (\ref{effact}) and (\ref{phieom}) at leading order in $G$.

We conclude that in the conformal limit of massless electrons, the pole in the trace 
sector of the $\lag TJJ\rag$ anomaly amplitude contributes to gravitational scattering 
amplitudes as would a scalar field coupled to the trace of the energy-momentum tensor of 
classical sources. The gravitationally coupled intermediate scalar can be understood 
as arising from collinear $e^+e^-$ correlated pairs in a total spin $0^+$ state in 
Fig. \ref{Fig:epair}. Although the result appears similar in some respects to a 
Brans-Dicke scalar \cite{JBD}, and indeed (\ref{newphieom}) is identical in form to 
that of a Brans-Dicke scalar with vanishing Brans-Dicke coupling ($\omega = 0$), because 
of the unusual off-diagonal kinetic term dictated by the structure of the trace anomaly, 
(\ref{effactT}) is certainly {\it not} a Brans-Dicke theory. Thus, although (\ref{newphieom}) 
tells us that $\varphi$ is sourced by the trace of matter stress tensor with gravitational 
strength, $\varphi$ cannot react back on matter except through $\psi'$ and hence 
$F_{\alpha\beta}F^{\alpha\beta}$, reproducing (\ref{SSimple}), from whence it
was derived.  Each of the two scalar fields each couples to a {\it different} source, 
with an off-diagonal propagator, $G_{\psi'\varphi}$. There is no direct coupling of 
the trace of matter stress tensors $T^{\prime\mu}_{\ \mu}$ to $T^{\nu}_{\ \nu}$ 
via a scalar exchange as there would be in a classical scalar tensor theory
of the Jordan-Brans-Dicke kind. Hence the phenomenology of (\ref{effactT})
will be quite different, and the observational limits on a Jordan-Brans-Dicke 
scalar \cite{Will} do not apply. Note also that the matrix elements of 
$F_{\alpha\beta}F^{\alpha\beta} = 2(\vec E^2 - \vec B^2)$ vanish for 
monochromatic photons on shell. Thus to leading order the scattering diagram 
in Fig. \ref{Fig:GravPhoTri} also does not contribute to photon scattering on shell. 
The contribution of the massless scalar pole to higher order or off shell physical 
processes and the prospects for detecting its effects experimentally
are important and interesting questions currently under investigation.

\subsection{The Effective Action of Low Energy Gravity}

With the foregoing detailed consideration of anomalies, massless poles and
their long distance effects, we consider finally the EFT of four dimensional
macroscopic gravity. This gravitational EFT is determined by the same general 
principles as in other contexts \cite{DGH}, namely by an expansion in powers of 
derivatives of local terms consistent with symmetry. Short distance effects are 
parameterized by the coefficients of local operators in the effective action, 
with higher order terms suppressed by inverse powers of an ultraviolet cutoff 
scale $M_{_{UV}}$. The effective theory need not be renormalizable, as indeed Einstein's 
theory is not, but is expected nonetheless to be quite insensitive to the details 
of the underlying microscopic degrees of freedom, because of decoupling \cite{DGH}. 
It is the decoupling of short distance degrees of freedom from the macroscopic 
physics that makes EFT techniques so widely applicable, and which we assume 
applies also to gravity.

As a covariant metric theory with a symmetry dictated by the Equivalence Principle, 
General Relativity may be regarded as just such a local EFT, truncated at second 
order in derivatives of the metric field $g_{ab}(x)$ \cite{Dono}. When quantum 
matter is considered, the stress tensor $T^a_{\ b}$ becomes an operator. 
Because the stress tensor has mass dimension four, containing up to quartic
divergences, the proper covariant renormalization of this operator requires 
fourth order terms in derivatives of the metric. However the effects of such 
higher derivative {\it local} terms in the gravitational effective action are 
suppressed at distance scales $L \gg L_{Pl}$ in the low energy EFT limit. 
Hence surveying only local curvature terms, it is often tacitly assumed that 
Einstein's theory contains all the low energy macroscopic degrees of freedom 
of gravity, and that General Relativity cannot be modified at macroscopic 
distance scales, much greater than $L_{Pl}$, without violating general 
coordinate invariance and/or EFT principles. As we have argued previously
in two dimensions, this presumption should be re-examined in the presence of 
quantum anomalies.

When a classical symmetry is broken by a quantum anomaly, the naive decoupling of 
short and long distance physics assumed by an expansion in local operators with 
ascending inverse powers of $M_{_{UV}}$ fails. In this situation even the low energy 
symmetries of the effective theory are changed by the presence of the anomaly, 
and some remnant of the ultraviolet physics survives in the low energy description. 
An anomaly can have significant effects in the low energy EFT because it is not 
suppressed by any large energy cutoff scale, surviving even in the limit 
$M_{_{UV}} \rightarrow \infty$. Any explicit breaking of the symmetry in the 
classical Lagrangian serves only to mask the effects of the anomaly, but in the 
right circumstances the effects of the non-local anomaly may still dominate the 
local terms. The axial anomaly in QCD, discussed in Sec. \ref{subsec:axial} has
low energy effects, unsuppressed by the EFT ultraviolet cutoff scale, 
$M_{_{UV}} \sim \Lambda_{QCD}$ in that case. Although the quark masses are 
non-zero, and chiral symmetry is only approximate in Nature, the chiral anomaly 
gives the dominant contribution to the low energy decay amplitude of 
$\pi^0\rightarrow 2\gamma$ in the standard model \cite{Adler,BarFriGM}, a 
contribution that is missed entirely by a local EFT expansion in pion fields. 
Instead the existence of the chiral anomaly requires the explicit addition to 
the local effective action of a {\it non-local} term in four physical dimensions 
to account for its effects \cite{WZ,DGH}. Although when an anomaly is present, 
naive decoupling between the short and long distance degrees of freedom fails, 
it does so in a well-defined way, with a coefficient that depends only on the 
quantum numbers of the underlying microscopic theory. 

The low energy effective action for gravity in four dimensions 
contains first of all, the local terms constructed from the
Riemann curvature tensor and its derivatives and contractions up to
and including dimension four. This includes the usual Einstein-Hilbert
action of General Relativity,
\be
S_{EH}[g] = \frac{1}{16\pi G} \int\, d^4x\,\sqrt{-g}\, (R-2\Lambda)  
\label{clfour}
\ee
as well as the spacetime integrals of the fourth order curvature
invariants,
\be
S_{local}^{(4)}[g] = \frac{1}{2} \int\, d^4x\,\sqrt{-g}\, \left(\alpha C_{abcd}C^{abcd}
+ \beta R^2\right),
\label{R2}
\ee
with arbitrary dimensionless coefficients $\alpha$ and $\beta$. There are two 
additional fourth order invariants, namely $E= ^*\hskip-.2cm R_{abcd}\,
^*\hskip-.1cm R^{abcd}$ and $\sq R$, which could be added to (\ref{R2}) as well, 
but as they are total derivatives yielding only a surface term and no local 
variation, we omit them. All the possible local terms in the effective action 
may be written as the sum,
\be
S_{local}[g] = \frac{1}{16\pi G}\int\,d^4x\,\sqrt{-g}\,(R-2\Lambda) +
S_{local}^{(4)} + \sum_{n=3}^{\infty} S_{local}^{(2n)}\,.
\label{locsum}
\ee
with the terms in the sum with $n\ge 3$ composed of integrals of local curvature 
invariants with dimension $2n \ge 6$, and suppressed by $M_{_{UV}}^{-2n + 4}$ at 
energies much less than $M_{_{UV}}$. Here $M_{_{UV}}$ is the ultraviolet cutoff scale 
of the low energy effective theory which we may take to be of order $M_{pl}$. The 
higher derivative terms with $n \ge 3$ are irrelevant operators in the infrared, 
scaling with negative powers under global rescalings of the metric, and may be 
neglected at macroscopic distance scales, at least with respect to the
classical scaling dimensions. On the other hand the two terms in 
the Einstein-Hilbert action $n= 0, 1$ scale positively, and are clearly 
relevant in the infrared. The fourth order terms in (\ref{R2}) are 
neutral under such global rescalings. 

The exact quantum effective action also contains non-local terms in general. 
All possible terms in the effective action (local or not) can be classified 
according to how they respond to global Weyl rescalings of the metric,
{\it i.e.} $\sigma=\sigma_0 =$ const. If the non-local terms are non-invariant 
under global rescalings, then they scale either positively or negatively under 
(\ref{confdef}). If $m^{-1}$ is some fixed length scale associated with the non-locality, 
arising for example by the integrating out of fluctuations of fields with mass $m$, 
then at much larger macroscopic distances ($mL \gg 1$) the non-local 
terms in the effective action become approximately local. The terms which 
scale with positive powers of $e^{\sigma_0}$ are constrained by general 
covariance to be of the same form as the $n=0,1$ Einstein-Hilbert terms 
in $S_{local}$, (\ref{clfour}). Terms which scale negatively with 
$e^{\sigma_0}$ become negligibly small as $mL \gg 1$ and are infrared 
irrelevant at macroscopic distances. This is the expected decoupling of 
short distance degrees of freedom in an effective field theory description, 
which are verified in detailed calculations of loops in massive field 
theories in curved space. The only possibility for contributions 
to the effective field theory of gravity at macroscopic distances,
which are not contained in the local expansion of (\ref{locsum}) arise 
from fluctuations not associated with any finite length scale, {\it i.e.} 
$m=0$. These are the non-local contributions to the low energy EFT which 
include those associated with the anomaly.

The anomaly effective action is associated with non-trivial cohomology of the
Weyl group in the space of metrics. Since $\Gamma_{WZ}$ (\ref{WZfour})
from which the effective action of the anomaly was derived satisfies WZ consistency,
{\it i.e.} is closed but not exact under the Weyl group, it is unique up to an 
arbitrary admixture of local trivial cocycles, which in physical terms are either
trivial because they are completely Weyl invariant effective actions obeying
\be
S_{inv} [e^{2\sigma} g] = S_{inv}[g]\,,
\label{Sinv}
\ee
and drop out of difference (\ref{Weylshift}), or they are purely local terms easily 
catalogued by ascending powers of the Riemann curvature tensor, its covariant
derivatives and contractions in (\ref{locsum}). Thus the classification of
terms according to their global Weyl scaling properties tells us that
the exact effective action of any covariant theory must be of the form \cite{MazMot},
\be
S_{exact}[g] = S_{local}[g] + S_{inv}[g] + S_{anom}[g]\,,
\label{Sexact}
\ee
with $S_{local}$ given by the expansion (\ref{locsum}), $S_{inv}$ the (generally 
non-local) Weyl invaraint terms satisfying (\ref{Sinv}), and $S_{anom}$ the 
anomaly action given by (\ref{allanom})-(\ref{SEF}). The higher dimension local 
terms in (\ref{locsum}) are strictly irrelevant in the IR, since they scale 
to zero with negative powers of $e^{\sigma_0}$ and may be neglected for physics 
far below the Planck scale, while the lower dimension local terms are nothing 
but the terms of the usual Einstein-Hilbert classical action (\ref{clfour}). 
These classical terms grow as positive powers of $e^{\sigma_0}$ under global 
dilations and are clearly IR relevant terms. Indeed the naive classical 
scaling of these terms are positive powers ($L^4$ and $L^2$) under 
rescaling of distance, and are clearly relevant operators of the low 
energy description. 

The local dimension four terms involving the Weyl tensor squared $C^2$ is fully 
locally Weyl invariant (while that involving $R^2$ is invariant only under global
Weyl rescalings) and among the many terms that can appear in $S_{inv}$. Because 
both of these are neutral under global dilations, scaling like $L^0$ we expect 
them to be marginally irrelevant in the IR (as well as in the UV). All the higher 
dimension local terms in the sum in (\ref{locsum}) for $n\ge 3$ scale to zero as 
$\sigma_0 \rightarrow \infty$ and are clearly strictly IR irrelevant. However
among all the possible terms that can be generated by quantum loops in the
exact effective action of gravity, the anomaly effective action is {\it unique}
in scaling logarithmically under the global Weyl group. Indeed we note that in 
the form (\ref{SEF}) the simple shift of the auxiliary field $\varphi$ by a spacetime 
constant, 
\be
\varphi \rightarrow \varphi + 2\sigma_0
\ee
corresponding to a global logarithmic variation of length scales
yields the entire dependence of $S_{anom}$ on the global Weyl rescalings
(\ref{confdef}), {\it viz.}
\bea
&& S_{anom}[g; \varphi, \psi] \rightarrow 
S_{anom}[e^{2\sigma_0} g; \varphi + 2 \sigma_0, \psi] \nn\\
&& \quad = S_{anom}[g; \varphi, \psi] + \sigma_0\!
\,\int\,d^4x\,\sqrt{-g}\left[ bF + b'\left(E - \frac{2}{3} \sq R\right)\right],\qquad\quad
\label{logscale}
\eea
owing to the strict invariance of the terms quadratic in the auxiliary fields under 
(\ref{confdef}) and eqs. (\ref{FEsig}). Hence $S_{anom}$ scales
logarithmically ($\sim\log L$) with distance under Weyl rescalings. 

Because of this infrared sensitivity to global rescalings, unlike local higher derivative 
terms in the effective action, which are either neutral or scale with negative powers 
of $L$, the anomalous terms should not be discarded in the low energy, large distance 
limit. Ordinarily, {\it i.e.} absent anomalies, the Wilson effective action should 
contain only {\it local} infrared relevant terms consistent with symmetry \cite{RG}. 
However, like the anomalous effective action generated by the chiral anomaly 
in QCD, the non-local $S_{anom}$ must be included in the low energy EFT to 
account for the anomalous Ward identities, even in the zero momentum limit, 
and indeed logarithmic scaling with distance (\ref{logscale}) indicates that 
$S_{anom}$ is an infrared relevant term. Even if no massless matter fields 
are assumed, the quantum fluctuations of the metric itself will generate a term 
of the same form as $S_{anom}$ in the infrared \cite{AMMc}. The scalar fields 
of the local form (\ref{allanom})-(\ref{SEF}) of $S_{anom}$ describe massless 
scalar degrees of freedom of low energy gravity, not contained in classical General 
Relativity. As we have seen, these massless scalars may be understood as correlated 
two-particle states of the underlying anomalous QFT, and show up as poles in gauge 
invariant physical scattering amplitudes in the EFT where the original quantum 
fields appear only in internal quantum loops. Thus the effective action of the 
anomaly $S_{anom}$ should be retained in the EFT of low energy gravity, which is 
specified then by the first two strictly relevant local terms of the classical 
Einstein-Hilbert action (\ref{clfour}), and the logarithmic $S_{anom}$, {\it i.e.}
\be
S_{eff}[g] = S_{_{EH}}[g] + S_{anom}[g;\varphi,\psi]
\label{Seff}
\ee
with $S_{anom}$ is given by (\ref{Sanom}) or in its local form by 
(\ref{allanom})-(\ref{SEF}). 

The complete classification of the terms in the exact effective action (\ref{Sexact})
into just three categories means that all possible infrared relevant terms in the low 
energy EFT, which are not contained in $S_{local}$ of (\ref{locsum}) must fall into 
$S_{anom}$, {\it i.e.} they must correspond to non-trivial co-cycles of the local Weyl 
group \cite{MazMot,MotVau}. The Weyl invariant terms in the exact effective action 
(\ref{Sinv}) are by definition insensitive to rescaling of the metric at large distances. 
Hence the (generally quite non-local) terms in $S_{inv}$ do not give rise to infrared 
relevant terms in the Wilson effective action for low energy gravity. By this classification 
of terms (local or non-local) according to their behavior under global Weyl rescalings, 
the Wilson effective action (\ref{Seff}) contains all the infrared relevant terms in low energy 
gravity for energies much less than $M_{pl}$. 

Note also that it would be {\it inconsistent} with the semi-classical Einstein eqs.
(\ref{scE}) to have as their source a stress tensor which is not covariantly conserved.
Thus in a clash of symmetries which a quantum anomaly presents, it is necessary
to choose the option that conformal invariance is broken, {\it not} general coordinate
invariance. The action $S_{anom}$ is invariant under general coordinate
transformations and under no conditions ({\it i.e.} in the presence of an horizon 
or not) is there a gravitational anomaly \cite{Bertl}. This is not a state dependent
condition dependent on whether an horizon exists or not, but a condition of consistency
of the semi-classical Einstein eqs. (\ref{scE}). Under the defining assumptions 
of general covariance and the EFT hypothesis of decoupling of physics associated 
with massive degrees of freedom, any infrared modifications of Einstein's theory 
generated by quantum effects is tightly constrained and the effective action 
(\ref{Seff}) becomes essentially unique. The addition of the anomaly term(s) and 
the scalar degrees of freedom $\varphi$ and $\psi$ they contain to the low energy 
effective action of gravity amounts to a non-trivial infrared modification of General 
Relativity, required by the existence of the trace anomaly, fully consistent with both 
quantum theory and the Equivalence Principle. 

\section{Macroscopic Effects of the Trace Anomaly}
\label{sec:anom}

Having a fully covariant effective action the most straightforward application is
to compute the covariantly conserved stress-energy tensor corresponding to it,
and study its effects in particular backgrounds. If the effects are significant it
will be necessary then to include the anomalous stress tensor as a source
for Einstein's eqs. to obtain new solutions, but at first one can evauate the
stress tensor in certain fixed backgrounds, such as the Schwarzschild black
hole geometry discussed in Sec. \ref{sec:BH} and the de Sitter geometry of Sec. 
\ref{sec:darkenergy}. This will provide the first evidence of the relevance
of the anomaly effective action and the scalars $\varphi$ and $\psi$ for
macroscopic gravity in the presence of horizons.

After these studies in fixed classical backgrounds one can consider next
dynamical effects of the fluctuations associated with the anomaly scalars,
their role in cosmology and their relevance to both the problem of 
gravitational collapse and cosmological dark energy.

\subsection{Anomaly Stress Tensor in Schwarzschild Spacetime}

From (\ref{allanom}) and (\ref{SEF}) the stress tensor of the anomalous 
effective action consists of two independent terms,
\be
T_{ab}^{(anom)} \equiv -\frac{2}{\sqrt{-g}} \frac{\delta
S_{anom}}{\delta g^{ab}}[g;\varphi,\psi] = b' E_{ab} + b F_{ab} 
\label{Tanom}
\ee
namely
\bea
E_{ab} &=&-2\, (\nabla_{(a}\varphi) (\nabla_{b)} \sq \varphi) 
+ 2\,\nabla^c \left[(\nabla_c \varphi)(\nabla_a\nabla_b\varphi)\right]  
- \frac{2}{3}\, \nabla_a\nabla_b\left[(\nabla_c \varphi)
(\nabla^c\varphi)\right]\nn\\ 
&&  
+ \frac{2}{3}\,R_{ab}\, (\nabla_c \varphi)(\nabla^c \varphi)  
- 4\, R^c_{\ (a}(\nabla_{b)} \varphi) (\nabla_c \varphi)
+ \frac{2}{3}\,R \,(\nabla_a \varphi) (\nabla_b \varphi)\nn\\ 
&& \hspace{-1.2cm} + \frac{1}{6}\, g_{ab}\, \left\{-3\, (\sq\varphi)^2 
+ \sq \left[(\nabla_c\varphi)(\nabla^c\varphi)\right] 
+ 2\left( 3R^{cd} - R g^{cd} \right) (\nabla_c \varphi)(\nabla_d
\varphi)\right\}\nn\\ 
&&
\hspace{-1cm} - \frac{2}{3}\, \nabla_a\nabla_b \sq \varphi 
- 4\, C_{a\ b}^{\ c\ d}\, \nabla_c \nabla_d \varphi
- 4\, R_{(a}^c \nabla_{b)} \nabla_c\varphi 
+ \frac{8}{3}\, R_{ab}\, \sq \varphi  
+ \frac{4}{3}\, R\, \nabla_a\nabla_b\varphi \nn\\
&&  \hspace{-1.5cm} - \frac{2}{3}\!\left(\nabla_{(a}R\right) \nabla_{b)}\varphi
+ \frac{1}{3}g_{ab}\left\{ 2\sq^2 \varphi + 6R^{cd} \,\nabla_c\nabla_d\varphi 
- 4R\sq \varphi + (\nabla^c R)\nabla_c\varphi\right\}
\label{Eab}
\eea
and
\bea
&& \hspace{-.6cm}F_{ab} \!=\! -2(\nabla_{(a}\varphi) (\nabla_{b)}\!\sq \psi) 
\!-\!2(\nabla_{(a}\psi) (\nabla_{b)}\!\sq \varphi)
\!+\! 2\nabla^c\! \left[(\nabla_c \varphi)(\nabla_a\!\nabla_b\psi)\! 
 +\! (\nabla_c \psi)(\nabla_a\!\nabla_b\varphi)\right]
\nn\\ 
&&  
\hspace{-5mm}- \frac{4}{3}\nabla_a\nabla_b\left[(\nabla_c \varphi)
(\nabla^c\psi)\right]
+ \frac{4}{3}R_{ab}(\nabla_c \varphi)(\nabla^c \psi) 
- 4R^c_{\ (a}\!\left[(\nabla_{b)} \varphi) (\nabla_c \psi)
\!+\! (\nabla_{b)} \psi) (\nabla_c \varphi)\right]\nn\\ 
&& 
\hspace{1cm} + \frac{4}{3}R \,(\nabla_{(a} \varphi) (\nabla_{b)} \psi) 
+ \frac{1}{3}g_{ab}\Big\{-3(\sq\varphi)(\sq\psi)
+ \sq \left[(\nabla_c\varphi)(\nabla^c\psi)\right] 
\nn\\ 
&&
\hspace{.3cm} \left. + 2\left( 3R^{cd} - R g^{cd} \right) (\nabla_c
\varphi)(\nabla_d \psi)\right\}- 4\, \nabla_c\nabla_d\left( C_{(a\ b)}^{\ \ c\ \ d}
\varphi \right)  - 2\, C_{a\ b}^{\ c\ d} R^{cd} \varphi \nn\\
&&
\hspace{-1mm}- \frac{2}{3}\, \nabla_a\nabla_b \sq \psi 
- 4\, C_{a\ b}^{\ c\ d}\, \nabla_c \nabla_d \psi
- 4\, R_{(a}^c (\nabla_{b)} \nabla_c\psi) 
+ \frac{8}{3}\, R_{ab}\, \sq \psi  
+ \frac{4}{3}\, R\, \nabla_a\nabla_b\psi \nn\\
&& \hspace{-4mm} - \frac{2}{3}\! \left(\nabla_{(a}R\right) \nabla_{b)}\psi + 
\frac{1}{3}g_{ab}\left\{\!2\sq^2 \psi + 6R^{cd} \,\nabla_c\nabla_d\psi
- 4R\sq \psi + (\nabla^c R)(\nabla_c\psi)\!\right\}.
\label{Fab}
\eea
Each of these two tensors are individually conserved and they have the local traces,
\bes
\bea
E^a_{\ a} &=& 2 \Delta_4 \varphi = E - \frac{2}{3} \sq R\,,\\
F^a_{\ a} &=& 2 \Delta_4 \psi = F = C_{abcd}C^{abcd}\,,
\label{Ftr}
\eea\label{EFtraces}\ees
corresponding to the two terms respectively in the trace anomaly 
in four dimensions in (\ref{tranom}) (with $\beta_i=0$). 

In the four dimensional Schwarzschild geometry the full contraction of the
Riemann tensor is given by (\ref{RieS}), and since $R_{ab} =0$  
the invariants $F\big\vert_{_S} = E\big\vert_{_S} = {48\,M^2}/{r^6}$.
(where we temporarily set $GM = M$ to simplify the expressions below).
A particular solution of either of the inhomogeneous Eqs. (\ref{auxeom}) 
is given then by $\bar \varphi (r)$, with
\be
\frac{d\bar\varphi}{dr}\Big\vert_{_S} = -\frac{4M}{3r^2f} \ln \left(\frac{r}{2M}\right)
-\frac{1}{2M}\left(1 + \frac{4M}{r} \right)
\label{partsoln}
\ee
and $f(r) = 1 - \frac{2M}{r}$, The general solution of (\ref{auxeom}) for 
$\varphi =\varphi(r)$ away from the singular points $r= (0, 2M, \infty)$ is easily 
found and may be expressed in the form 
\bea
&&\hspace{-5mm}\frac{d\varphi}{dr}\Big\vert_{_S} = \frac{d\bar\varphi}{dr}\Big\vert_{_S} 
+ \frac{2M c_{_H}}{r^2f} + \frac{q-2}{4M^2r^2f}\! \int_{2M}^r\!drr^2\ln f 
+ \frac{c_\infty}{2M}\left(\!\frac{r}{2M} + 1 + \frac{2M}{r}\!\right) \nn\\
&&\hspace{-3mm}=\frac{q-2}{6M}\left(\frac{r}{2M} + 1 + \frac{2M}{r}\right) 
\ln f - \frac{q}{6r}\left[\frac{4M}{r-2M} 
\ln \left(\frac{r}{2M}\right) + \frac{r}{2M} + 3 \right] \nn\\
&& \qquad  - \frac{1}{3M} - \frac{1}{r} + \frac{2M c_{_H}}{r (r-2M)} 
+ \frac{c_\infty}{2M}\, \left(\frac{r}{2M} + 1 + \frac{2M}{r}\right)
\label{phipS}
\eea
in terms of the three dimensionless constants of integration,
$c_{_H}$, $c_\infty$, and $q$. This expression has the limits,
\bes
\bea
&&\frac{d\varphi}{dr}\Big\vert_{_S}\! \!\rightarrow\! \frac{c_{_H}}{r-2M}
+ \frac{q-2}{2M} \ln\left(\frac{r}{2M} - 1\right) - \frac{1}{2M}\!
\left(3 c_\infty - c_{_H} - q - \frac{5}{3}\right) + \dots,\qquad r\rightarrow 2M;
\label{Schlima}\\
&&\hspace{-7mm}\frac{d\varphi}{dr}\Big\vert_{_S}\! \rightarrow\!\frac{c_\infty r}{4M^2} + 
\frac{2c_\infty - q}{4M} + \frac{c_\infty}{r} - \frac{2M}{3r^2}\,q\, 
\ln \left(\frac{r}{2M}\right) + \frac{2M}{r^2}\left[c_{_H} - \frac{7}{18}
(q - 2)\right] + \dots,\quad r \rightarrow \infty .
\label{Schlimb}
\eea\ees
Hence $c_{_H}$ controls the leading behavior as $r$ approaches the 
horizon, while $c_\infty$ controls the leading behavior as $r\rightarrow \infty$, 
which is the same as in flat space. The leading behavior at the horizon is 
determined by the homogeneous solution to (\ref{auxeom}), $c_{_H} \ln f = c_{_H}
\ln (-K^aK_a)^{\frac{1}{2}}$ where $K = \partial_t$ is the timelike Killing
field of the Schwarzschild geometry for $r>2M$. Because of these singular
behaviors in (\ref{Schlimb}), (\ref{phipS}) is clearly a solution of
(\ref{EFtraces}) in the distributional sense, {\it i.e.} containing possible
$\delta$ functions or derivatives thereof at the origin, at the horizon and 
at infinity.

To the general spherically symmetric static solution (\ref{phipS}) we may add 
also a term linear in $t$, {\it i.e.} we replace $\varphi(r)$ by
\be
\varphi (r,t) = \varphi (r) + \frac{p} {2M}\,t\,.
\label{tSch}
\ee
Linear time dependence in the auxiliary fields is the only allowed
time dependence that leads to a time-independent stress-energy.
The conformal transformation to flat space near the horizon
corresponds to the particular choice $c_{_H} = \pm p = 1$,
leaving the subdominant terms in (\ref{Schlima}) parameterized
by $q$ and $c_\infty$ undetermined.

Since the equation for the second auxiliary field $\psi$ is identical to that 
for $\varphi$, its solution for $\psi = \psi(r,t)$ is of the same form as 
(\ref{phipS}) and (\ref{tSch}) with four new integration constants, 
$d_{_H}, d_\infty$, $q'$ and $p'$ replacing $c_{_H}, c_\infty$, $q$,
and $p$ in $\varphi(r,t)$. Adding terms with any higher powers of $t$ or 
more complicated $t$ dependence produces a time dependent stress-energy tensor. 
Inspection of the stress tensor terms in (\ref{Tanom}) also shows that it does not 
depend on either a constant $\varphi_0$ or $\psi_0$ but only the derivatives of 
both auxiliary fields in Ricci flat metrics such as Schwarzschild spacetime. For 
that reason we do not need an additional integration constant for either of the 
fourth order differential equations (\ref{auxeom}). 

With the general spherically symmetric solution for $\varphi(r,t)$ and 
$\psi(r,t)$, we can proceed to compute the stress-energy tensor (\ref{Tanom})
in a stationary, spherically symmetric quantum state. For example the Boulware 
state is characterized as that state which approaches the flat space vacuum as 
rapidly as possible as $r\rightarrow \infty$ \cite{Boul}. In the flat space 
limit this means that the allowed $r^2$ and $r$ behavior in the auxiliary
fields ($r$ and constant behavior in their first derivatives) must be set
to zero. Inspection of the asymptotic form (\ref{Schlimb}) shows that
this is achieved by requiring
\bes
\bea
c_{\infty} = d_{\infty} &=& 0\qquad {\rm and}\\
q=q' &=& 0\,.\qquad {\rm (Boulware)}
\eea\label{Boulcond}\ees
If we set $p=p'=0$ as well, in order to have a static ansatz for the 
Boulware state, then the remaining two constants $c_{_H}$ and $d_{_H}$ 
are free parameters of the auxiliary fields, which lead to a stress-energy 
which diverges as $r \rightarrow 2M$ on the horizon. Matching the leading
divergence of the stress-energy in the Boulware state (\ref{TBoul}) by 
adjusting $c_{_H}$ and $d_{_H}$ appropriately, one then has a one 
parameter fit to the numerical data of \cite{JenMcOtt}. The results of this fit of 
(\ref{Tanom}) with $c_{_H}$ and $d_{_H}$ treated as free parameters are 
illustrated in Figs. \ref{fig:TBOtt} to \ref{fig:TBOthth}, for all three non-zero components 
of the stress tensor expectation value of a massless, conformally coupled scalar field 
in the Boulware state. The (approximate) best fit values plotted were obtained 
with $c_{_H} = -\frac{7}{20}$ and $d_{_H} = \frac{55}{84}$. 

\begin{figure}[htp]
\hskip -3mm
\includegraphics[height=3in,width=5in] {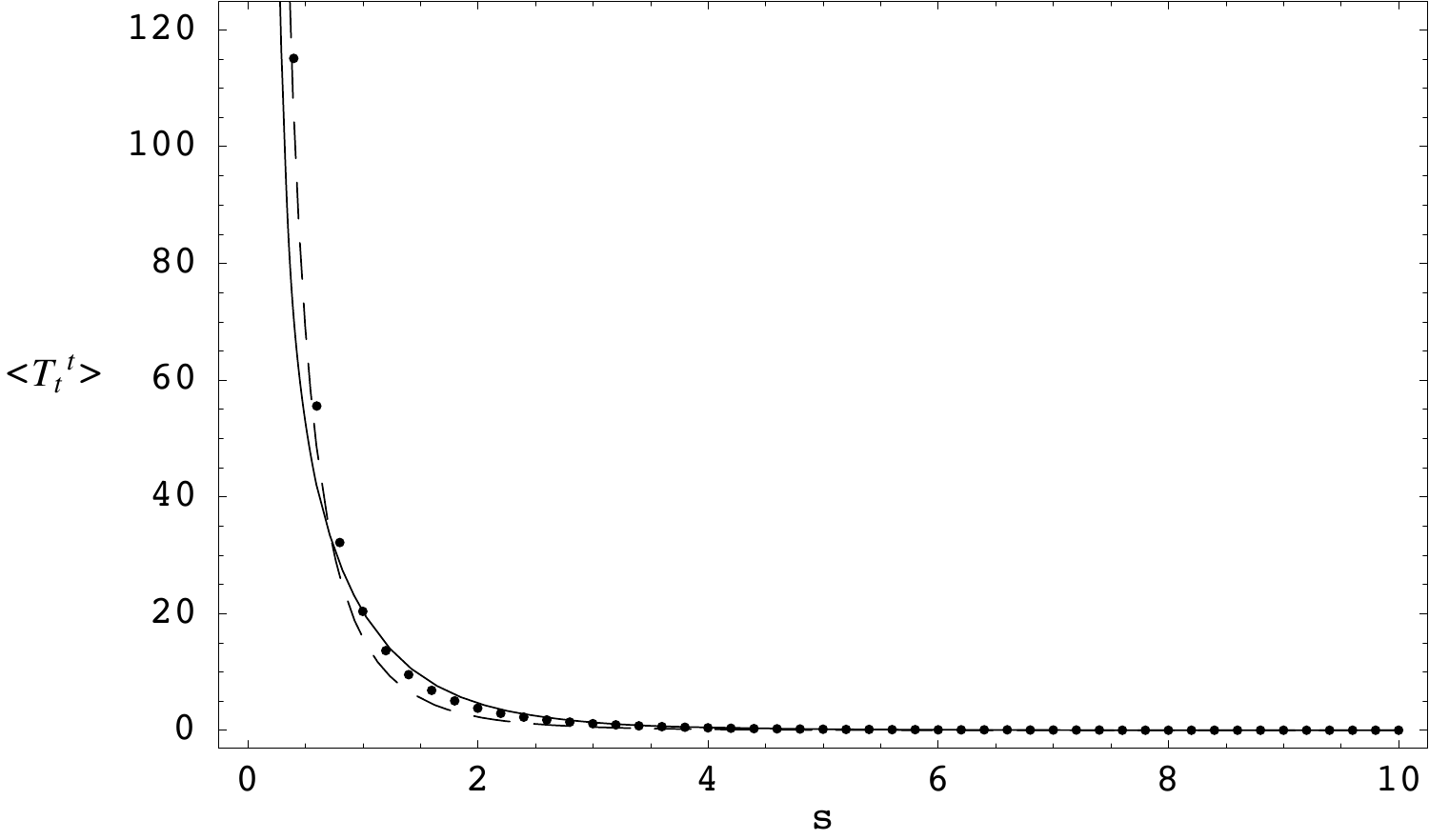}
\caption{The expectation value $\langle T_t^{\ t} \rangle$ of a conformal 
scalar field in the Boulware state in Schwarzschild spacetime, as a function 
of $s=\frac{r-2M}{M}$ in units of $\pi^2 T_H^4/90$. The solid curve is Eq. 
(\ref{Tanom})-(\ref{Fab}) with (\ref{partsoln}), (\ref{phipS}), (\ref{Boulcond}) 
and $c_{_H} = -\frac{7}{20}, d_{_H} = \frac{55}{84}$. The dashed curve is 
the analytic approximation of  \cite{BroOtt}, and the points are the numerical 
results of \cite{JenMcOtt}.}
\label{fig:TBOtt}
\end{figure}

\begin{figure}[htp]
\hskip -3mm
\includegraphics[height=3in,width=5in]{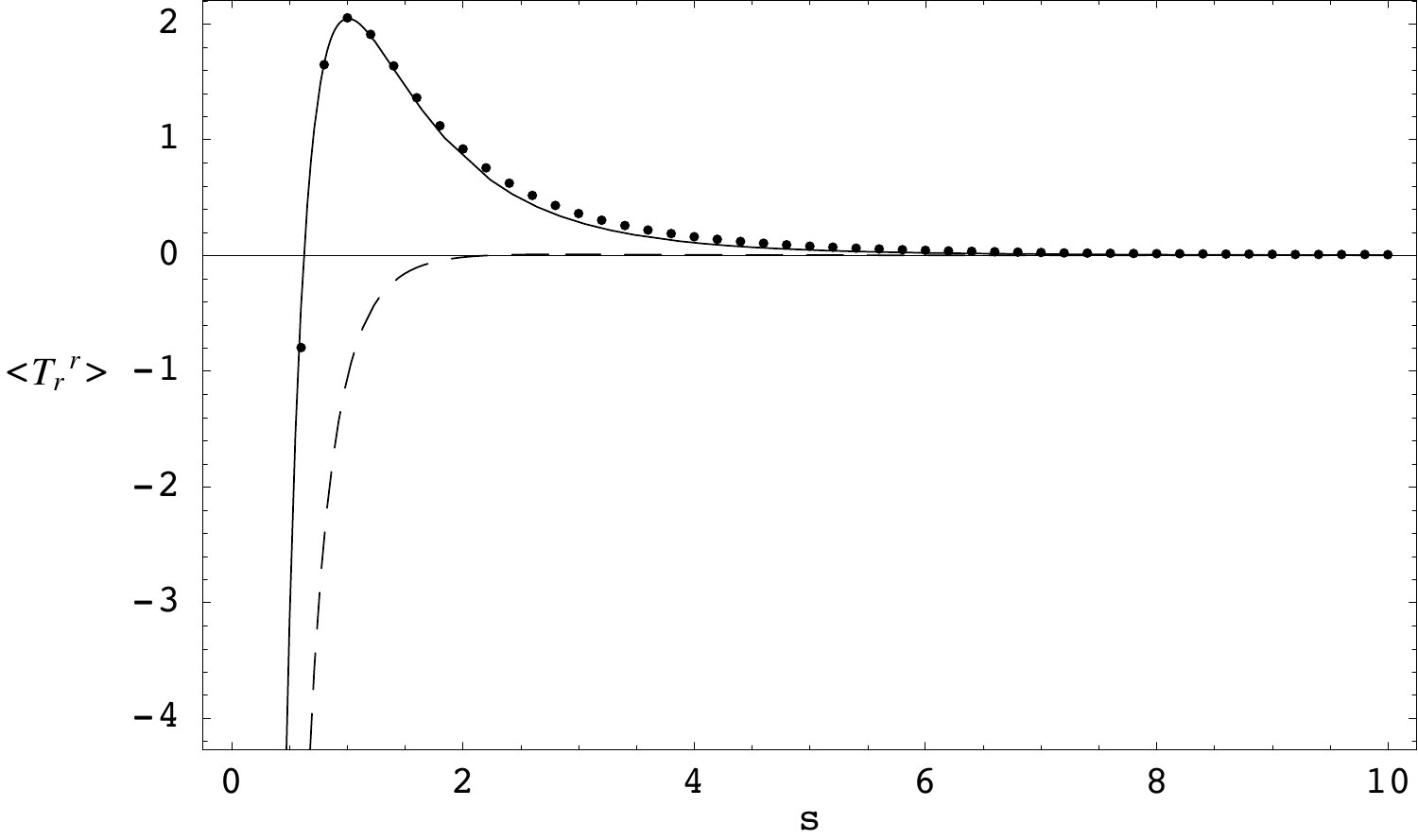}
\vspace{-1mm}
\caption{The radial pressure $\langle T_r^{\ r} \rangle$ of a 
conformal scalar field in the Boulware state in Schwarzschild spacetime.
The axes and solid and dashed curves and points are as in Fig. \ref{fig:TBOtt}.
Note the better agreement of the anomaly stress tensor (solid curve) with
the numerical results of \cite{JenMcOtt}, compared to the analytic 
approximation of \cite{BroOtt}.}
\label{fig:TBOrr}
\vspace{1cm}
\end{figure}

\begin{figure}[htp]
\hskip -3mm
\includegraphics[height=3in,width=5in]{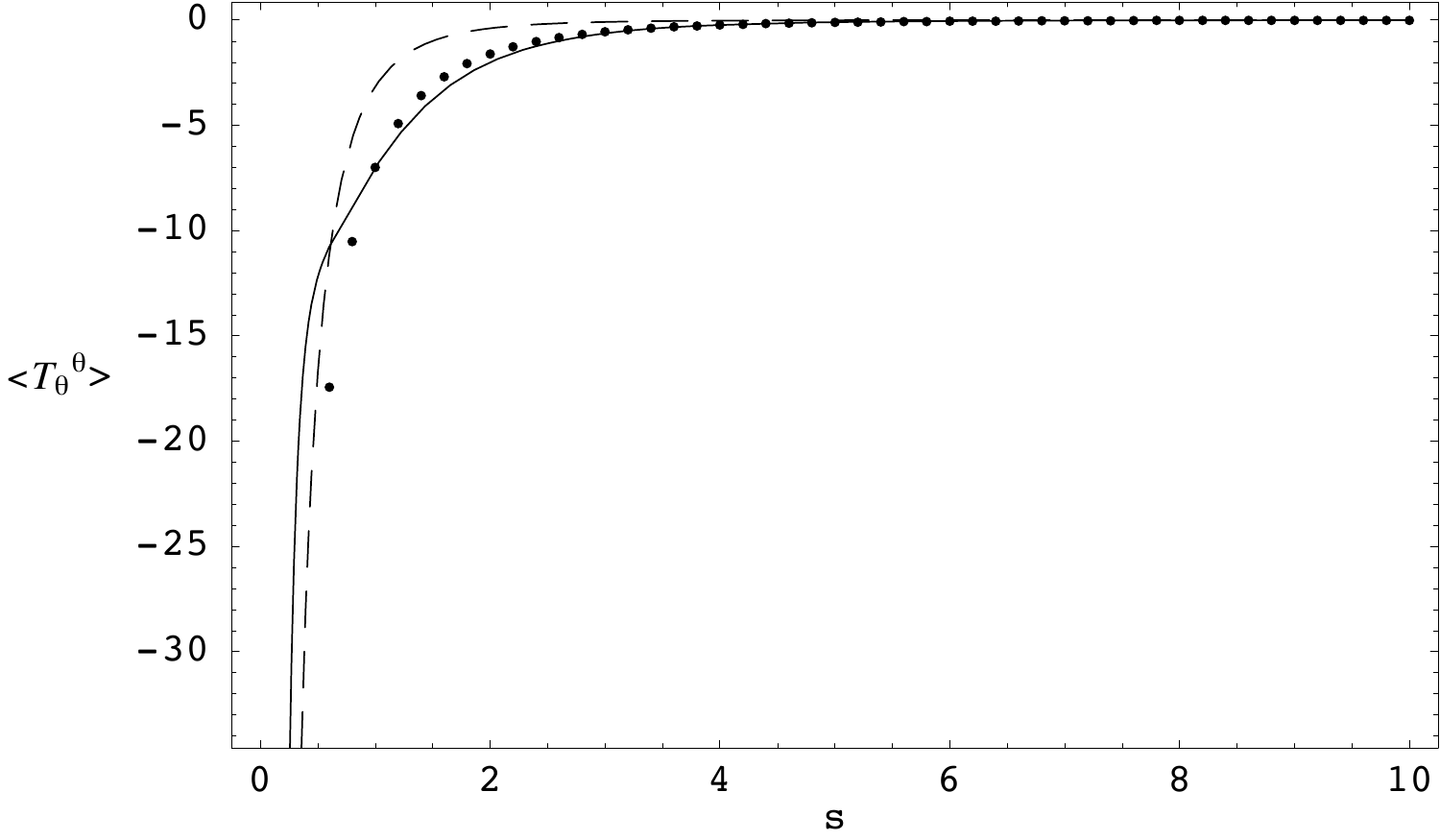}
\vspace{-1mm}
\caption{The tangential pressure $\langle T_\theta^{\ \theta} \rangle$ of 
a conformal scalar field in the Boulware state in Schwarzschild spacetime.
The axes and solid and dashed curves and points are as in Fig. \ref{fig:TBOtt}.}
\label{fig:TBOthth}
\end{figure}

For comparison purposes, we have plotted also the analytic approximation 
of Page, Brown, and Ottewill \cite{BroOtt,Pag82,BrOtP} (dashed curves in 
Figs. \ref{fig:TBOtt} to \ref{fig:TBOthth}). We observe that the two parameter fit 
with the anomalous stress tensor in terms of the auxiliary $\varphi$ and $\psi$ 
fields is more accurate than the approximation of refs. \cite{BroOtt,Pag82,BrOtP} 
for the Boulware state.

The next important point to emphasize is that the stress-energy diverges on 
the horizon in an entire family of states for generic values of the eight 
auxiliary field parameters $(c_{_H}, q, c_{\infty}, p; d_{_H}, q', d_{\infty}, p')$, 
in addition to the Boulware state. Hence in the general allowed parameter space 
of spherically symmetric macroscopic states, horizon divergences of the 
stress-energy are quite generic, and not resticted to the Boulware state. 
On the other hand, the condition that the stress-energy on
the horizon be finite gives four conditions on these eight parameters, in 
order to cancel the four possible divergences $s^{-2}, s^{-1},\ln^2 s,\ln s$. 
The simplest possibility with the minimal number of conditions on the 
auxiliary field parameters is via
\bes
\bea
&& (2b + b') c_{_H}^2 + p (2b p' + b'p) = 0 \qquad (s^{-2})\\
&& (b + b') c_{_H} = bd_{_H} \qquad \qquad (s^{-1})
\label{fincondb}\\
&& \qquad q=q' = 2 \qquad\quad (\ln^2s \ \ {\rm and} \ \ \ln s)\,. 
\eea\label{fincond}\ees
It is clear that these finiteness conditions on the horizon are incompatible
with the conditions for fall off at infinity (\ref{Boulcond}), in their
values of $q$ and $q'$. Thus the effective action and behavior of the
anomaly scalar fields and stress tensor illustrate at a glance the general
topological obstruction discussed in Sec. \ref{sec:BH} for vacuum behavior 
of $\lag T^a_{\ b}\rag_{_R}$ at both the horizon and infinity simultaneously.
With conditions (\ref{Boulcond}) the anomaly scalar fields give a very weak 
Casimir-like long-range interaction between massive bodies, that falls off 
very rapidly with distance (at least as fast as $r^{-7}$).

By taking different values of the parameters $(c_{_H}, q, c_{\infty}, p; d_{_H}, q', d_{\infty}, p')$
of the homogeneous solutions to (\ref{auxeom}), corresponding to different states of
the underlying quantum theory, it is possible to approximate the stress tensor of the
Hartle-Hawking and Unruh states as well \cite{MotVau,BFS}. In states which are regular 
on the horizon there is no particular reason to neglect the Weyl invariant terms 
$S_{inv}$ in the exact effective action, and the stress-energy tensor it produces 
would be expected to be comparable in magnitude to that from $S_{anom}$. In 
this case of bounded stress tensors, the contributions from both $S_{anom}$ and 
$S_{inv}$ are both of order $M^{-4}$ and negligibly small on macroscopic scales 
in any case. However in states such as the previous Boulware example, the
diverging behavior of the stress tensor near the horizon is captured accurately 
by the terms in (\ref{Tanom}) arising from the anomaly, which have the same 
generically diverging behaviors as the quantum field theory expectation value 
$\lag T^a_{\ b}\rag_{_R}$. This general behavior has been found in the 
Reissner-Nordstr\o m case of electrically charged black holes as well \cite{AndMotVau}.
We find no {\it a priori} justification for excluding generic states with stress-energy 
tensors that grow without bound as the horizon is approached. In any such states the 
backreaction of the stress-energy on the geometry will be substantial in this region 
and lead to large backreaction effects near the horizon.

\subsection{Anomaly Stress Tensor in de Sitter Spacetime}

In conformally flat spacetimes with $g_{ab} = e^{2\sigma}\eta_{ab}$, one can
choose $\varphi = 2 \sigma$ and $\psi = 0$ to obtain the stress tensor
of the state conformally transformed from the Minkowski vacuum. In this state
$F_{ab}$ vanishes, and $\varphi$ can be eliminated completely
in terms of the Ricci tensor with the result \cite{MazMot,BroCas},
\be
E_{ab}= \frac{2}{9} \nabla_a\nabla_b R + 2 R_a^{\ c}R_{bc} - \frac{14}{9}R R_{ab}
+ g_{ab} \left(-\frac{2}{9}\sq R - R_{cd}R^{cd} + \frac{5}{9} R^2\right)\,.
\label{Ecflat}
\ee
Thus all non-local dependence on boundary conditions of the auxiliary fields
$\varphi$ and $\psi$ drops out in conformally flat spacetimes for the
state conformally mapped from the Minkowski vacuum. In the special case
of maximally $O(4,1)$ symmetric de Sitter spacetime, $R_{ab} = 3H^2 g_{ab}$
with $R = 12H^2$ a constant, 
\be
E_{ab}\Big\vert_{dS} = 6 H^4\, g_{ab}\,.
\ee
Hence we obtain immediately the expectation value of the stress tensor
of a massless conformal field of any spin in the Bunch-Davies (BD) state 
in de Sitter spacetime,
\be
T_{ab}\Big\vert_{BD,dS} = 6b' H^4 g_{ab} = -\frac{H^4}{960\pi^2}\,g_{ab}\,
(N_s + 11 N_f + 62 N_v)\,,
\label{BDdS}
\ee
which is determined completely by the trace anomaly.

States in de Sitter space which are not maximally $O(4,1)$ symmetric are easily 
studied by choosing diiferent solutions of (\ref{auxeom}) for $\varphi$ and $\psi$.
For example, if de Sitter spacetime is expressed in the spatially flat coordinates
(\ref{FLRW}) and $\varphi = \varphi(\tau)$, we obtain from (\ref{Tanom}) the 
stress-energy in spatially homogeneous, isotropic states. Since in de Sitter
spacetime the $\Delta_4$ operator factorizes,
\be
\Delta_4\Big\vert_{dS} \varphi = \left(\sq - 2H^2\right) (\sq \varphi) = 12H^4\,.
\label{deseom}
\ee
it is straightforward to show that the general solution to this
equation with $\varphi = \varphi(\tau)$ is
\be
\varphi(\tau) = 2 H \tau + c_0 + c_{-1} e^{-H\tau} + c_{-2} e^{-2H\tau} 
+ c_{-3} e^{-3H\tau} \,.
\label{destau}
\ee
When the full solution for $\varphi(\tau)$ of (\ref{destau}) is substituted into 
(\ref{Eab}) we obtain additional terms in the stress tensor which are not de Sitter 
invariant, but which fall off at large $\tau$, as $e^{-4H\tau}$. The stress tensor of 
this time behavior is traceless and corresponds to the redshift of massless modes 
with the equation of state, $p= \rho/3$. 

States of lower symmetry in de Sitter spacetime may be found by considering static 
coordinates (\ref{dSstat}), in terms of which the operator $\Delta_4$ again
separates. Inserting the ansatz $\varphi = \varphi(r)$ in (\ref{deseom}), the general 
$O(3)$ spherically symmetric solution regular at the origin is easily found:
\be
\varphi(r)\big\vert_{dS} = \ln\left(1-H^2r^2\right) + c_0 + \frac{q}{2} 
\ln\left(\frac{1-Hr}{1+Hr}\right)
+ \frac{2c_{_H} - 2 - q}{2Hr} \ln\left(\frac{1-Hr}{1+Hr}\right)\,.
\label{genphids}
\ee
A possible homogeneous solution proportional to $1/r$ has been discarded, 
since it is singular at the origin. An arbitrary linear time dependence $2Hpt$ 
could also be added to $\varphi (r)$, {\it i.e.} 
$\varphi(r) \rightarrow \varphi(r,t) = \varphi(r) + 2Hpt$. 
The particular solution,
\be
\varphi_{_{BD}}(r,t) = \ln\left(1-H^2r^2\right) + 2Ht = 2H\tau\,.
\label{phiBD}
\ee
is simply the previous solution for the Bunch-Davies state we found
in the homogeneous flat coordinates (\ref{FLRW}). 

In the general $O(3)$ symmetric state centered about the origin of
the coordinates $r=0$, the stress tensor is generally dependent on
$r$. In fact, it generally diverges as the observer horizon $r= H^{-1}$ 
is approached, much as in the Schwarzschild case considered
previously. From (\ref{genphids}) we observe that
\be
\varphi(r)\big\vert_{dS} \rightarrow 
\left[c_{_H} + \left(c_{_H} - 1 - \frac{q}{2}\right) (1-Hr)
+ \dots \right] \ln\left(\frac{1-Hr}{2}\right)
+ \ {\cal O}(1-Hr)\,,
\label{hordeS}
\ee
as $Hr \rightarrow 1$, so that the integration constant $c_{_H}$ controls the 
most singular behavior at the observer horizon $r=H^{-1}$. 

The second auxiliary field $\psi$ satisfies the homogeneous equation,
\be 
\Delta_4 \psi = 0\,,
\label{psidel4}
\ee
which has the general spherically symmetric solution linear in $t$,
\be
\psi(r,t)\big\vert_{dS} = d_0 + 2Hp't + \frac{q'}{2} 
\ln\left(\frac{1-Hr}{1+Hr}\right)
+ \frac{2d_{_H} - q'}{2Hr} \ln\left(\frac{1-Hr}{1+Hr}\right)\,.
\label{genpsids}
\ee
Note that the constant $d_{_H}$ enters this expression differently than
$c_{_H}$ enters the corresponding Eq. (\ref{genphids}), due to the
inhomogeneous term in (\ref{deseom}), which is absent from the $\psi$
equation. Since the anomalous stress tensor is independent of $c_0$ and $d_0$,
it depends on the six parameters $(c_{_H}, d_{_H}, q, q', p , p')$
in the general stationary $O(3)$ invariant state. The simplest way
to insure no $r^{-2}$ or $r^{-1}$ singularity of the stress tensor at 
the origin is to choose $q=q'=0$. With $q=q'=0$ there are no sources 
or sinks at the origin and the zero flux condition,
\be
T^r_{\ t} = -\frac{4H^2}{r^2}\,\left( bpq' + bp'q + b'pq\right) = 0
\label{dSflux}
\ee
is satisfied automatically, for any $p$ and $p'$. Because of the subleading 
logarithmic behavior (\ref{hordeS}) of $\varphi$ on the observer horizon at 
$r= H^{-1}$, there will $\ln^2 (1-Hr)$ and $\ln (1-Hr)$ divergences in the 
other components of the stress tensor (\ref{Tanom}) in general. These
divergences are removed when $q=q'=0$ and $c_{_H} = 1$.
All the divergences of the stress-energy at both the origin and 
the observer horizon are cancelled if the four conditions,
\bes
\bea
&& 2b\left(d_{_H} + pp'\right)= b'\left(1-p^2\right)\\
&&c_{H} = 1 \qquad {\rm and}\\
&&q=q'=0
\eea\label{findes}\ees
are satisfied. These are satisfied by the Bunch-Davies state 
with $p=\pm 1$ and $d_{_H} = p'=0$. In fact, with conditions (\ref{findes}) 
the tensor $F_{ab}$ of (\ref{Fab}) vanishes identically, the $\psi$
field drops out entirely, and the full anomalous stress tensor is 
given by (\ref{Eab}), which takes the Bunch-Davies form (\ref{BDdS}).

If the first two of the four conditions (\ref{findes}) are relaxed, then the 
stress-energy remains finite at the origin, but becomes divergent at $r=H^{-1}$. 
This is the generic case. It includes in particular the analog of the static 
Boulware vacuum \cite{Boul} in de Sitter space. There is no analog of the Unruh 
state \cite{Unr} in de Sitter spacetime, since by continuity a flux through 
the future or past observer horizon at $r=H^{-1}$ would require a source or sink 
of flux at the origin $r=0$, a possibility we have excluded by (\ref{dSflux}) 
above.

The important conclusion from these studies of the stress tensor (\ref{Tanom}) 
obtained from the anomaly effective action is that rather than being a pathology 
of a single state the divergent behavior of the stress tensor on the horizon is 
{\it generic}. The general spherically symmetric solutions of the linear scalar field 
eqs. (\ref{auxeom}) are easy found, which allows an overview of a variety of
quantum states in the underlying field theory which would normally require
a laborious effort to study individually, and for each spin field separately. This 
overview of possible states shows that at least four conditions (\ref{fincond})
on the set of eight parameters is necessary to eliminate divergences on the 
future horizon of a Schwarzschild black hole, and a similar set of conditions 
(\ref{findes}) are necessary to remove divergences on the de Sitter horizon. 
This suggests that in the generic state of the gravitational collapse problem 
the backreaction effects will be large, and alter significantly the classical 
picture of a black hole horizon, which is the source of the paradoxes 
discussed in Sec. \ref{sec:BH}, and that the black hole and cosmological horizon 
singularities may be related.

\subsection{Conformal Phase of 4D Gravity and Infrared Running of $\Lambda$}
\label{Lamrun}

In order to understand the {\it dynamical} effects of the kinetic terms in the 
anomaly effective action, one can consider simplest case of the quantization
of the conformal factor in the Wess-Zumino action (\ref{WZfour}) in the case
that the fiducial metric is flat, {\it i.e.} $g_{ab} = e^{2 \sigma}\eta_{ab}$. 
Then the Wess Zumino effective action (\ref{WZfour}) simplifies to
\bea
\Gamma_{WZ} [\eta; \sigma] = -\frac{Q^2}{16\pi^2} \int d^4 x \ (\sq \sigma)^2 \,,
\label{flata}
\eea
where 
\be
Q^2 \equiv -32 \pi^2 b'\,.
\label{Qdef}
\ee
This action quadratic in $\sigma$ is the action of a free scalar field, albeit with 
a kinetic term that is fourth order in derivatives. The propagator for this kinetic 
term is $(p^2)^{-2}$ in momentum space, which is a logarithm in position space, 
\be
G_{\sigma} (x,x') = -\frac{1}{2Q^2} \, \ln \left[\mu^2 (x-x')^2\right]\,,
\ee
the same behavior as that for a massless field in two dimensions.
Of course this is no accident but rather 
a direct consequence of the association with the anomaly of a conformally invariant 
differential operator, $\sq$ in two dimensions and $\Delta_4$ in four dimensions, 
a pattern which continues in all higher even dimensions. Because of this logarithmic 
propagator a similar sort of infrared fluctuations, conformal fixed point and dressing 
exponents as those obtained in two dimensional gravity \cite{AntMot}.

The classical Einstein-Hilbert action for a conformally flat
metric $g_{ab} = e^{2 \sigma}\eta_{ab}$ is
\be
\frac{1}{8\pi G}\ \int d^4 x \ \left[ 3 e^{2 \sigma} 
(\partial_a \sigma)^2 - \Lambda e^{4 \sigma} \right]\,,
\label{cEin}
\ee
which has derivative and exponential self-interactions in $\sigma$.
It is remarkable that these complicated interactions can be treated 
systematically using the the fourth order kinetic term of (\ref{flata}). 
In fact, these interaction terms are renormalizable and their anomalous 
scaling dimensions due to the fluctuations of $\sigma$ can be computed 
in closed form \cite{AntMot,Odint,AMMd}. Direct calculation of the conformal 
weight of the Einstein curvature term shows that it acquires an 
anomalous dimension $\beta_2$ given by the quadratic relation,
\be
\beta_2 = 2 + \frac{\beta_2^2}{2Q^2}\,.
\label{betE}
\ee
In the limit $Q^2 \rightarrow \infty$ the fluctuations of $\sigma$ are 
suppressed and we recover the classical scale dimension of the coupling 
$G^{-1}$ with mass dimension $2$. Likewise the cosmological term in 
(\ref{cEin}) corresponding to the four volume acquires an anomalous 
dimension given by 
\be
\beta_0 = 4 + \frac{\beta_0^2}{2Q^2}\,.
\label{betL}
\ee
Again as $Q^2 \rightarrow \infty$ the effect of the fluctuations of
the conformal factor are suppressed and we recover the classical scale
dimension of $\Lambda/G$, namely four. The solution of the quadratic
relations (\ref{betE}) and (\ref{betL}) determine the scaling dimensions 
of these couplings at the conformal fixed point at other values of $Q^2$. 
This can be extended to local operators of any non-negative integer
mass dimension $p$, with associated couplings of mass dimension $4-p$, by 
\be
\beta_p = 4-p + \frac{\beta_p^2}{2Q^2}\,.
\label{betn}
\ee
In order to obtain the classical scale dimension $4-p$ in the limit
$Q^2 \rightarrow \infty$ the sign of the square root is determined
so that
\be
\beta_p = Q^2 \left[ 1 - \sqrt{1 - \frac{(8-2p)}{Q^2}}\right]\,,
\ee
valid for $Q^2 \ge 8 - 2p$ for all $p\ge 0$, and thus $Q^2 \ge 8$. 
These scaling dimensions were computed both by covariant and canonical
operator methods. In the canonical method we also showed that the anomalous 
action for the conformal factor does not have unphysical ghost or tachyon
modes in its spectrum of physical states \cite{AMM}.

In the framework of statistical mechanics and critical phenomena the
quadratic action (\ref{flata}) describes a Gaussian conformal fixed 
point, where there are no scales and conformal invariance is
exact. The positive corrections of order $1/Q^2$ (for $Q^2 > 0$)
in (\ref{betE}) and (\ref{betL}) show that this fixed point is
stable in the infrared, that is, both couplings $G^{-1}$ and
$\Lambda/G$ flow to zero at very large distances. Because both 
of these couplings are separately dimensionful, at a conformal fixed
point one should properly speak only of the dimensionless
combination $\hbar G \Lambda/ c^3 = \lambda$. By normalizing
to a fixed four volume $V= \int d^4 x$ one can show that the finite
volume renormalization of $\lambda$ is controlled by the anomalous 
dimension,
\be
2 \delta - 1 \equiv 2\, \frac{\beta_2}{\beta_0} - 1 = 
\frac{\sqrt{1 - \frac{8}{Q^2}} - \sqrt{1 - \frac{4}{Q^2}}}
{1 + \sqrt{1 - \frac{4}{Q^2}}} \le 0\,.
\label{scal}
\ee
This is the anomalous dimension that enters the infrared
renormalization group volume scaling relation \cite{AMMc},
\be
V \frac{d}{d V} \lambda = 4\, (2 \delta - 1)\, \lambda\,.
\label{renL}
\ee
The anomalous scaling dimension (\ref{scal}) is negative for all 
$Q^2 \ge 8$, starting at $1 - \sqrt 2 = -0.414$ at $Q^2 =8$
and approaching zero as $- 1/Q^2$ as $Q^2 \rightarrow \infty$. 
This implies that the dimensionless cosmological term $\lambda$ has 
an infrared fixed point at zero as $V\rightarrow \infty$. Thus the 
cosmological term is dynamically driven to zero as $V\rightarrow \infty$
by infrared fluctuations of the conformal part of the metric described 
by (\ref{flata}).

We emphasize that no fine tuning is involved here and no
free parameters enter except $Q^2$, which is determined by the trace
anomaly coefficient $b'$ by (\ref{Qdef}). Once $Q^2$ is assumed
to be positive, then $2 \delta - 1$ is negative, and $\lambda$ is
driven to zero at large distances by the conformal fluctuations 
of the metric, with no additional assumptions. 

The result (\ref{renL}) does rely on the use of (\ref{flata}) or its 
curved space generalization (\ref{Sanom}) as the free kinetic term in the 
effective action for gravity, treating the usual Einstein-Hilbert terms as 
interactions or marginal deformations of the conformal fixed point. This 
conformal fixed point represents a new phase of gravity, non-perturbative 
in any expansion about flat space. In this phase conformal invariance
is restored and the mechanism of screening $\lambda$ due to quantum effects 
proposed in \cite{AntMot} is realized. 

Identifying the fluctuations responsible for driving $\lambda$ to zero 
within a framework based on quantum field theory and the Equivalence Principle, 
free of {\it ad hoc} assumptions or fine tuning is an important first step
towards a full solution of the cosmological constant problem. However, this
is not yet a complete or testable cosmological model. Near the conformal fixed 
point the inverse Newtonian constant $G^{-1}$ is also driven to zero when 
compared to some fixed mass scale $m$ \cite{AMMd}. This is clearly different 
from the situation we observe in our local neighborhood. Under what conditions 
can (\ref{Sanom}) play a decisive role in realistic cosmological models?
How would the conformal fixed point behavior of the scaling relations
such as (\ref{renL}) be detectable by cosmological observations?

\section{Cosmological Consequences of the Anomaly Effective Action}
\label{sec:cosmo}

Absent a full cosmological model, the simplest possibility is to consider 
the essentially kinematic effects of conformal invariance in cosmology
and predictions for observables such as the Cosmic Microwave Background (CMB),
without specifying the dynamics or spacetime history of the universe which
gave rise to that conformal invariance. Later in this section we take
a first step to a cosmological model by considering the semi-classical 
linear response of quantum fluctuations generated by the anomaly terms
in the effective action around de Sitter space. 

\subsection{Conformal Invariance in de Sitter Space}

To illustrate how conformal invariance on flat spatial sections of FLRW
cosmology can be a natural result of a de Sitter phase in the evolution
of the universe, consider the $O(4,1)$ de Sitter symmetry group and its
ten Killing vectors, $\xi_a^{(\alpha)}$. In coordinates (\ref{FLRW}).
the Killing eq.,
\be
\nabla_a \xi_b^{(\alpha)} + \nabla_b \xi_a^{(\alpha)} = 0\,,\qquad \alpha = 1,\dots,10\,,
\label{Killing}
\ee
becomes
\bes
\bea
&&\partial_{\tau}\xi_{\tau}^{(\alpha)} = 0\,,\\
&&\partial_{\tau} \xi_i^{(\alpha)} + \partial_i\xi_{\tau}^{(\alpha)}
- 2H \xi_i^{(\alpha)} = 0\,,\\
&&\partial_i\xi_j^{(\alpha)} + \partial_j\xi_i^{(\alpha)} 
- 2Ha^2 \delta_{ij} \xi_{\tau}^{(\alpha)} = 0\,.
\label{cKv}
\eea
\label{KilldS}
\ees
For $\xi_{\tau} =0$ we have the three translations, $\alpha = T_j$,
\be
\xi_{\tau}^{(T_j)} = 0\,,\qquad \xi_i^{(T_j)} = a^2 \delta_{\ i}^j\,,\qquad j = 1, 2, 3\,,
\ee
and the three rotations, $\alpha =R_\ell$,
\be
\xi_{\tau}^{(R_\ell)} = 0\,,\qquad \xi_i^{(R_\ell)} = a^2 \epsilon_{i\ell m}x^m\,,\qquad \ell = 1, 2, 3\,.
\ee
This accounts for $6$ of the $10$ de Sitter isometries which are self-evident in
the flat FLRW coordinates with $\xi_{\tau} =0$. The $4$ additional solutions of (\ref{KilldS}) have
 $\xi_{\tau} \neq 0$. They are the three special conformal transformations
of $\mathbb{R}^3$, $\alpha = C_n$,
\be
\xi_{\tau}^{(C_n)} = -2H x^n,\quad \xi_i^{(C_n)} = H^2a^2( \delta_{\ i}^n \delta_{jk}x^jx^k 
- 2 \delta_{ij}x^jx^n)
- \delta_{\ i}^n,\quad n = 1, 2, 3,
\label{confspec}
\ee
and the dilation, $\alpha = D$,
\be
\xi_{\tau}^{(D)} = 1\,,\qquad \xi_i^{(D)} = H a^2\,\delta_{ij} x^j\,.
\label{dilat}
\ee
This last dilational Killing vector is the infinitesimal form of the finite dilational symmetry,
\bes
\bea
&&\vec x \rightarrow \lambda \vec x\,,\\
&& \eta \rightarrow \lambda \eta\,\\
&& a(\tau) \rightarrow \lambda^{-1} a(\tau)\,,\\
&&\tau \rightarrow \tau - H^{-1}\ln \lambda
\eea
\ees
of de Sitter space. The existence of this symmetry explains why Fourier modes of
different $|\vec k|$ leave the de Sitter horizon at a shifted FLRW time $\tau$, so in
an eternal de Sitter space, in which there is no preferred $\tau$, one expects
a scale invariant spectrum.

The existence of the three conformal modes of $\mathbb{R}^3$ (\ref{confspec}) implies in 
addition that any $O(4,1)$ de Sitter invariant correlation function must decompose
into representations of the conformal group of three dimensional flat space.
Fundamentally this is because the de Sitter group $O(4,1)$ is the conformal
group of flat Euclidean $\mathbb{R}^3$, as eqs. (\ref{KilldS})-(\ref{dilat}) shows explicitly.
Moreover, because of the exponential expansion in de Sitter space, the 
decomposition into representations of the conformal group become simple at 
distances large compared to the horizon scale $1/H$.

If we consider a massive scalar field in the Bunch-Davies state in de Sitter space
(with $m^2 = \mu^2 + \xi R = \mu^2 + 12\xi H^2$), its propagator
function $G (z(x,x'); m^2)$ satisfies
\bea
&&\left(-\sq + m^2\right) G(z(x,x'); m^2) =\nn\\
&& -H^2 \left[ z(1-z) \frac{d^2}{dz^2} + 2 (1 -2z) \frac{d}{dz} -
\frac{m^2}{H^2}\right]G(z; m^2) = \delta^4(x, x'),\qquad\quad
\label{massiveprop}
\eea
where $z(x,x')$ is the de Sitter invariant distance function given by
\bea
&&1 - z(x,x') = \frac{1}{4\eta\eta'}\, \left[ - (\eta - \eta')^2 + (\vec x - \vec x')^2 \right]\nn\\
&&\qquad = \frac{H^2}{4} a(\eta)a(\eta')\,  \left[ - (\eta - \eta')^2 
+ (\vec x - \vec x')^2 \right],
\label{distcflat}
\eea
in the conformally flat coordinates of (\ref{conflat}). Since for $x\neq x'$, (\ref{massiveprop}) 
is a standard form of the hypergeometric eq., it has the solution,
\be
G(z; m^2) = \frac{H^2}{16\pi^2}\, \Gamma(\alpha) \Gamma(\beta)\, F(\alpha,\beta; 2; z)
\label{deSmprop}
\ee
in terms of the hypergeometric function $_2F_1 = F$, with parameters
\bes\bea
&&\alpha = \frac{3}{2} + \nu \,,\\
&&\beta = \frac{3}{2} - \nu\,.
\eea
\label{defalpbet}\ees
where $\nu$ is defined by (\ref{defnu}). The particular solution (\ref{deSmprop})
and its normalization is dictated by the requirement that $(-\sq + m^2)G(x,x')$ 
yield a delta function of unit magnitude at $x=x'$ in (\ref{massiveprop}) and 
possess no other singularities.

To explicitly see the conformal behavior in the massive case
consider (\ref{deSmprop}) in the limit $a(\tau), a(\tau') \gg 1$, that is,
after many e-foldings of the exponential de Sitter expansion \cite{astph}. From
(\ref{distcflat}) in this limit,
\be
1- z(x,x') \rightarrow \frac{H^2}{4} a(\tau)a(\tau') \, (\vec x - \vec x')^2 \rightarrow \infty
\label{dSexp}
\ee
for fixed $\vec x, \vec x'$. Using the asymptotic form of the hypergeometric function
in (\ref{deSmprop}) for $z \rightarrow -\infty$, we obtain \cite{astph}
\be
G(z(x,x'); M^2)  \rightarrow A_+\,\vert\vec x - \vec x'\vert^{- 3 + 2\nu} + A_-\,
\vert\vec x - \vec x'\vert^{-3 - 2\nu}
\label{confmass}
\ee
a sum of simple power law scaling behaviors in the distance $|\vec x - \vec x'|$, with coefficients
\be
A_{\pm}(\tau,\tau') = \frac{1}{2H\pi^2} \left(\frac{H}{2}\right)^{\pm 2\nu}
\frac{\Gamma(\frac{3}{2} \mp \nu)\Gamma(\pm 2\nu)}
{\Gamma(\frac{1}{2} \pm \nu)}\exp \left[\left(-\frac{3}{2} \pm \nu\right) H(\tau +\tau')\right]\,.
\ee
Thus even massive non-conformal fields in de Sitter space exhibit conformal power law
scaling behavior on the flat $\mathbb{R}^3$ spatial sections after sufficiently long exponential 
expansion in a de Sitter inflationary phase. The corrections to (\ref{confmass}) are
integer power law terms which are a Taylor series in $z^{-1} \ll 1$ given by (\ref{dSexp}).
Moreover if $m^2 < 9H^2/4$ then only the leading power law term $|\vec x - \vec x'|^{-3 + 2 \nu}$
survives in the limit. This would identify the CFT scaling  exponent to be $3/2 - \nu$
if the fluctuations responsible for the CMB were generated in the late time de Sitter
expansion by a field of mass $m$ such that $\nu$ defined in (\ref{defnu}) is real.
The conformal behavior of de Sitter invariant correlation functions is an example of 
a kind of dS/CFT correspondence \cite{MazMot}, and the result of the mathematical {\it isomophism} between the conformal group of the flat $\mathbb{R}^3$ sections and 
the four-dimensional de Sitter group $SO(4,1)$ expressed by the solutions of the Killing 
equation.

Note that in this picture the fluctuations of a massless, minimally coupled scalar
field with $m=0$ or gravitons themselves have $\nu =3/2$ and $\Delta = \nu - 3/2 = 0$,
as should be expected for the fluctuations of a dimensionless variable such as
the metric tensor, and a nearly conformally invariant spectrum is generated
on the FLRW flat spatial sections. This presumes that the de Sitter isometry
group is broken in a specific way, to a standard spatially homogeneous, isotropic
FLRW cosmology. In slow roll inflation this breaking of de Sitter invariance
is achieved by the addition of a scalar field, the inflaton, with a nearly 
flat potential \cite{MukFelBra,LidL}. The fluctuations of this field generate 
fluctuations in its stress tensor which act as a source for scalar metric fluctuations 
through Einstein's eqs. as the system rolls slowly down the potential and out of 
de Sitter space everywhere in space uniformly, maintaining overall spatial 
homogeneity and isotropy and the form (\ref{FLRW}) of the FLRW line element. 
Since gravity is treated purely classically, in the limit of a strictly constant 
inflaton potential or in the absence of any inflaton field at all there would be 
no coupling of the scalar inflaton modes to the metric and hence the amplitude 
of the CMB power spectrum would vanish. For this reason also in this picture 
the non-Gaussian bi-spectrum is expected to have a very small amplitude, higher 
order in the slow roll parameters \cite{BKMRio}.

Now the effective action of the trace anomaly has the potential to modify this
picture in at least two fundamental ways. First, the scalar degrees of freedom
$\varphi$ and $\psi$ are present in the effective action and will generate
scalar fluctuations in a de Sitter epoch, independently of any inflaton field
or potential. Thus, the deviations from a strict Harrison-Zel'dovich spectrum
and primordial non-Gaussianities need not be controlled by any small slow
roll parameters. Second, as discussed in Sec. \ref{Lamrun} the anomaly 
effective action gives rise to dynamical dark energy, and the possibility of a
departure from global FLRW kinematics usually assumed in cosmological models. 

\subsection{Conformal Invariance and the CMB} 

Let us consider first the consequences of conformal invariance on flat
FLRW sections, independently of any detailed model. Our studies of 
fluctuations in de Sitter space suggest that the fluctuations responsible 
for the screening of $\lambda$ take place at the horizon scale. In that 
case then the microwave photons in the CMB reaching us from their surface 
of last scattering should retain some imprint of the effects of these 
fluctuations. It then becomes natural to extend the classical notion of 
scale invariant cosmological perturbations, pioneered by Harrison and 
Zel'dovich \cite{HZ} to full conformal invariance. In that case the 
classical spectral index of the perturbations should receive corrections 
due to the anomalous scaling dimensions at the conformal phase \cite{sky}. 
In addition to the spectrum, the statistics of the CMB should reflect 
the non-Gaussian correlations characteristic of conformal invariance. 

Scale invariance was introduced into physics in early attempts to describe the 
apparently universal behavior observed in turbulence and second order phase 
transitions, which are independent of the particular short distance dynamical 
details of the system. The gradual refinement and development of this simple 
idea of universality led to the modern theory of critical phenomena, one of 
whose hallmarks is well-defined logarithmic deviations from naive scaling 
relations \cite{RG}. A second general feature of the theory is the specification 
of higher point correlation functions of fluctuations according to 
the requirements of conformal invariance at the critical point \cite{pol}. 

In the language of critical phenomena, the observation of Harrison and 
Zel'dovich that the primordial density fluctuations should be characterized by 
a spectral index $n=1$ is equivalent to the statement that the observable giving 
rise to these fluctuations has engineering or naive scaling dimension $p = 2$. 
This is because the density fluctuations $\delta \rho$ are related to the metric 
fluctuations by Einstein's equations, $\delta R \sim G\delta\rho$, which is second 
order in derivatives of the metric. Hence, the two-point spatial correlations 
$\lag \delta\rho(x) \delta\rho(y)\rag \sim \lag \delta R(x) \delta R(y)\rag$
should behave like $|x-y|^{-4}$, or $|k|^1$ in Fourier space, according to simple 
dimensional analysis.

One of the principal lessons of the modern theory of critical phenomena is that 
the transformation properties of observables under conformal transformations 
at the fixed point is {\it not} given by naive dimesnional analysis. 
Rather one should expect to find well-defined logarithmic deviations 
from naive scaling, corresponding to a (generally non-integer) dimension 
$\Delta$. The deviation from naive scaling is the ``anomalous'' dimension 
of the observable due to critical fluctuations, which may be quantum or statistical 
in origin. Once $\Delta$ is fixed for a given observable, the requirement of conformal 
invariance determines the form of its two- and three-point correlation functions up 
to an arbitrary amplitude, without reliance on any particular dynamical model. 

Consider first the two-point function of any observable ${\cal O}_{\Delta}$
with dimension $\Delta$. Conformal invariance requires \cite{RG,pol}
\be
\langle{\cal O}_{\Delta} (x_1) {\cal O}_{\Delta} (x_2)\rangle
\sim \vert x_1-x_2 \vert^{-2\Delta}
\label{ODel}
\ee
at equal times in three dimensional flat spatial coordinates. In
Fourier space this gives
\be
G_2(k) \equiv\langle\tilde{\cal O}_{\Delta} (k) \tilde 
{\cal O}_{\Delta} (-k)\rangle \sim \vert k \vert^{2\Delta - 3} \,.
\label{G2}
\ee
Thus, we define the spectral index of this observable by
\be
n \equiv 2 \Delta - 3\ .
\label{index}
\ee
In the case that the observable is the primordial density fluctuation 
$\delta\rho$, and in the classical limit where its anomalous dimension 
vanishes, $\Delta \rightarrow 2$, we recover the 
Harrison-Zel'dovich spectral index of $n=1$.

In order to convert the power spectrum of primordial density 
fluctuations to the spectrum of fluctuations in the CMB at large angular 
separations we follow the standard treatment \cite{peeb} relating the
temperature deviation to the Newtonian gravitational potential 
$\varphi$ at the last scattering surface, $\frac{\delta T}{T} \sim \delta 
\varphi$, which is related to the density perturbation in turn by 
\be
\nabla^2 \delta\varphi = 4\pi G\, \delta\rho \ .
\label{lap}
\ee
Hence, in Fourier space, 
\be
\frac{\delta T}{T} \sim \delta \varphi \sim 
\frac{1}{k^2}\frac{\delta\rho}{\rho}\ ,
\ee
and the two-point function of CMB temperature fluctuations is
determined by the conformal dimension $\Delta$ to be
\bea
&&C_2(\theta) \equiv \left\lag\frac{\delta T}{T}(\hat r_1)
\frac{\delta T}{T}(\hat  r_2)\right\rag
\sim \nn\\
&&\int d^3 k\left(\frac{1}{k^2}\right)^2 G_2(k) e^{i k\cdot 
r_{12}}
\sim \Gamma (2-\Delta) (r_{12}^2)^{2 - \Delta}\ ,
\label{C2}
\eea
where $r_{12} \equiv (\hat r_1 - \hat r_2)r$
is the vector difference between the two positions from which the 
CMB photons originate. They are at equal distance $r$ from the observer 
by the assumption that the photons were emitted at the last scattering 
surface at equal cosmic time. Since $r_{12}^2 = 2 (1- \cos \theta)r^2$, 
we find then
\be
C_2(\theta) \sim \Gamma (2-\Delta) (1-\cos\theta)^{2 - \Delta}
\label{Ctheta} 
\ee
for arbitrary scaling dimension $\Delta$.  

Expanding the function $C_2(\theta)$ in multipole moments,
\be
C_2(\theta) = \frac{1}{4\pi} \sum_{\ell} (2\ell + 1)
c_{\ell}^{(2)}(\Delta) P_{\ell} (\cos \theta)\ ,
\label{c2m}
\ee
\be
c_{\ell}^{(2)}(\Delta) \sim \Gamma(2-\Delta) \sin\left[ \pi 
(2-\Delta)\right]
\frac{\Gamma (\ell - 2 + \Delta)}{\Gamma (\ell + 4 -\Delta)}\ ,
\ee
shows that the pole singularity at $\Delta =2$ appears only in the $\ell = 
0$ monopole moment. This singularity is just the reflection of the fact 
that the Laplacian in (\ref{lap}) cannot be inverted on constant functions, 
which should be excluded. Since the CMB  anisotropy is defined by removing 
the isotropic monopole moment (as well as the dipole moment), the $\ell =0$ 
term does not appear in the sum, and the higher moments of the anisotropic 
two-point correlation function are well-defined for $\Delta$ near $2$. 
Normalizing to the quadrupole moment $c_2^{(2)}(\Delta)$, we find
\be
c_{\ell}^{(2)}(\Delta) = c_2^{(2)}(\Delta) 
\frac{\Gamma (6 - \Delta)}{\Gamma (\Delta) } 
\frac{\Gamma (\ell - 2 + \Delta)}{\Gamma(\ell + 4 - \Delta)}\ ,
\ee
which is a standard result \cite{peeb}. Indeed, if $\Delta$ is replaced by 
$p = 2$ we obtain $\ell (\ell + 1)  c_{\ell}^{(2)}(p) = 6 c_2^{(2)} (p)$, 
which is the well-known  predicted behavior of the lower moments ($\ell \le 30 $) 
of the CMB anisotropy where the Sachs-Wolfe effect should dominate.

Turning now from the two-point function of CMB fluctuations to higher point 
correlators, we find a second characteristic prediction of conformal invariance, 
namely non-Gaussian statistics for the CMB. The first correlator sensitive to this 
departure from gaussian statistics is the three-point function of the observable 
${\cal O}_{\Delta}$, which takes the form \cite{pol}
\be
\langle{\cal O}_{\Delta} (x_1) {\cal O}_{\Delta} (x_2) 
{\cal O}_{\Delta} (x_3)\rangle
\sim |x_1-x_2|^{-\Delta} |x_2- x_3|^{-\Delta} |x_3 - x_1|^{-\Delta}\ ,
\ee 
or in Fourier space,\footnote{Note that (\ref{three}) corrects two minor typographical
errors in eq. (16) of Ref. \cite{sky}}
\bea
&&\hspace{-7mm}G_3 (k_1, k_2) \sim \int d^3 p\ |p|^{\Delta -3}\,  |p + 
k_1|^{\Delta -3}\, 
|p- k_2|^{\Delta -3}\ \sim \ 
\frac{\Gamma\left( 3 - \textstyle{\frac{3\Delta}{2}}\right)}
{\left[\Gamma\left(\frac{3-\Delta}{2}\right)\right]^3}\,\times\nn\\
&&\hspace{-5mm}\int_0^1\! du\int_0^1\! dv\,
\frac{\left[u(1-u)v\right]^{\frac{1-\Delta}{2}}
(1-v)^{-1 + \frac{\Delta}{2}}}
{\left[u(1\!-\!u)(1\!-\!v)k_{_1}^2 + v(1\!-\!u)k_{_2}^2 + uv(k_{_1}\! + 
k_{_2})^2\right]^{3-\frac{3\Delta}{2}}}\,.\!\!
\label{three}
\eea
This three-point function of primordial density fluctuations gives rise to
three-point correlations in the CMB by reasoning precisely analogous as that
leading from Eqns.~(\ref{G2}) to (\ref{C2}). That is,
\bea
&& C_3(\theta_{12}, \theta_{23}, \theta_{31}) 
\equiv \left\lag\frac{\delta T}{T}(\hat r_1)
\frac{\delta T}{T}(\hat  r_2)\frac{\delta T}{T}(\hat  
r_3)\right\rag \nn \\
&& \qquad\sim \int \frac{d^3 k_1\,d^3k_2}
{k_1^2\, k_2^2\, (k_1 + k_2)^2}\ G_3 (k_1, k_2)\, e^{i k_1\cdot r_{13}} 
e^{ik_2\cdot r_{23}}
\label{C3}
\eea
where $r_{ij}\equiv ({\hat r}_i-{\hat r_j})r$ and 
$r_{ij}^2=2(1-\cos\theta_{ij}) r^2$. 

In the general case of three different angles, this expression for the non-Gaussian
three-point correlation function (\ref{C3}) is quite complicated, although it can be 
rewritten in parametric form analogous to (\ref{three}) to facilitate numerical evaluation. 
In the special case of equal angles $\theta_{ij}=\theta$, it follows from its global
scaling behavior that the three-point correlator is 
\be
C_3(\theta)\sim (1-\cos\theta)^{\frac{3}{2}(2-\Delta)}\ .
\ee
Expanding the function $C_3(\theta)$ in multiple moments as in (\ref{c2m})
with coefficients $c_{\ell}^{(3)}$, and normalizing to the quadrupole moment,
we find
\be
c_{\ell}^{(3)}(\Delta) =c_{2}^{(3)}(\Delta)
\frac{\Gamma (4+\frac{3}{2}(2-\Delta))}{\Gamma (2-\frac{3}{2}(2-\Delta))}
\frac{\Gamma (\ell-\frac{3}{2}(2-\Delta))}{\Gamma(\ell+2+\frac{3}{2}(2-\Delta))}\ .
\label{cl3}
\ee
In the limit $\Delta \rightarrow 2$, we obtain $\ell(\ell+1)c_{\ell}^{(3)}=6c_2^{(3)}$, 
which is the same result as for the moments $c_{\ell}^{(2)}$ of the two-point correlator 
but with a different quadrupole amplitude. The value of this quadrupole normalization 
$c_2^{(3)}(\Delta)$ cannot be determined by conformal symmetry considerations alone,
and requires more detailed dynamical information about the origin of conformal
invariance in the spectrum.

For higher point correlations, conformal invariance does not 
determine the total angular dependence. Already the four-point 
function takes the form,
\be
\langle{\cal O}_{\Delta} (x_1) {\cal O}_{\Delta} (x_2) 
{\cal O}_{\Delta} (x_3) {\cal O}_{\Delta} (x_4)\rangle
\sim \frac{A_4}{{\prod_{i<j} r_{ij}^{2\Delta/3}} }\ ,
\ee
where the amplitude $A_4$ is an arbitrary function of the two 
cross-ratios, 
$r_{13}^2 r_{24}^2/r_{12}^2 r_{34}^2$ and 
$r_{14}^2 r_{23}^2/r_{12}^2 r_{34}^2$.
Analogous expressions hold for higher $p$-point functions. However in the 
equilateral case $\theta_{ij}=\theta$, the coefficient amplitudes 
$A_p$ become constants and the angular dependence is again completely 
determined, with the result,
\be
C_p(\theta)\sim (1-\cos\theta)^{\frac{p}{2}(2-\Delta)}\,.
\ee
The expansion in multiple moments yields coefficients $c_{\ell}^{(p)}$ 
of the same form as in Eqn.~(\ref{cl3}) with $3/2$ replaced by $p/2$.
In the limit $\Delta =2$, we obtain the universal $\ell$-dependence 
$\ell(\ell+1)c_{\ell}^{(p)}=6c_2^{(p)}$.

Again it bears emphasizing that these results depend upon the hypothesis 
of conformal invariance {\it on the flat spatial sections} of FLRW 
geometries, but otherwise makes no dynamical assumptions, such as in
scalar field inflaton models. Because of the possibilities of large
backreaction on the horizon in de Sitter space, this assumption of
global spatial homogeneity and isotropy implied by the FLRW line
element (\ref{FLRW}) may not in fact be realized. We study these horizon
backreaction effects to linear order in the fluctuations next.

\subsection{Linear Response in de Sitter Space}
\label{sec:linres}

The first step to a cosmological model based on the effective action
(\ref{Seff}) going beyond the essentially purely kinematic considerations
of the previous subsection is to study the dynamical effects of (\ref{Seff})
in de Sitter space. This one can do by performing a linear response
analysis of perturbations about de Sitter space with the anomaly scalars
$\varphi$ and $\psi$, in place of any {\it ad hoc} inflaton field.

Because of the $O(4,1)$ maximal symmetry, with $10$ Killing generators
(\ref{KilldS}), the maximally symmetric Bunch-Davies (BD) state in de Sitter 
spacetime is the natural one about which to consider perturbations. This
is the state usually considered in cosmological perturbation theory. The linear 
response approach requires a self-consistent solution of equations (\ref{scE}), 
around which we perturb the metric and stress tensor together. In the BD state 
$O(4,1)$ de Sitter symmetry implies that $\lag T^a_{\ b}\rag$ is also proportional 
to $\delta^a_{\ b}$, which therefore guarantees that de Sitter space is a
self-consistent solution of the semi-classical eqs. (\ref{scE}). The self-consistent 
value of the scalar curvature $R$, including the quantum contribution from 
$\langle T^a_{\ b}\rangle$ is determined by the trace of (\ref{scE}), {\it i.e.}
\bes\bea
&&-R + 4 \Lambda = 8 \pi G b' E = \frac{4 \pi Gb'}{3}\, R^2
\qquad {\rm or}\qquad \\
&&\qquad \frac{R}{12} = H^2 = \frac{\Lambda}{3} \left( 1 - \frac{16\pi Gb'\Lambda}{3}
+ \dots \right)\,,
\eea
\label{selfcons}\ees
in an expansion around the classical de Sitter solution with
$G\Lambda |b'|\ll 1$. Thus, in this limit the stress tensor source 
of the semi-classical Einstein's equations (\ref{scE}) in the BD state (\ref{BDdS})
gives a small finite correction to the classical cosmological term in the
self-consistent de Sitter solution. The trace of (\ref{BDdS}) is exactly $b'E = b'R^2/6$ 
used in determining the self-consistent scalar curvature including the quantum 
BD corrections in (\ref{selfcons}). The solutions for the anomaly scalar fields,
\bes
\bea
&&\bar \varphi = 2 \ln a = 2H\tau\,,\\
&& \bar\psi = 0\,,
\eea
\label{auxBD}\ees
correspond exactly to the BD state, and it is consistent to expand the semi-classical
Einstein eqs. (\ref{scE}) around de Sitter space using the effective action and stress
tensor of the anomaly scalar fields with background values (\ref{auxBD}).

With the self-consistent BD de Sitter solution  (\ref{BDdS}), (\ref{selfcons}),
and (\ref{auxBD}), one may consider the linear response variation of
the semi-classical Einstein eqs. (\ref{scE}) \cite{AndMolMot,DSAnom}
\be 
\delta\left\{R^a_{\ b} - \frac{R}{2} \delta^a_{\ b} + \Lambda
\delta^a_{\ b}\right\} = 8\pi G \, \delta\lag T^a_{\ b}\rag_{_R}
\, ,
\label{linresgen}
\ee
with the source the anomaly generated stress tensor (\ref{Tanom}). 
Considering the general form of the exact quantum effective action (\ref{Sexact}),
its variation includes three kinds of terms, corresponding to the local
terms up to fourth order in derivatives of the metric, Weyl invariant
non-local terms, and those terms coming from the anomaly generated effective 
action (\ref{allanom})-(\ref{SEF}). Parameterizing the local fourth order geometric 
terms in (\ref{R2}) with finite coefficients $\alpha_{_R}$ and $\beta_{_R}$,
(\ref{linresgen}) with (\ref{Sexact}) gives then
\bea
&&\delta R^a_{\ b} -\frac{\delta R}{2}\, \delta^a_{\ b} = 8\pi G\ 
\delta\left[(T^a_{\ b})^{loc} + (T^a_{\ b})^{inv} + (T^a_{\ b})^{anom}\right]  \nn\\
&& \hspace{-1cm}= 8\pi G \left[ - \alpha_{_R}\,\delta A^a_{\ b} - \beta_{_R}\,
\delta B^a_{\ b} + (T^a_{\ b})^{inv} + b'\, \delta E^a_{\ b} + b\, \delta F^a_{\ b}\right],
\label{linaux}
\eea
where all terms to linear order in $\delta g_{ab} = h_{ab}$ and in the variations
of the auxiliary fields,
\bes
\bea
&& \delta \varphi \equiv \varphi - \bar\varphi \equiv \phi\\
&& \delta\psi \equiv \psi - \bar\psi = \psi \,,
\eea
\label{auxvar}
\ees\vspace{-5mm}

\noindent
are to be retained. All indices are raised and lowered by the background de Sitter
metric $g_{ab}$ at linear order in the perturbations. Since $(T^a_{\ b})^{inv}$
is the variation of a Weyl invariant action, it has zero trace.

Since the terms coming from the fourth order local invariants $C_{abcd}C^{abcd}$ 
and $R^2$, namely $\delta A^a_{\ b}$ and $\delta B^a_{\ b}$ are higher order in 
derivatives  than the Einstein-Hilbert terms, they are important only in the extreme 
ultraviolet regime at energies of order $M_{Pl}$, where in any case one should 
not trust the semi-classical effective theory. These local terms were explicitly 
analyzed in Ref. \cite{DSAnom} and do not affect any physics on the cosmological 
horizon scale $H \ll M_{Pl}$. In particular, from the trace of (\ref{linaux}) one obtains 
the purely local eq.,
\be
\left[\left(3\beta_{_R} + \frac{b}{3}\right) \sq + \left(\frac{1}{16\pi G} 
+ 2 b' H^2\right)\right] 
\delta R = 0\,.
\label{delReq}
\ee
Because of the $G^{-1} = M_{Pl}^2$ term, the only non-trivial solutions of  (\ref{delReq}) 
are on Plank scale, and outside the range of applicability of the EFT approach. 
Hence we restrict attention to only the remaining solutions of (\ref{linaux}) and
(\ref{delReq}) that satisfy
\be
\delta R = 0\,.
\label{delR0}
\ee
Notice that by so doing we are excluding the trace anomaly driven
inflationary solutions studied by Starobinsky \cite{Starob}.

In general the non-local Weyl invariant terms in the exact quantum effective action 
$S_{inv}$ are difficult to calculate and are not known. However in the case
of conformally invariant matter/radiation fields in the conformaly flat de Sitter
geometry, it is possible to keep track of these non-local terms, and calculate
the linearized stress tensor they provide $\delta (T^a_{\ b})^{inv}$ completely
from the corresponding quantity in the conformally related flat space \cite{HorWal,Starob81}.
By keeping track of these terms in the de Sitter linear response analysis, we will then 
be able to provide a non-trivial check on our general argument based on the
classification of terms in (\ref{Sexact}) according to their response under
global Weyl transformations that these terms play no role in the infrared, and
the Wilson effective action for low energy gravity consists of the classical
Einstein-Hilbert and anomaly terms only in (\ref{Seff}).

The last two terms in the last line of (\ref{linaux}) arise from the anomaly action 
and give rise to new physical effects in cosmology. The variation of the stress 
tensors $E^a_{\ b}$ and $F^a_{\ b}$ depend on both the variations of the metric
and the variations of the auxiliary fields. The variation of the auxiliary
field equations (\ref{auxeom}) gives
\bes\bea
&&\hspace{-1.5cm}\delta (\sq^2 \varphi) - \frac{R}{6}\, \delta (\sq \varphi) =
\left(-\sq + 2H^2\right) \delta (-\sq \varphi) =
- 2 (\nabla_a\nabla^b \bar\varphi) \delta R^a_{\ b},
\label{varyphi}\\
&&\hspace{-1.5cm}\delta (\sq^2 \psi) - \frac{R}{6}\, \delta (\sq \psi) =
\left(-\sq + 2H^2\right) (-\sq)\psi = 0\,.  \label{varypsi}
\eea
\label{varyaux}\ees
The first of these two equations shows that at linear order there is mixing between
the fluctuations of the anomaly scalar field $\varphi$ and the metric perturbation
$\delta g_{ab}= h_{ab}$ around de Sitter space. This mixing is algebraically
simplest to study by making a suitable gauge choice, although of course the
results in the end must be gauge invariant. Here we present only the results of
the gauge invariant analysis \cite{DSAnom}.

The metric perturbations which are scalar with respect to the background three-metric
$g_{ij} = a^2 \eta_{ij}$ can be parameterized in terms of four functions, $({\cal A, B, C, E})$
in the form
\cite{Bard,Stew},
\bes\bea
h_{tt}&=&-2{\cal A} \\
h_{tj}&=&a \partial_j {\cal B} \rightarrow ia k_j {\cal B} \\
h_{ij}&=& 2a^2 \left[\eta_{ij}\, {\cal C} + \left( \frac{\eta_{ij}}{3}\,k^2
- k_ik_j\right) {\cal E}\right]\, \, .
\eea\label{linmet}\ees
in momentum space. As is well known, only two linear combinations of these four functions 
are gauge invariant (for $\vec k \neq 0$).  The two gauge invariant metric perturbation variables 
may be taken to be
\bes
\bea
&& \Upsilon_{_{\!\!\cal A}}
\equiv {\cal A} + \partial_{\tau}(a {\cal B})- \partial_{\tau}(a^2 \partial_{\tau}{\cal E})\,,\\
&&\Upsilon_{_{\!\cal C}} \equiv {\cal C}
- \frac{\nabla^2}{3\,}\, {\cal E} + (\partial_{\tau} a) {\cal B} 
- a (\partial_{\tau}a) \partial_{\tau}\cal E\,,
\eea\label{defUpsAC}\ees
which correspond to the gauge invariant Bardeen-Stewart potentials denoted by $\Phi_A$ and $\Phi_C$ in 
Refs. \cite{Bard,Stew}. For $\vec k = 0$ there is only one gauge invariant metric combination, namely
$\partial_{\tau} \Upsilon_{_{\!\cal C}} - H \Upsilon_{_{\!\!\cal A}}$.

The variation for $\delta R$ can be written in the form  
\bea
&&\delta R = - \sq h + \nabla_a\nabla_b h^{ab} - R^{ab}h_{ab} \nn\\
&&\hspace{-1cm}=  6 (\partial^2_{\tau}\Upsilon_{_{\!\cal C}} - H \partial_{\tau}\Upsilon_{_{\!\!\cal A}})
+ 24 H  (\partial_{\tau}\Upsilon_{_{\!\cal C}} - H \Upsilon_{_{\!\!\cal A}})
- \frac{2\,}{a^2} \nabla^2\,\Upsilon_{_{\!\!\cal A}} 
- \frac{4\,}{a^2} \nabla^2\,\Upsilon_{_{\!\cal C}}\,,
\label{delRg}
\eea
where $\nabla^2$ is the Laplacian on the flat $\mathbb{R}^3$ FLRW sections.
Hence condition (\ref{delR0}), $\delta R = 0$, is gauge invariant, and equivalent to
\be
\left(\frac{\partial}{\partial \tau} + 4H \right)
\left(\frac{\partial \Upsilon_{_{\!\cal C}}}{\partial\tau}  
-H \Upsilon_{_{\!\!\cal A}}\right)  
- \frac{\nabla^2}{3a^2} \Upsilon_{_{\!\!\cal A}} - \frac{2\nabla^2}{3\,a^2}
\Upsilon_{_{\!\cal C}}=0\,,
\label{delRg6}
\ee
which provides one constraint between the two gauge invariant potentials $\Upsilon_{_{\!\!\cal A}}$ 
and $\Upsilon_{_{\!\cal C}}$. This means that there remains only {\it one} gauge invariant 
metric function to be determined by linear response in this scalar sector. As proven 
in \cite{DSAnom} the information about this remaining metric degree of freedom is contained 
completely in the $\tau\tau$ component of the linear response equations (\ref{linaux}).
Thus, recalling the condition (\ref{delR0}), one can define the gauge invariant quantity
\be
q \equiv -\frac{2a^2}{H^2} \delta G^{\tau}_{\ \tau} =  
-\frac{2a^2}{H^2} \delta R^{\tau}_{\ \tau} = 
\frac{12a^2}{H}\left(\frac{\partial\Upsilon_{_{\!\cal C}}}{\partial \tau}- H\Upsilon_{_{\!\!\cal A}}\right)
-\frac{4}{H^2}
\,\nabla^2\,\Upsilon_{_{\!\cal C}}\,,
\label{qUps}
\ee
which appears in the $\tau\tau$ component of the linear response eq. (\ref{linaux}),
and which contains the only remaining gauge invariant information in
the sector of scalar metric perturbations (\ref{linmet}) after condition (\ref{delR0})
has been imposed.

It is easily checked and verified in Ref. \cite{DSAnom} that the quantity
\be
\Phi \equiv \phi + 2 (\partial_{\tau} a){\cal B}- 2 a (\partial_{\tau} a)(\partial_{\tau}{\cal E})
\label{Phidef}
\ee
is gauge invariant. This is similar to the gauge invariant variable that
can be constructed from the scalar field in scalar inflaton models
of slow roll inflation \cite{MukFelBra}. The second anomaly scalar
field $\psi$ is already gauge invariant. Defining also the explicitly
gauge invariant  quantities,
\be
u \equiv {\cal D}_0\,\Phi + \frac{6}{H} \, \frac{\partial \Upsilon_{_{\!\cal C}}}{\partial\tau}
 - \frac{2}{H} \, \frac{\partial \Upsilon_{_{\!\!\cal A}}}{\partial\tau} 
- 8\, \Upsilon_{_{\!\!\cal A}}\,.
\label{udef}
\ee
and
\be
v \equiv {\cal D}_0 \,\psi =  \frac{1}{H^2} \left(\frac{\partial^2}{\partial \tau^2} +  
H\frac{\partial}{\partial\tau}  - \frac{\nabla^2}{a^2}\right)\psi\,,
\label{vdef}
\ee
where the differential operator ${\cal D}_n$ is defined for arbitrary integer $n$ by
\be
{\cal D}_n \equiv  \frac{1}{H^2} \left(\frac{\partial^2}{\partial \tau^2} 
+ (2n+1) H\frac{\partial}{\partial\tau} 
+ n(n+1) H^2 - \frac{\nabla^2}{a^2}\right)\,,
\label{diffD}
\ee
one finds that the linear response equations around de Sitter space can be written in
terms of the gauge invariant variables $u$, $v$, $\delta R$, and $q$.
With the condition $\delta R =0$ imposed, the equations for $u$ and $v$ are
found from (\ref{varyaux}) and become simply
\bes\bea
{\cal D}_2\,u = \frac{1}{H^2}\left( \frac{\partial^2}{\partial\tau^2}+ 
5H \frac{\partial}{\partial\tau}
+ 6H^2 - \frac{\nabla^2}{a^2}\right)u &=& 0\,,\\
{\cal D}_2 \, v = \frac{1}{H^2} \left( \frac{\partial^2}{\partial\tau^2}+ 
5H \frac{\partial}{\partial\tau}
+ 6H^2 - \frac{\nabla^2}{a^2}\right) v &=& 0\,,
\eea
\label{auxginv}\ees
while that for $q$ is the somewhat more complicated, because of the non-local
Weyl invariant terms in $\delta (T^a_{\ b})^{inv}$. For the case of a conformal 
matter/radiation field in de Sitter space these terms can be computed exactly 
\cite{HorWal,Starob81}, with the result \cite{DSAnom}
\bea
&&\left(1 - \bar\beta_{_R} - \frac{5\varepsilon}{3}\right) q
= \varepsilon H\tau\, ({\cal D}_1q)  - \bar\alpha_{_R}\,({\cal D}_1q)
+  \varepsilon' \frac{k^2}{H^2} u - \frac{\varepsilon}{3} \frac{k^2}{H^2} v\nn\\
&&\qquad \qquad- \frac{\varepsilon}{2a^2 } \int_{\eta_0}^{\eta}\, d\eta'\,
K(\eta -\eta'; k ;\mu)[a^2\,({\cal D}_1q)]_{\eta'} \,,
\label{qginv}
\eea
in Fourier space where  $-\nabla^2 \rightarrow k^2$, and $[a^2\,({\cal D}_1q)]$ 
in the integrand is evaluated at $\eta'$. The $u$ and $v$ terms coming
from the stress tensor of the anomaly and obeying the eqs. (\ref{auxginv})
are particular realizations of the possible state dependent terms mentioned
in Ref. \cite{BirDav}. The last term in (\ref{qginv}) is the non-local 
Weyl invariant term from $\delta (T^a_{\ b})^{inv}$ that involves the 
kernel $K$ is defined in conformal time $\eta$ by
\be 
K (\eta-\eta'; k; \mu) \equiv
\int_{-\infty}^{\infty}\frac{d\omega}{2\pi}\, e^{-i\omega
(\eta-\eta')} \,\ln\left[\frac{ -\omega^2 + k^2 - i\epsilon\,
{\rm sgn}\,\omega} {\mu^2}\right]\,.
\label{Kt}
\ee
It depends on an arbitrary renormlization scale $\mu$, and derives from 
the Weyl invariant term $S_{inv}$ in the exact quantum one-loop action 
around de Sitter space. The small (order $\hbar$) 
$\bar\alpha_{_R},\varepsilon, \varepsilon'$ parameters are defined by 
\bes
\bea
\bar\alpha_{_R} \equiv && 16 \pi  GH^2\alpha_{_R}\,,\\
\bar \beta_{_R} \equiv && 192 \pi  GH^2\beta_{_R}\,\\
\varepsilon \equiv && 32 \pi GH^2 b \,,\\
\varepsilon' \equiv && -\frac{32\pi }{3} GH^2 b'\,.
\eea\label{smparam}\ees
Let us make some observations about the {\it exact} linear response eqs.
(\ref{auxginv})-(\ref{qginv}) for conformal quantum fields about de Sitter space.
First it is important to point out that although the anomaly scalar eqs. 
(\ref{auxeom}) are fourth order in time derivatives, in fact only the gauge 
invariant quantities $u$ and $v$ satisfying the {\it second order} eqs. 
(\ref{auxginv}) couple to the linearized Einstein eq. (\ref{qginv}) to linear 
order. In other words, half of the solutions of the fourth order eqs. (\ref{linaux}) 
are annihilated by the operator ${\cal D}_0$ in (\ref{udef})-(\ref{vdef}) and 
{\it decouple} entirely. Only the other half of the solutions with non-zero $(u,v)$ 
satisfying (\ref{auxginv}) couple to the physical metric perturbations. This seems 
to be the analog of the elimination of the ghost degrees of freedom of the fourth
order action $\Gamma_{WZ}$ of (\ref{WZfour}) by the diffeomorphism constraints 
found previously in its quantization on $\mathbb{R}\otimes \mathbb{S}^3$ \cite{AMM}. 

Secondly, although the variations of the fourth order invaraints in $S_{local}$
are fourth order in time derivatives, they also appear in (\ref{qginv}) only 
with the {\it second order} differential operator (\ref{diffD}) on the gauge 
invariant quantity $q$. This is a consequence of the fact that the classical 
Einstein eqs. contain no dynamical metric perturbations in the scalar sector 
(\ref{linmet}) at all, and would be completely constrained ({\it i.e.} $q=0$) 
if it were not for the quantum corrections on the right side of (\ref{qginv}) 
which vanish in the limit $\hbar \rightarrow 0$. When $\hbar \neq 0$, the 
character of the eqs. changes discontinuously in that there are new non-trivial 
gauge invariant solutions of the second order equations (\ref{auxginv})-(\ref{qginv}).

Next we showed in Ref. \cite{DSAnom} that most of the complications of the exact 
linear response eq. (\ref{qginv}) are {\it irrelevant} for macroscopic physics. 
First, because all the dimensionless parameters defined in (\ref{smparam}) are 
very small compared to unity for $H << M_{Pl}$, the corrections in the parentheses 
on the left side of (\ref{qginv}) may be neglected. Then we analyzed the general 
homogeneous solutions of (\ref{auxginv}) and (\ref{qginv}), {\it i.e.} with
$u=v=0$ but including the effect of the non-local term involving $K$. If $u=v=0$ 
the non-trivial homogeneous solutions of (\ref{qginv}) involve oscillations on the 
Planck scale $M_{Pl}$, and hence lie outside the range of applicability of the effective 
action of low energy gravity. Hence all these homogeneous solutions of (\ref{qginv}) 
may be consistently neglected in the low energy theory for $k \ll M_{Pl}$, just
as the non-trivial solutions of (\ref{delReq}) with $\delta R \neq 0$ lie outside 
the range of applicability of the semi-classical effective theory and should
be discarded as unreliable. With the homogeneous solutions of (\ref{qginv}) thereby 
discarded for the same reason, the only non-trivial solutions remaining are the 
{\it inhomogeneous} ones for which either $u \neq 0$ or $v \neq 0$ (or both). For 
these solutions all the terms on the right side of (\ref{qginv}) linear in $q$ are 
suppressed by powers of $H/M_{Pl} \ll 1$ or $k_{phys}/M_{Pl} = k/aM_{Pl} \ll 1$ 
compared to the $u,v$ terms. 
 
Thus, as far as the predictions valid for macroscopic gravity are concerned
with physical wavelengths of perturbations much greater than the Planck
length, one may replace the complicated, non-local exact (\ref{qginv}) 
by the much simpler and fully determined
\be
\delta R^{\tau}_{\ \tau} = -\frac{H^2}{2a^2}\, q =  -\frac{\nabla^2}{2a^2} 
\left( -\varepsilon'  u + \frac{\varepsilon}{3} v\right) =
-\frac{16\pi GH^2}{3}\, \frac{\nabla^2}{a^2}\,(b'u +bv)\,,
\label{qinhom}
\ee
so that $\delta R^{\tau}_{\ \tau}\neq 0$ if and only if driven by the 
non-trivial solutions of the additional degrees of freedom provided by 
the anomaly scalar field equations (\ref{auxginv}). Hence we have verified 
explicitly by this analysis of the full linear response of conformal 
matter/radiation field fluctuations around de Sitter space that the $S^{(4)}_{local}$ 
and $S_{inv}$ terms in the exact quantum effective action do not influence 
low energy or macroscopic physics, consistent with the general classification
and Weyl scaling arguments of Sec. \ref{sec:EFT}. Instead we could have 
started with only the low energy effective action (\ref{Seff}) and obtained
(\ref{qinhom}) much more directly, which is correct at scales $k_{phys} \ll M_{Pl}$. 
Conversely, without inclusion of the anomaly generated terms one would miss 
entirely the physics associated with the degrees of freedom $u$ and $v$ 
and the solutions (\ref{auxginv}) in the scalar sector of cosmological 
perturbation theory not present in the purely classical theory for which 
$\delta R^{\tau}_{\ \tau} =0$ in this scalar sector.

The general solution of (\ref{auxginv}) for either $u$ or $v$ in
FLWR coordinates is easily found in Fourier space, namely
\be
u_{\pm} = v_{\pm} =  \frac{1}{a^2} \,\exp \left ( \pm \frac{ik}{Ha}\right)
e^{i \vec{k} \cdot \vec x}
= H^2\eta^2 \,e^{\mp i k\eta + i \vec k \cdot \vec x}\,.
\label{vsoln}
\ee
Thus, the auxiliary fields of the anomaly action yield the non-trivial gauge 
invariant solutions (\ref{qinhom}) with (\ref{vsoln}) for the linearized Ricci 
tensor perturbations $\delta R^{\tau}_{\ \tau} = -\delta R^i_{\ i}$ in the
scalar sector, which clearly are not present in the purely classical
Einstein theory without the anomaly action. Since with (\ref{delR0})
\be
\delta R^{\tau}_{\ \tau} = \delta R^i_{\ i} = 
-6H \left(\frac{\partial \Upsilon_{_{\!\cal C}}}{\partial\tau} 
 -H \Upsilon_{_{\!\!\cal A}}\right) + \frac{2\nabla^2}{a^2}\Upsilon_{_{\!\cal C}}\,,
\label{delRt}
\ee
using the solution for $\delta R^{\tau}_{\ \tau}$ (\ref{qinhom}) in (\ref{delRt}), 
differentiating and substituting in (\ref{delRg6}), one can solve for 
$\Upsilon_{_{\!\cal C}}$, to obtain
\be
-\frac{\nabla^2}{a^2}\Upsilon_{_{\!\cal C}}
= 8 \pi GH^2 \left(H\frac{\partial}{\partial \tau} + 4H^2 + 
\frac{\nabla^2}{3a^2}\right)(b'u + bv)\,,
\ee
where we have canceled a common factor of $\nabla^2$ from both sides
(valid for $\vec k \neq 0$).  This can be rewritten as
\be
-\nabla^2 \Upsilon_{_{\!\cal C}}
= 8 \pi GH^2 \left(H\frac{\partial}{\partial \tau} + 2H^2 + 
\frac{\nabla^2}{3a^2}\right)\left[a^2(b'u + bv)\right]\,.
\ee
Since from (\ref{vsoln}),
\be
a^2(b'u + bv) =  \left( c_+ e^{-ik\eta} + c_-e^{-i k\eta} \right) e^{i\vec k \cdot \vec x}\,,
\ee
with $c_{\pm}$ constants (depending on $b,b'$), at late times $\eta \rightarrow 0$, 
\be
-\nabla^2 \Upsilon_{_{\!\cal C}} \rightarrow  16 \pi GH^4\, 
( c_+ + c_-)\,e^{i \vec k \cdot \vec x}
\quad {\rm as} \quad \tau \rightarrow \infty\,.
\ee
Thus the Bardeen-Stewart potential $\Upsilon_{_{\!\cal C}}$ describing the gauge
invariant linearized perturbations of the de Sitter geometry, generated by the
stress tensor of the conformal anomaly in the scalar sector remains non-vanishing 
at late times {\it for every} $\vec k$ (while $\Upsilon_{_{\!\!\cal A}}$ and 
$\delta R^{\tau}_{\ \tau}$ falls off with $a^{-2}$).

Being solutions of (\ref{auxginv}) which itself is independent of the Planck scale, 
these new solutions due to the effective action of the anomaly vary instead 
on arbitrary scales determined by the wavevector $\vec k$, and are therefore 
genuine modes of the semi-classical effective theory. This is a non-trivial result 
since these modes appear in the {\it tracefree} sector of the semi-classical Einstein 
equations, with $\delta R=0$, and hence cannot be deduced directly from the local 
form of the trace anomaly itself, but only with the help of the covariant action 
functional (\ref{Sanom}) and the additional scalar degrees of freedom which the 
local form of this action implies. These additional modes, which couple to the scalar 
sector of metric perturbations in a gauge invariant way are due to a quantum effect
because the auxiliary scalar fields from which they arise are part of the
one-loop effective action for conformally invariant quantum fields, rather than 
a classical action for an {\it ad hoc} scalar inflaton field usually considered 
in inflationary models \cite{MukFelBra,LidL}. This demonstrates the relevance of the 
anomaly action for describing physical infrared fluctuations in the effective
semi-classical theory of gravity, on macroscopic or cosmological scales
unrelated to the Planck scale.

The Newtonian gravitational constant $G$ and the Planck scale enter
Eq. (\ref{qinhom}) only through the small coupling parameters $\varepsilon$
and $\varepsilon'$ between the anomaly scalar fields and the perturbation of the
geometry $\delta R^{\tau}_{\ \tau}$, which are defined in (\ref{smparam}). 
Thus in the limit of either flat space, or arbitrarily weak coupling 
$GH^2 \rightarrow 0$ the modes due to the anomaly scalars decouple 
from the metric perturbations at linear order.  As in the case of the gravitational 
scattering of Sec. \ref{sec:EFT} this explains why the anomaly scalars are 
so weakly coupled and difficult to detect directly in the flat space limit. 

The effective action (\ref{Seff}) expanded to quadratic order
about the self-consistent de Sitter solution (\ref{selfcons}) in terms of the
gauge invariant variables $u$, $v$, $\Upsilon_A$, and $\Upsilon_C$ is
\bea
&& S_{eff}^{(2)}\big\vert_{dS} = S_G
+ b^\prime \int d^3 \vec{x}\, d\tau \,a^3 \, \left[ - \frac{H^4 {u}^2}{2} 
+  \frac{H^2u\, \delta R}{3}  \right] \qquad\qquad\qquad\quad\nn \\
&& + b\!\int\! d^3 \vec{x}\, d\tau \,a^3\! \left\{ \!- H^4 uv + \frac{H^2v\, \delta R}{3}
+ \frac{4 \ln a}{3\, a^4}  \left[ \vec\nabla^2 (\Upsilon_A -  \Upsilon_C )\right]^2\right\}
\label{S2dS} 
\eea
where
\bea
&&S_G = \frac{1}{8 \pi G} \int d^3 \vec{x}\, d\tau \,a^3 
\left[ - 3\, \left( \frac{\partial \Upsilon_C} {\partial \tau} \right)^2
+    6\,H\,\Upsilon_A\, \frac{\partial \Upsilon_C}{\partial \tau} \right.\qquad\qquad \nn\\
&& \qquad\qquad\quad \left. +   \frac{2}{\,a^2}\ (\vec{\nabla}\Upsilon_A)
\cdot (\vec{\nabla}\Upsilon_C)
+   \frac{(\vec{\nabla} \Upsilon_C)^2}{a^2} -3H^2\,\Upsilon^2_A  \right] 
\eea
is the Einstein-Hilbert part of the action, and $\delta\!R$ is given by (\ref{delR0}). Varying
(\ref{S2dS}) with respect to $\Phi, \psi, \Upsilon_A$ and $\Upsilon_C$, and setting 
$\delta\! R= 0$ yields the low energy form of the gauge invariant linear response 
equations (\ref{auxginv}) and (\ref{qinhom}). The $\delta R$ terms in (\ref{S2dS}) 
cannot be set to zero until after the variations of $S_{dS}^{(2)}$ are performed since they 
generate the coupling between the $u, v$ and metric variables $\Upsilon_A, \Upsilon_C$. 
Note that the variations of (\ref{S2dS}) must be performed with respect to the original 
set of gauge invariant variables $\Phi, \psi, \Upsilon_A, \Upsilon_C$, in order to obtain
eqs. (\ref{udef})-(\ref{auxginv}) and (\ref{qinhom}). The linearized equations for the 
anomaly scalar fields so obtained are equivalent to (\ref{varyaux}), when it is 
recognized that
\be
\Delta_4 \psi = (-\sq + 2H^2)(-\sq) \psi = H^4 {\cal D}_2({\cal D}_0 \psi)\,,
\label{Del4factor}
\ee
{\it i.e.} that the fourth order conformal differential operator $\Delta_4$
associated with the Euler-Gauss-Bonnet invariant of the trace anomaly
factorizes into two second order differential operators in two {\it different}
ways in de Sitter space. As a practical matter this means that any solution
of the fourth order eq. (\ref{psidel4}), for which ${\cal D}_0 \psi \neq 0$,
automatically is a solution of (\ref{auxginv}) for $u$ or $v$. The disappearance
from the action (\ref{S2dS}) of any solution of the scalar field eqs. 
(\ref{linaux}) that is in the kernel of ${\cal D}_0$ is again noteworthy.

\subsection{Cosmological Horizon Modes}

The solutions (\ref{qinhom})-(\ref{vsoln}) corresponds to a linearized stress 
tensor perturbation of the form
\be
\delta \lag T^{\tau\tau} \rag_{_R} = \frac{H^2\,q}{16 \pi Ga^2} \propto
\,\frac{H^2k^2}{a^4}\, e^{\mp i k\eta + i \vec{k} \cdot \vec x}\,,
\label{varTttq}
\ee
Thus each mode of fixed $\vec k$ redshifts like $a^{-4}$, as a mode of a classical 
conformal radiation field. 

Not apparent from this form of this stress tensor perturbation for a
given $\vec k$ mode in FLRW coordinates is their possible relevance to the 
vacuum polarization effects near the horizon discussed in previous sections. 
To address this we need the other components of the stress tensor perturbation, 
and the behavior in the static coordinates of de Sitter space (\ref{dSstat}).

The other components of the stress tensor perturbation (\ref{varTttq}) in
FLRW coordinates can be found by a general tensor decomposition for
scalar perturbations analogous to Eqs.  (\ref{linmet}) for the metric. That is,
the general perturbation of the stress tensor $\delta \lag T^{ab} \rag_{_R}$
in the scalar sector can be expressed in terms of $\delta \lag T^{\tau\tau} \rag_{_R}$
plus three additional functions. These three functions are determined
by the conditions of covariant conservation,
\be
\nabla_b\, \delta \lag T^{ab} \rag_{_R} = 0\,
\ee
for $a=\tau$ and $a=i$ (two conditions), plus the tracefree condition,
\be
\delta \lag T^a_{\ a} \rag_{_R} = 0\,,
\ee
imposed as a result of the $\delta R = 0$ condition. Choosing the arbitrary
proportionality constant in (\ref{varTttq}) to be unity, a straightforward
calculation using these conditions and the Christoffel coefficients
in the flat FLRW coordinates gives \cite{DSAnom} 
\bes
\bea
&& \delta \lag T^{\tau\tau} \rag_{_R} = H^2  \frac{k^2}{a^4} \,
e^{\mp i k\eta + i \vec k \cdot \vec x} = 
- \frac{H^2}{a^4}\, \vec \nabla^2\,
e^{\mp i k\eta + i \vec k \cdot \vec x}\, ,\\
&&\delta \lag T^{\tau i} \rag_{_R} =
\pm H^2\,\frac{k^i k}{a^5}\, e^{\mp i k\eta
+ i \vec k \cdot \vec x}
= \frac{H^2}{a^4}\, \frac{\partial^2}{\partial x^i \partial \tau} \,
e^{\mp i k\eta + i \vec k \cdot \vec x}\, ,\\
&&\delta \lag T^{ij} \rag_{_R} = H^2\, \frac{k^ik^j}{a^6}
\, e^{\mp i k\eta + i \vec k \cdot \vec x} =
- \frac{H^2}{a^6}\, \frac{\partial^2}{\partial x^i \partial x^j}\,
e^{\mp i k\eta + i \vec k \cdot \vec x}\,,\qquad
\eea\label{TkFRW}\ees
\noindent
for the other components of the stress tensor variation for these modes
in the flat FLRW  coordinates of de Sitter space. If one averages this form over
the spatial direction of $\vec k$, a spatially homogeneous, isotropic stress
tensor is obtained with pressure $p = \rho/3$. In FLRW coordinates
this averaging describes incoherent or mixed state thermal perturbations
of the stress tensor which are just those of massless radiation.

Next we use (\ref{RWcoor}) and (\ref{staticoor}) to read off the coordinate transformation
from FLRW flat coordinates $\tau$ and $\varrho = |\vec x|$ to static $t$ and $r$ 
coordinates, given by
\bes
\bea
&& r = a\, |\vec x | \equiv a \varrho = \varrho\, e^{H \tau}\,,\\
&& t = \tau - \frac{1}{2H} \ln \left(1 - H^2 \varrho^2 e ^{2H\tau}\right)\,.
\eea
\ees
The inverse transformations are
\bes
\bea
\varrho \equiv |\vec x|= \frac{r\,e^{-H t}}{\sqrt{1-H^2r^2}}\,,\\
\tau = t + \frac{1}{2H} \ln (1-H^2r^2)\,.
\eea
\ees
The Jacobian matrix of this $2 \times 2$ transformation of bases is
\be
\frac{\partial (t,r)}{\partial(\tau,\varrho)} \equiv \left( \begin{array}{cc}
\frac{\partial t}{\partial \tau} & \frac{\partial t}{\partial \varrho}\\
\frac{\partial r}{\partial \tau} & \frac{\partial r}{\partial \varrho}
\end{array}\right) = \left(\begin{array}{cc}
\frac{1}{1-H^2r^2} & \frac{Hr^2}{\varrho(1-H^2r^2)}\\
Hr & r/\varrho \end{array}\right)
\label{Jacob}
\ee
Using these relations, one may express the action of the differential
operators in Eq. (\ref{auxginv}) in terms of the static coordinates
(\ref{dSstat}) instead. Since
\be
-\frac{\vec \nabla^2}{a^2} = \frac{1}{a^2}\left[-\frac{1}{\varrho^2} 
\frac{\partial}{\partial \varrho}
\left( \varrho^2 \frac{\partial}{\partial \varrho}\right) 
+ \frac{L^2}{\varrho^2}\right]\,,
\ee
where $- L^2$ is the scalar Laplacian on the sphere $\mathbb{S}^2$, 
a straightforward calculation using (\ref{Jacob}) shows that
\bea
&&- \frac{\vec \nabla^2}{a^2}\,v = -\frac{1}{(1-H^2r^2)^2}
\left[ H^2r^2 \frac{\partial^2}{\partial t^2} 
+ H(3-H^2r^2) \frac{\partial}{\partial t}\right]v\nn\\
&& \qquad - \frac{2Hr}{1-H^2r^2} \frac{\partial^2 v}{\partial t\partial r}
-\frac{1}{r^2} \frac{\partial}{\partial r} \left(r^2 \frac{\partial v}{\partial r}\right)
+ \frac{L^2}{r^2}\, v\,,
\label{uinT}
\eea
while
\bea
&&H^2 v = \left(\frac{\partial^2}{\partial \tau^2} + H\frac{\partial}{\partial \tau}
- \frac{\vec \nabla^2}{a^2}\right) \psi = H^2 {\cal D}_0 \psi\nn\\
&&\hspace{-1.6cm}=\frac{1}{1-H^2r^2}\!\left( \frac{\partial^2 \psi}{\partial t^2} 
- 2H\frac{\partial \psi}{\partial t}\right)\!
- (1-H^2r^2)\frac{1}{r^2} \frac{\partial}{\partial r}\!
\left(r^2 \frac{\partial\psi}{\partial r}\right)\!
+ \frac{L^2}{r^2}\psi,
\label{vop}
\eea
and
\bea
&&{\cal D}_2 v = 0= \left(\frac{\partial^2}{\partial \tau^2} 
+ 5 H\frac{\partial}{\partial \tau} + 6H^2 - \frac{\vec \nabla^2}{a^2}\right) v\qquad\qquad  \nn\\
&&\qquad=\frac{1}{1-H^2r^2}\!\left( \frac{\partial^2 v}{\partial t^2} 
+  2H\frac{\partial v}{\partial t}\right) - (1-H^2r^2)\frac{\partial^2v}{\partial r^2}\nn\\
&& \qquad\qquad\quad +2\left( 3H^2r - \frac{1}{r}\right)\frac{\partial v}{\partial r} 
+ 6H^2 v + \frac{L^2}{r^2}v\,,
\label{D2v}
\eea
thus converting these differential operators from flat FLRW to de Sitter static 
coordinates (\ref{dSstat}).

In (\ref{genphids}) and (\ref{genpsids}) the general solutions
of the homogeneous eq. (\ref{psidel4}) as functions of the static $r$
in de Sitter space are given. Of the four,
\be
\frac{1}{Hr}\ln\left(\frac{1-Hr}{1+Hr}\right)\,,\quad
\ln\left(\frac{1-Hr}{1+Hr}\right)\,,
\quad 1\,, \quad \frac{1}{r}\,.
\label{psistatic}
\ee
the second and third solutions are solutions of the second order eq. for
a minimally coupled scalar field, {\it i.e.} they satisfy $\sq \psi =0$,
while the first and last solution satisfy the second order eq, for a
conformally coupled scalar field, {\it i.e.} they satisfy $(-\sq + 2H^2) \psi = 0$.
The fourth solution $1/r$ is singular at the origin and so was not considered in 
Sec. \ref{sec:anom}. In any case this solution and the constant solution in
(\ref{psistatic}) give vanishing contribution to $v$ in (\ref{vop}) while the
first and second solutions give for $ v$,
\be
\frac{4}{1-H^2r^2}\,, \quad \frac{4}{H r}\,\frac{1+H^2r^2}{1-H^2r^2} \,,
\label{vstatic2}
\ee
respectively. The second gives a singular contribution to $v$ and the stress tensor
at $r=0$, so we consider only the first solution. Note that if we allow also for a term 
linear in $t$ in the list (\ref{psistatic}) of homogeneous solutions to (\ref{psidel4}), 
then from (\ref{vop}) it would also produce the identical form for $v$ as the first member of 
(\ref{vstatic2}). Thus the general spherically symmetric static solution to the 
homogeneous eq. (\ref{psidel4}), which because of (\ref{Del4factor}) is also 
a solution of (\ref{D2v}), which is non-singular at the origin is
\be
\frac{2}{3}(b'u + bv) \propto \frac{1}{1-H^2r^2}\,.
\label{vstatic}
\ee
Choosing the arbitrary normalization constant of proportionality to
be unity, then from (\ref{qinhom}), (\ref{uinT}) and (\ref{vstatic}), we obtain
\be
\delta \lag T^{\tau\tau}\rag_{_R} = 
-H^2\frac{\nabla^2}{a^2} \left(\frac{1}{1-H^2r^2}\right) = -\frac{H^4}{(1-H^2r^2)^2}
\left( \frac{3}{2} + \frac{2H^2 r^2}{1-H^2r^2}\right).
\label{divTR}\ee
This shows that a linear superposition of solutions (\ref{vsoln}) of the linear response 
equations in static coordinates can lead to gauge invariant perturbations which 
{\it diverge} on the de Sitter horizon.

To see what stress tensor (\ref{divTR}) corresponds to in the static $(t, r)$ coordinates,
we use the form of the other components in (\ref{TkFRW}) in FLRW coordinates,
(\ref{Jacob}), and the transformation relation for tensors,
\be
T^{tt} = \left(\frac{\partial t}{\partial \tau }\right)^2 T^{\tau\tau} +
2 \left(\frac{\partial t}{\partial \tau}\right)
\left(\frac{\partial t}{\partial x^i}\right)T^{\tau i}
+ \left(\frac{\partial t}{\partial x^i}\right)
\left(\frac{\partial t}{\partial x^j}\right) T^{ij}\,,
\ee
with (\ref{Jacob}), (\ref{uinT}), to obtain
\be
\delta \lag T^{t}_{\ t}\rag_{_R} = -(1- H^2r^2)\,
\delta \lag T^{t t}\rag_{_R} =  \frac{6\, H^4}{(1-H^2r^2)^2} \,.
\label{divstatT}
\ee
Here use has also been made of the identities,
\be
\frac{\partial t}{\partial x^i} =
\frac{\partial t}{\partial \varrho} \frac{\partial \varrho}{\partial x^i}
= \frac{\partial t}{\partial \varrho} \frac{x_i}{\varrho}\,,
\ee
and
\bes
\bea
&& \frac{\partial}{\partial \tau} = \frac{\partial r}{\partial \tau}
\frac{\partial}{\partial r} + \frac{\partial t}{\partial \tau}
\frac{\partial}{\partial t} = Hr\,\frac{d}{dr}\,,\\
&& x_i \partial^i =  x^i \partial_i = \varrho \frac{\partial}{\partial\varrho} =
\varrho \left( \frac{\partial r}{\partial \varrho} \frac{\partial}{\partial r}
+ \frac{\partial t}{\partial \varrho} \frac{\partial}{\partial t}\right)
= r \frac{d}{dr}\\
&&x_ix_j \partial^i\partial^j = (x^i\partial_i)(x^j\partial_j)
- x^i\partial_i = r \frac{d}{dr}\left(r \frac{d}{dr}\right)
- r\frac{d}{dr}\qquad
\eea\ees
valid when operating on functions of $r$ only. Likewise we find
\be
\delta \lag T^r_{\ r}\rag_{_R} = \delta \lag T^{\theta}_{\ \theta}\rag_{_R} =
\delta \lag T^{\phi}_{\ \phi}\rag_{_R} = -\frac{2\, H^4}{(1-H^2r^2)^2}
= - \frac{1}{3} \delta \lag T^{t}_{\ t}\rag_{_R}\,,
\label{statT}
\ee
corresponding to a perfect fluid with $p = \rho/3$, but now in {\it static} 
coordinates. As we know from the conservation eq. (\ref{cons}) in static 
coordinates, and (\ref{dSthermal}), this eq. of state leads to a quadratic 
divergence in the metric factor $(1-H^2r^2)^{-2}$ as $r \rightarrow r_{_H}$. 
This is the exact analog of the large blueshift factor discussed previously 
for the Schwarzschild case, (\ref{TBoul}). Because of the buildup of the effects 
of the anomaly scalar fields on the cosmological horizon in de Sitter space, 
they may be called {\it cosmological horizon modes}.

The form of the stress tensor (\ref{divstatT}), (\ref{statT}) is the form of a finite
temperature fluctuation away from the Hawking-de Sitter temperature $T_{_H} = H/2\pi$
of the Bunch-Davies state in static coordinates \cite{AndHis}. Since the equation
the solutions (\ref{psistatic}) satisfy is the same as (\ref{psidel4}), it follows
that there exist linear combinations of the solutions  (\ref{vsoln}) found in
Sec. \ref{sec:linres} which give (\ref{vstatic}) and the diverging behavior
of the linearized stress tensor on the horizon, corresponding to this global
fluctuation in temperature over the volume enclosed by the de Sitter horizon.
Since the solution $\psi = pt$ also generates a solution of (\ref{auxginv})
for $v$ of the same form as the first member of (\ref{vstatic2}), we can expect
that perturbations of the quantum state described by a non-zero
time derivative (in static time coordinates) within one causal horizon volume 
at some initial time $t=0$ will give rise to very large stress tensors
of the form of (\ref{divstatT}) and (\ref{statT}) on tha horizon of
that causal region at later times.

Note that in static coordinates the stress tensor $p= \rho/3$ does not
involve averaging over directions of $\vec k$, but a particular {\it coherent} 
linear superposition over modes (\ref{vsoln}) with different $\vec k$ in order
to obtain a particular isotropic but spatially inhomogeneous solution of
(\ref{vstatic}). This selects a preferred origin and corresponding horizon
in static coordinates (\ref{dSstat}). The fluctuations in Hawking-de Sitter
temperature thus preserve an $O(3)$ subgroup of the de Sitter isometry group
$O(4,1)$. Clearly the origin chosen at $r=0$ is completely arbitrary,
and a particular $O(3)$ subgroup is chosen at random by the perturbation. 
This is similar to a bubble nucleation process in a homogeneous thermal
ensemble: no origin for the bubble nucleation center is preferred over any
other, but some definite particular origin around which the bubble nucleates
is selected by the fluctuations. In other words, the de Sitter group $O(4,1)$ 
is spontaneously broken to $O(3)$ by any random fluctuation of temperature 
away from its Hawking-de Sitter value, centered on some arbitrary but definite 
point of de Sitter space at an initial time $t=0$.

The free energy functional relevant to this spontaneous symmetry
breaking of de Sitter symmetry is the quadratic effective action (\ref{S2dS}), 
in which the last $\ln a$ term acts as a the {\it negative} of a potential term 
(for $b > 0$ and $\ln a  > 0$) favoring a non-zero  and spatially inhomogeneous 
$\nabla^2 (\Upsilon_{_{\!\!\cal A}} - \Upsilon_{_{\!\cal C}})$ perturbation away 
from de Sitter space. The $-b' u^2$ term behaves similarly (for $b' < 0$). In 
FLRW coordinates the action is time dependent. However, in static coordinates 
the volume measure is time independent and it is clear that a solution for $u$ 
of the form (\ref{vstatic}) makes the Euclidean continued action under 
$t \rightarrow it$ {\it arbitrarily negative} from its divergent behavior 
(\ref{vstatic}) at the horizon. Thus perturbations of this kind should be 
energetically favored, and {\it destabilize} global de Sitter space in favor 
of a finite patch up to the horizon, centered on a fixed origin. To understand 
what becomes of the near horizon region clearly requires going beyond 
the linearized expansion around the de Sitter background. 

Referring back to the full anomalous effective action (\ref{Seff})-(\ref{SEF}), 
one observes that the $C_{\alpha\beta\mu\nu}C^{\alpha\beta\mu\nu} \varphi$ 
term acts as a {\it negative} potential term in general as well, favoring the 
breaking of Weyl invariance unless $\varphi$ vanishes, which is impossible
in de Sitter space because of the non-vanishing source term for $\varphi$
in (\ref{auxeom}). Note also that the Euler-Gauss-Bonnet source term
$E$ is non-vanishing for any solution of the vacuum Einstein's eqs.
with $\Lambda \neq 0$. Only $\Lambda_{eff} = 0$ (and $\sq R = 0$) can 
give a vanishing source for $\varphi$, and eliminate the $C^2 \varphi$ 
term in the effective action. Thus the effective action of the trace
anomaly provides a possible way to distinguish vacua with different
values of the vacuum energy in gravity, and in the absence
of any boundary terms, selecting $\Lambda_{eff} =0$.

To follow the diverging behavior (\ref{divstatT}), (\ref{statT})  all the way 
to the horizon one would clearly require a linear combination of the solutions 
(\ref{vsoln}) with large Fourier components. However once $8\pi G$ times the 
perturbed stress tensor in (\ref{divstatT}) becomes of the same order as the 
classical background Ricci tensor $H^2$, the linear response theory breaks 
down and non-linear backreaction effects must be taken into account. The
perturbation becomes of the same order as the background at
$r = r_{_H} - \Delta r$ near the horizon, where
\be
\Delta r \sim L_{Pl}\,,
\ee
or because of the line element (\ref{dSstat}), at the proper distance 
from the horizon of
\be
\ell \sim \sqrt{r_{_H}L_{Pl}} \gg L_{Pl}\,.
\label{ellest}
\ee
This correspond to a maximum $k_{phys} \sim 1/\ell \ll M_{Pl}$,
where the semi-classical description may still be trusted. At the distance (\ref{ellest})
from the horizon, the state dependent contribution to the stress-energy tensor becomes
comparable to the classical de Sitter background curvature, the linear approximation
breaks down, and non-linear backreaction effects may be expected. 

\section{Gravitational Condensate Stars}
\label{sec:grava}

The paradoxes of black hole physics and the possibility of large backreaction
effects on the Schwarzschild horizon were discussed in Sec. \ref{sec:BH}. The problem 
of dark energy and corresponding possible large backreaction effects on the
de Sitter horizon were discussed in Secs. \ref{sec:darkenergy} and \ref{sec:cosmo}. 
A considerable technical machinery and intuition associated with quantum
effects near horizons has been elaborated. The effective action of the
trace anomaly contains massless degrees of freedom which can become significant 
and potentially lead to large backreaction effects on the classical geometry close
to horizons. In addition the fluctuations of the massless 
scalar degrees of freedom associated with the conformal factor of the metric 
described by the anomaly action allow the vacuum energy to become dynamical 
in which case it may vary in both space and time, unlike in the classical Einstein
theory. 

To bring all these ideas and results together it is attractive to consider 
the possibility of a resolution of both the black hole and cosmological
dark energy problems at one stroke, by matching an interior de Sitter
geometry with one positive value of the vacuum energy to an exterior
Schwarzschild geometry with another smaller value of the vacuum energy
(which we can take to be zero). The matching should take place in a small 
but finite region at the position of the common de Sitter and Schwarzschild 
boundary $r_{_H} \simeq r_{_S}$. A spherically symmetric vacuum ``bubble" 
of this kind was suggested several years ago, and called a {\it gravitational 
condensate star} or {\it gravastar} \cite{gstar,PNAS,QFEXT03}.
Similar ideas in which the black hole event horizon is replaced by a quantum
critical surface were proposed in Refs. \cite{CHLS01,Lau}.

In the gravastar model, we replace the effective action and the stress tensor
it generates by an easier to treat effective fluid description, and reconsider
the Einstein eqs. with a perfect fluid source as a simplified model of the quantum 
effects to be described finally in the full EFT. In an effective mean field treatment 
for a perfect fluid at rest in the coordinates (\ref{sphsta}), any static, spherically 
symmetric collapsed object must satisfy the Einstein eqs. (\ref{Einsab}), together 
with the conservation eq. (\ref{cons}) which ensures that the other components of 
the Einstein eqs. are satisfied. In the general spherically symmetric situation the 
tangential pressure, $p_{\perp} \equiv T^{\theta}_{\ \theta} = T^{\phi}_{\ \phi}$ 
is not necessarily equal to the radial normal pressure $p = T^r_{\ r}$. However, 
the simplest possibility which illustrates the main idea is to take $p_{\perp} = p$,
(except possibly at the boundaries between layers). In that case, we have three first order eqs. for four unknown functions of $r$, {\it viz.} 
$f, h, \rho$, and $p$. The system becomes closed when an eq. of state for the 
fluid, relating $p$ and $\rho$ is specified. If we define the mass function $m(r)$ by 
\be
h(r) = 1 - \frac{2Gm(r)}{r}\,,
\ee
so that (\ref{Einsa}) becomes
\be
\frac {dm}{dr} = 4 \pi r^2\rho\,,
\ee
and eliminate $f$ between (\ref{Einsb}) and (\ref{cons}), we obtain
\be
\frac{dp}{dr} = - \frac{G(\rho + p) (m + 4 \pi r^3 p)}{r(r-  2Gm)}\,,
\ee
which is the TOV equation of hydrostatic equilibrium \cite{TOV}.

Because of the considerations of the previous sections, we allow for three different 
regions with the three different eqs. of state,
\be
\begin{array}{clcl}
{\rm I.}\ & {\rm Interior:}\ & 0 \le r < r_1\,,\ &\rho = - p \,,\\
{\rm II.}\ & {\rm Thin\ Shell:}\ & r_1 < r < r_2\,,\ &\rho = + p\,,\\
{\rm III.}\ & {\rm Exterior:}\ & r_2 < r\,,\ &\rho = p = 0\,.
\end{array}
\label{regions}
\ee
In the interior region $\rho = - p$ is a constant from (\ref{cons}). This is an effective
cosmological ``constant" in the interior. Let us call the constant $\rho_{_V} = 3H_0^2/8\pi G$. 
If we require that the origin is free of any mass singularity then the interior is determined 
to be a region of de Sitter spacetime in static coordinates, {\it i.e.}
\be
{\rm I.}\qquad f(r) = C\,h(r) = C\,(1 - H_0^2\,r^2)\,,\quad 0 \le r \le r_1\,,
\label{deS}
\ee
where $C$ is an arbitrary constant, corresponding to the freedom to redefine the
interior time coordinate.

The unique solution in the exterior vacuum region which approaches flat spacetime
as $r \rightarrow \infty$ is a region of Schwarzschild spacetime (\ref{Sch}),
{\it viz.}
\be
{\rm III.} \qquad f(r) = h(r) = 1 - \frac{2GM} {r}\,,\qquad  r_2 \le r\,.
\ee
The integration constant $M$ is the total mass of the object.

The only non-vacuum region is region II. Let us define the dimensionless
variable $w$ in this section by $w\equiv 8\pi G r^2 p$, so that
eqs. (\ref{Einsab})-(\ref{cons}) with $\rho = p$ may be recast in the form,
\be
\frac{dr}{r} = \ \frac{dh}{1-w-h}\,,
\label{ueqa}
\ee
\be
\frac{dh}{h} = -\frac{1-w-h}{1 + w - 3h}\, \frac{dw}{w}\,,
\label{ueqb}
\ee
together with $p f \propto wf/r^2$ a constant. Eq. (\ref{ueqa}) is equivalent to 
the definition of the (rescaled) Tolman mass function by $h = 1 - 2m(r)/r$ and 
$d m(r) = 4\pi G\, \rho r^2\, dr = w\, dr/2$ within the shell. Eq. (\ref{ueqb}) 
can be solved only numerically in general. However, it is possible to obtain an 
analytic solution in the thin shell limit, $0 < h \ll 1$, for in this limit we 
can set $h$ to zero on the right side of (\ref{ueqb}) to leading order,
and integrate it immediately to obtain
\be
h \equiv 1- \frac{2Gm}{r} \simeq  \epsilon\ \frac{(1 + w)^2}{w \ }\ll 1\,,
\label{hshell}
\ee
in region II, where $\epsilon$ is an integration constant.
Because of the condition $h \ll 1$, we require $\epsilon \ll 1$, with $w$
of order unity. Making use of eqs. (\ref{ueqa})-(\ref{ueqb}) and (\ref{hshell}) we have
\be
\frac{dr}{r} \simeq - \epsilon\, dw\, \frac{(1 + w)}{w^2}\,.
\label{req}
\ee
Because of the approximation $\epsilon \ll 1$, the radius $r$ hardly changes within
region II, and $dr$ is of order $\epsilon \,dw$. The final unknown function $f$ is given
by $f = (r/r_1)^2 (w_1/w) f(r_1) \simeq  (w_1/w) f(r_1)$ for small $\epsilon$, showing that
$f$ is also of order $\epsilon$ everywhere within region II and its boundaries.

At each of the two interfaces at $r=r_1$ and $r=r_2$ the induced three dimensional
metric must be continuous. Hence $r$ and $f(r)$ are continuous at the interfaces,
and
\be
f(r_2) \simeq \frac{w_1}{w_2} f(r_1) = \frac{C w_1}{w_2} (1-H_0^2 r_1^2) 
= 1 - \frac{2GM}{r_2}\,.
\ee
To leading order in $\epsilon \ll 1$ this relation implies that
\be
r_1 \simeq \frac{1}{H_0} \simeq 2GM \simeq r_2\,.
\label{HMiden}
\ee
Thus the interfaces describing the phase boundaries at $r_1$ and $r_2$ are very close to
the classical event horizons of the interior de Sitter and exterior
Schwarzschild geometries, while the full solution has no horizon at all.

The significance of $0< \epsilon \ll 1$ is that both $f$ and $h$ are of order
$\epsilon$ in region II, but are nowhere vanishing. Hence there is no event horizon,
and $t$ is a global time. A photon experiences a very large, ${\cal O}(\epsilon^{-\frac{1}{2}})$
but finite blue shift in falling into the shell from infinity.
The proper thickness of the shell between these interface boundaries is
\be
\ell = \int_{r_1}^{r_2}\, dr\,h^{-\frac{1}{2}}  \simeq r_{_S}
\epsilon^{\frac{1}{2}}\int_{w_2}^{w_1} dw\, w^{-\frac{3}{2}}
\sim \epsilon^{\frac{1}{2}} r_{_S}\,,
\label{thickness}
\ee
and very small for $\epsilon \rightarrow 0$. Because of (\ref{HMiden}) the
constant vacuum energy density in the interior is just the total mass $M$
divided by the volume, {\it i.e.} $\rho_{_V} \simeq 3M/4\pi r_{_S}^3$,
to leading order in $\epsilon$. The energy within the shell itself,
\be
E_{\rm II} = 4\pi \int_{r_1}^{r_2}\rho\,r^2 dr \simeq \epsilon M
\int_{w_2}^{w_1}\frac{dw}{w}\,(1 + w)\sim \epsilon M
\ee
is extremely small.

We can estimate the size of $\epsilon$ and $\ell$ by consideration
of the expectation value of the quantum stress tensor in the static exterior
Schwarzschild spacetime. In the static vacuum state corresponding to no incoming
or outgoing quanta at large distances from the object, {\it i.e.} the Boulware
vacuum \cite{Boul}, the stress tensor near $r = r_{_S}$ is the negative of the stress 
tensor of massless radiation at the blue shifted temperature, $T_{loc} = T_{_H}/\sqrt {f(r)}$
and diverges as $T_{loc}^4 \sim f^{-2}(r)$ as $r\rightarrow r_{_S}$. The location
of the outer interface occurs at an $r$ where this local stress-energy $\propto M^{-4}
\epsilon^{-2}$, becomes large enough to affect the classical Schwarzschild curvature
$\sim M^{-2}$, {\it i.e.} when
\be
\epsilon \sim \frac{M_{pl}} {M} \simeq 10^{-38}\ \left(\frac {M_{\odot}}{M}\right)\,,
\label{epsest}
\ee
where $M_{Pl}$ is the Planck mass $\sqrt{\hbar c/G} \simeq 2 \times 10^{-5}$ gm.
Thus $\epsilon$ is indeed very small for a stellar mass object, justifying the
approximation {\it a posteriori}. With this semi-classical estimate for $\epsilon$
we find
\be
\ell \sim \,\sqrt {L_{_{Pl}}\, r_{_S}} \simeq \, 3 \times 10^{-14}\,
\left(\frac {M} {M_{\odot}}\right)^{\frac{1}{2}}{\rm cm.}
\label{meanell}
\ee
This is the same estimate that was obtained in (\ref{ellest}), for where the
vacuum polarization effects described by the stress tensor of the anomaly scalar 
fields becomes of the same order as the interior de Sitter curvature. Although still 
microscopic, the thickness of the shell is very much larger than the Planck
scale $L_{Pl} \simeq 2 \times 10^{-33}$ cm. The energy density and pressure in the shell
are of order $M^{-2}$ and far below Planckian for $M\gg M_{Pl}$, so that the geometry can be
described reliably by Einstein's equations in both regions I and II. 

One may think of $\ell$ as the analog of the {\it skin depth} of a metal arising from 
its finite conductivity that cuts off the divergence in the Casimir stress tensor near 
a curved boundary \cite{CanDeu,BorMohMil}, or of the {\it healing length} of an inhomogeneous
Bose-Einstein condensate \cite{Gro}, or finally as the thickness of a {\it boundary layer}
or stationary shock front in hydrodynamics \cite{Ands}. In all of these examples
from other areas of physics a lowest order macroscopic description of a bulk medium 
must be supplemented with some first order information about microscopic interactions
in order to describe a rapid spatial crossover between two regions in which
the bulk macroscopic description is completely adequate. In an EFT language this
means that certain higher derivative interaction terms in the mean field eqs.
related to fluctuations about the mean which are negligible in the bulk medium
must be taken into account in the surface crossover layer. Although these
higher derivative terms are present only because of some underlying microscopic 
degrees of freedom, the scale at which they become important are typically much
larger than the fundamental microscopic or atomic scale (here $L_{Pl}$), so that an
EFT approach is still possible.

Although $f(r)$ is continuous across the interfaces at $r_1$ and $r_2$, in our simple
model the discontinuity in the eqs. of state does lead to discontinuities in $h(r)$ and 
the first derivative of $f(r)$ in general. Defining the outwardly directed unit normal vector 
to the interfaces, $n^b= \delta_r^{\ b} \sqrt{h(r)}$, and the extrinsic curvature $K^a_{\ b} = \nabla_an^b$, the Israel junction conditions \cite{Isrj} determine the surface stress energy 
$\eta$ and surface tension $\sigma$ on the interfaces to be given by the discontinuities 
in the extrinsic curvature through
\cite{Isrj}
\bes\bea
&&[K_t^{\ t}] = \left[ \frac {\sqrt h}{2f} \frac {df}{dr} \right] = 4\pi G(\eta - 2\sigma)\,,
\label{junca}\\
&&\left[K_{\theta}^{\ \theta}\right] = [K_{\phi}^{\ \phi}] = \left[ \frac {\sqrt h}{r}\right]
= - 4\pi G\eta\,.
\label{juncb}
\eea\label{juncab}\ees
Since $h$ and its discontinuities are of order $\epsilon$, the energy density in the surfaces
$\eta \sim \epsilon^{\frac{1}{2}}$, while the surface tensions are of order $\epsilon^{-\frac{1}{2}}$.
The simplest possibility for matching the regions is to require that the surface energy densities
on each interface vanish. From (\ref{juncb}) this condition implies that $h(r)$
is also continuous across the interfaces, which yields the relations,
\bes\bea
h(r_1) &=& 1-H_0^2 r_1^2 \simeq \epsilon \,\frac{(1+ w_1)^2}{w_1}\,,
\label{matcha}\\
h(r_2) &=& 1 - \frac{2GM}{r_2} \simeq  \epsilon \,\frac{(1+ w_2)^2}{w_2}\,,\\
\frac{f(r_2)}{h(r_2)} &=& 1 \simeq \frac{w_1}{w_2} \frac{f(r_1)}{h(r_2)} = C
\left(\frac{1+w_1}{1+w_2}\right)^2
\label{matchb}
\eea\ees
From (\ref{req}) $dw/dr <0$, so that $w_2 < w_1$ and $C < 1$. In this case of 
vanishing surface energies $\eta = 0$ the surface tensions are determined by 
(\ref{juncab}) to be
\bes\bea
&&\quad \sigma_1 \simeq -\frac{1}{32\pi G^2M}
\frac{(3 + w_1)}{(1 + w_1)}\left(\frac{w_1}{\epsilon}\right)^{\frac{1}{2}}\,,
\label{surfa}\\
&&\quad \sigma_2 \simeq \frac{1}{32\pi G^2M}  
\frac{w_2}{(1 + w_2)}\left(\frac{w_2}{\epsilon}\right)^{\frac{1}{2}}.
\label{surfb}
\eea\label{surf}\ees
to leading order in $\epsilon$ at $r_1$ and $r_2$ respectively. The negative surface tension
at the inner interface is equivalent to a positive tangential pressure, which implies
an outwardly directed force on the thin shell from the repulsive vacuum within. The positive
surface tension on the outer interfacial boundary coresponds to the more familiar case of an
inwardly directed force exerted on the thin shell from without.

The entropy of the configuration may be obtained from the Gibbs relation,
$p + \rho = sT + n\mu$, if the chemical potential $\mu$ is known in each region.
In the interior region I, $p + \rho = 0$ and the excitations are the usual
transverse gravitational waves of the Einstein theory in de Sitter space.
Hence the chemical potential $\mu$ may be taken to vanish and
the interior has zero entropy density $s=0$, consistent with a single
macroscopic condensate state, $S= k_{_B} \ln W(E) = 0$ for $W(E)=1$.
In region II there are several alternatives depending upon the nature
of the fundamental excitations there. The $p=\rho$ eq. of state may come
from thermal excitations with negligible $\mu$ or it may come from a
conserved number density $n$ of gravitational quanta at zero temperature.
Let us consider the limiting case of vanishing $\mu$ first.

If the chemical potential can be neglected in region II, then the entropy of the shell
is obtained from the eq. of state, $p = \rho = (a^2/8\pi G) (k_{_B} T/\hbar)^2$. The
$T^2$ temperature dependence follows from the Gibbs relation with $\mu =0$, together
with the local form of the first law $d\rho = T ds$. The Newtonian constant $G$ has
been introduced for dimensional reasons and $a$ is a dimensionless constant.
Using the Gibbs relation again the local specific entropy density
$s(r) = a^2k_{_B}^2 T(r)/4\pi\hbar^2 G = a(k_{_B}/\hbar)(p/2\pi G)^{\frac{1}{2}}$
for local temperature $T(r)$. Converting to our previous variable $w$, we find
$s = (ak_{_B}/4\pi\hbar Gr)\,w^{\frac{1}{2}}$ and the entropy of the fluid within the shell is
\be
S =4\pi \int_{r_1}^{r_2}s\,r^2\,dr\,h^{-\frac{1}{2}}\simeq
\frac{ak_{_B}r_{_S}^2}{\hbar G}\
\epsilon^{\frac{1}{2}} \ \ln \left(\frac{w_1}{w_2}\right)\,,
\label{entsh}
\ee
to leading order in $\epsilon$. Using (\ref{thickness}) and (\ref{meanell}), this is
\be
S \sim a\, k_{_B}\frac{M\ell}{\hbar}
\sim 10^{57}\ a\,k{_{_B}}\,\left(\frac {M} {M_{\odot}}\right)^{\frac{3}{2}}\ll S_{_{BH}}\,.
\label{entest}
\ee
The maximum entropy of the shell and therefore of the entire configuration is some $20$
orders of magnitude smaller than the Bekenstein-Hawking entropy (\ref{SBH})
for a solar mass object, and of the same order of magnitude as a typical progenitor 
of a few solar masses. The scaling of (\ref{entest}) with $M^{\frac{3}{2}}$ is also 
in agreement with our general estimate (\ref{relstar}) for a relativistic star, and
the same order of magnitude as that for a supermassive star with $M > 50\,M_{\odot}$, 
whose pressure is dominated by radiation pressure \cite{ZelNov}. Thus the formation of the 
gravastar from either a solar mass or supermassive stellar progenitor does not require 
an enormous generation or removal of entropy. Since the entropy is of the same order
as that of a typical stellar progenitor, there is no information paradox and no significant 
entropy shedding needed to produce a cold gravitational vacuum or `grava(c)star' remnant. 
Because of the absence of an event horizon, the gravastar does not emit Hawking radiation.
Since $w$ is of order unity in the shell while $r \simeq r_{_S}$, the {\it local} temperature
of the fluid within the shell is of order $T_{_H} \sim \hbar/k_{_B}GM$. The strongly
redshifted temperature observed at infinity is of order $\sqrt\epsilon\, T_{_H}$,
which is very small indeed. Hence the rate of any thermal emission from the shell is
negligible. There is no negative specific heat and no instability since the
total bulk rest mass energy of the configuration remains essentially constant as the
thin shell cools.

If we do allow for a positive chemical potential within the shell, $\mu >0$, then
the temperature and entropy estimates just given become upper bounds, and it is 
possible to approach a zero temperature ground state with zero entropy.  As the 
shell does cool, the entropy decreases very slowly from its initial value
({\ref{entest}) and eq. of state of the shell must be replaced by a more accurate
quantum stress tensor, not accounted for in this simple model. The non-singular 
final state of ultimate gravitational collapse is then a cold, completely dark object 
sustained against any further collapse solely by quantum zero point pressure
of the interior. 

Realizing this alternative requires that a quantum gravitational 
vacuum phase transition intervene before the classical event horizon can form. 
This is exactly what the fluctuations of the anomaly scalar fields $\varphi$ and $\psi$
described in Sec. \ref{sec:anom} can provide. In a more realistic model based on
the EFT of the anomaly (\ref{Seff}), with (\ref{Tanom})-(\ref{Fab}), the rapid change 
of these fields near the horizon and their stress tensor can provide the suitable 
boundary layer which replaces the phenomenological eq. of state $p=\rho$ in 
region II, and the somewhat artificial sharp interface boundaries with the surface 
stresses (\ref{surf}). The entire surface layer will still be of order $\ell$ in (\ref{meanell}) 
in physical thickness, and if treated as very thin, be completely consistent with the
classical formula (\ref{diffSm}) with an actual surface tension carried by
the physical interface boundary. This is now under investigation.

Incidentally, because of the necessarily non-vanishing surface tensions 
at the interface, $p_{\perp} \neq p$ there and the gravastar solution also 
explicitly evades the Buchdahl lower bound on the radius of a compact object 
$R > 9GM/4$, since this bound is derived under the assumption that the radial 
and tangential pressures are everywhere equal (and positive) inside the object
\cite{WeinG,Buch,CatFabVis}.

Since the exterior spacetime is Schwarzschild until distances of order of the diameter 
of an atomic nucleus from $r=r_{_S}$, a gravastar cannot be distinguished from a black
hole by present observations of X-ray bursts \cite{Abr}. However, the shell with its 
maximally stiff eq. of state $p=\rho$, where the speed of sound is equal to the speed 
of light, could be expected to produce explosive outgoing shock fronts in the
process of formation. Active dynamics of the shell may produce other effects that
would distinguish gravastars from black holes observationally, possibly providing
a more efficient particle accelerator and central engine for energetic astrophysical 
sources. The spectrum of gravitational radiation from a gravastar should bear the 
imprint of its fundamental frequencies of vibration, and hence also be quite different 
from a classical black hole.

The interior de Sitter region with $p = -\rho$ may be interpreted also as a 
cosmological spacetime, with the horizon of the expanding universe replaced by
a quantum phase interface. The possibility that the value of the vacuum energy density
in the effective low energy theory can depend dynamically on the state of a gravitational
condensate may provide a new paradigm for cosmological dark energy in the universe.
The proposal that other parameters in the standard model of particle physics may depend
on the vacuum energy density within a gravastar has been discussed by Bjorken \cite{BJ}.
A stable bubble of positive vacuum zero point energy also realizes Dirac's idea for an
extensible model of the electron \cite{Direlec,Boyer}, but in the case of the attractive 
gravitational force instead (so that the repulsive quantum zero point force 
can be balanced by rather than added to the classical self-force of the extended body).
Similar suggestions for removing the singularity in gravitational collapse were
also made by Sakharov and Gliner \cite{SakGlin}.

In the original paper on gravastars \cite{gstar}, the stability of the configuration
was studied in the same hydrodynamic approximation used to construct it. In each of
the three regions (\ref{regions}), fluctuations are completely stable, for the
clear physical reasons that the de Sitter and Schwarzschild geometries are
stable to fluctuations and the fluid making up the this shell also has a physical
eq. of state. In this hydrodynamic model the positions of the interfaces at $r_1$ and $r_2$ 
are fixed by the externally provided information from the estimate (\ref{ellest}).
Thus the exact positions of $r_1$ and $r_2$ are not fixed dynamically, and there
are zero modes of neutral stability corresponding to shifts of $r_1$ and $r_2$ 
\cite{gstar,PNAS}. This can only be remedied by a fully dynamical theory in which
the gravastar solution is obtained by the variation of a well-defined action
functional. This we did not possess in $2001$, but do have now have in the effective
theory of the Einstein-Hilbert term plus the covariant action functional (\ref{allanom})
with (\ref{SEF}) generated by the anomaly. The subsequent studies of the solutions
of the equations of motion of the anomaly scalar field $\varphi$ and $\psi$ in both
the Schwarzschild and de Sitter geometries \cite{MotVau}, reviewed in Sec. \ref{sec:EFT},
show that the stress tensor of the anomaly scalars (\ref{Eab})-(\ref{Fab}) can become
large near both the black hole and cosmological horizons and provide exactly the
dynamical stresses that were hypothesized in \cite{gstar,PNAS}. Work is 
now currently in progress to find a gravastar solution to the effective field equations 
including the anomaly terms. As an extremum of a well-defined action principle, 
the dynamical stability of the solution can then be studied without any additional 
assumptions.

Since the appearance of the papers \cite{gstar,PNAS}, a number of other authors have 
addressed the issue of gravastar stability in a number of different ways
\cite{VisWilt,Dym,BCar,Lobo,dBHorIlKlV,ChirRez07,CarPCanCav,ChirRez08,RCdSW},
all requiring some phenomenological assumptions and parameters. Not surprisingly,
the results obtained depend upon the values of these parameters, although generally
a wide class of stable gravastar-like solutions have been found.
The authors of Ref. \cite{CarPCanCav} claimed to have found a more generic
instability of the ergoregion of rotating non-black hole compact objects.
However it was shown in Ref. \cite{ChirRez08} that this conclusion can be
avoided by making somewhat less restrictive assumptions than those of \cite{CarPCanCav}.

More to the point, it is by no means certain that a rotating gravastar possesses
an ergoregion at all. Our discussion of the stress tensor dependence upon the
norm of the timelike Killing field $(K_aK^a)^{\frac{1}{2}}$ in Sec. \ref{sec:BH}
and the global aspects of the trace anomaly in Secs. \ref{sec:EFT} and \ref{sec:anom}
lead to the expectation that large effects in the expectation value of the stress
tensor are to be expected where $K^a$ first becomes null. In the non-rotating
Schwarzschild geometry this is the horizon at $r=r_{_S}$. However, in the Kerr
geometry the unique timelike Killing vector at large $r$ first becomes null at
the outer boundary of the ergosphere. Hence if the region of large quantum effects
is here, the interior ergosphere region will be changed and likely not exist
at all, removing the question of stability of this region. 

Clearly this question and the entire subject of rotating gravastars, their 
possible magnetic fields and detailed interactions with matter, relevant for
realistic collapsed stars and astrophysical observations require further study 
before definite answers can be given or predictions made. It has been pointed 
out in Ref. \cite{Abr} that astrophysical observations or non-observations of 
X-ray bursts or other emisions from compact objects are not definitive for or
against the existence of a physical surface. The interpretations of this
data also depend heavily upon models of any surface and its possible
interactions with matter. An interesting possibility for detecting emission
from the surface based on a model of these interactions has been given
in \cite{BarChap}. A more model independent prediction of a physical
surface for a compact object is that it will have normal modes of vibration,
and a characteristic discrete energy spectrum which should be discernible in
gravitational wave observations. For this reason it is important to develop
both the theory and the observational techniques for detecting a gravitational
wave spectrum that may differ from that expected from pure classical GR
\cite{YPret}.

\section{Summary and Outlook}

Although quantum effects can often be quite subtle, and it is often assumed 
that in gravity they play a role only at the Planck scale, the challenges 
presented to current theory by quantum effects in black hole and cosmological 
spacetimes suggest otherwise. There are two macroscopic systems where quantum 
effects may in fact be crucial, namely the final state of gravitational
collapse and in accounting for cosmological dark energy of the universe itself. 
These are the principal challenges presented to the reconcilation of
General Relativity with quantum mechanics on macroscopic scales.

In this article we have reviewed the status of the horizons in the
classical Schwarzschild and de Sitter solutions of Einstein's equations. 
The paradoxes of black hole physics in particular argue for a careful 
analysis of the possible importance of large backreaction effects of 
quantum vacuum polarization in the vicinity of the event horizon. Large 
quantum effects of this kind do not violate the Principle of Equivalence, 
but simply serve to emphasize that quantum coherence and entanglement 
effects can depend on gauge invariant but non-local integrals of gauge 
potentials such as (\ref{Wloop}) in electrodynamics or (\ref{Knorm}) 
and (\ref{tform}) in gravity. Physically this is because quantum matter 
has wavelike properties and cannot be absolutely localized to a point. 
Mathematically, it is expressed by the fact that gauge theories 
including gravity are theories of gauge connections with generally 
non-trivial fiber bundle structure. Thus singular coordinate frame 
transformations of the kind often considered in classical GR ignore 
the possibility of the new degrees of freedom associated with these 
improper ``gauge" transformations, and may be unwarranted in quantum 
theory.

Based on a number of examples from both flat space QED and QCD, and gravity 
in both $2$ and $4$ dimensions, the central role of anomalies was discussed. 
Anomalies play a special role in EFT's since they provide an essentially
unique way to violate the naive decoupling of ultraviolet from infrared 
sectors of a Lorentz invariant or generally covariant theory. This is because
they are at the same time local in the divergence of the classically
conserved current, but also imply a non-local structure in the effective
action. The non-locality is associated with the propagation of massless 
correlated pair states, analogous to Cooper pairs in a superconductor or 
collective modes in other media. Both the chiral anomalies of QED and QCD 
and the trace anomaly in a gravitational background lead generically to 
additional propagating degrees of freedom than is apparent from the original 
classical or single particle Lagrangian. The spectral density in these 
propagating anomaly channels obey ultraviolet sum rules and hence remain
physical states, even away from the strictly conformal on-shell limit. The
states in the trace anomaly channel couples only to fourth order curvature
invariants and therefore only very weakly to matter, so that they would
not have been detected so far in terrestrial or solar system tests of
gravity. 

The $0^+$ states in trace anomaly amplitudes can be described as local
scalar fields in an effective field theory description. The logarithmic
scaling with distance of the effective action of the trace anomaly
(\ref{allanom})-(\ref{SEF}) implies that it remains a (marginally)
relevant operator in the EFT of low energy gravity, and should be
explicitly appended to the classical Einstein-Hilbert terms to take
account of quantum effects of pair correlations due to the trace anomaly. 
This has the consequence that General Relativity receives quantum 
corrections relevant at macroscopic distances. 

Due to the topological character of the Euler-Gauss-Bonnet invariant in
the trace anomaly, the scalar degrees of freedom in the anomaly effective
action are sensitive to non-local or global aspects of the underlying
quantum theory, and this is the fundamental reason why they can play
a decisive role near classical event horizons, where they have
macroscopically large effects. As the horizon is approached, the large
blueshift of frequencies overwhelms all finite mass and energy scales,
so that all fields become essentially massless, and the behavior is 
conformal, accurately described by the conformal anomaly effective 
action. This has been shown quantitatively by comparison of the 
stress tensor of the anomaly (\ref{Tanom})-(\ref{Fab}) in terms 
of the two scalar fields $\varphi$ and $\psi$ to the renormalized
expectation value of the stress tensor of quantum matter fields 
and their vacuum polarization effects computed by standard methods.
The important qualitative conclusion of the study of the anomaly
stress tensor in the Schwarzschild and de Sitter cases is that the
diverging behavior proportional to $(-K_aK^a)^{-2} = (-g_{tt})^{-2}$
is quite generic and appears in a wide range of states other than
the Boulware state. A fine tuning of the state is necessary to remove
all divergences of the stress tensor on the horizon in each case,
which does not seem to be {\it a priori} warranted.

Since the anomaly scalar fields have kinetic terms, they must also be 
treated as dynamical fields in their own right. In fact, their fluctuations 
induce the running of the classical cosmological term, and provide a 
mechanism for the vacuum energy to be dynamical, and dependent upon 
infrared effects, rather than a constant. The value $G \Lambda = 0$ is an 
infrared fixed point of the renormalization group flow. Thus a dynamical 
relaxation of the vacuum energy to zero is possible by this mechanism of 
anomaly scalar field fluctuations \cite{AntMot,NJP,QFEXT09}. 

The restoration of conformal invariance these fluctuations imply may
be observable in the Cosmic Microwave Background. One form this
conformal invariance could show up is in non-Gaussian correlations 
of a specific angular form (\ref{C3}), assuming that conformal
invariance is realized on the flat spatial sections of a spatially
hologeneous, isotropic FLRW cosmology. Another way of realizing
conformal invariance in de Sitter space is associated with the breakdown 
of the $O(4,1)$ isometry group to $O(3)$, relevant for cosmological
dark energy in the present epoch. These different realizations
lead to different cosmological models and different predictions
for the non-Gaussianities which may allow us to differentiate models.

In de Sitter space the fluctuations of the anomaly scalars imply scalar
degrees of freedom from fundamental theory without the {\it ad hoc}
introduction of an inflaton field. Their fluctuations grow large at the
de Sitter horizon as at the Schwarzschild horizon. This suggests
that the horizon of a black hole should be viewed instead as the locus 
of a quantum phase transition induced by the anomaly scalar degrees of 
freedom, where the effective value of the gravitational vacuum energy density 
can change. In the EFT including the trace anomaly terms, $\Lambda_{eff}$ 
becomes a dynamical condensate whose value depends upon the macroscopic 
boundary conditions at the horizon. By taking a positive value in the interior,
the effective ``repulsive" dark energy core can remove the black hole 
singularity of the classical Schwarzschild solution, replacing it with 
an horizon volume of de Sitter space. The resulting {\it gravitational 
condensate star} or {\it gravastar} model resolves all black hole paradoxes, 
and provides a testable non-singular quantum alternative to black holes as 
the final state of complete gravitational collapse. 

The cosmological dark energy of our universe may be a macroscopic finite 
volume condensate whose value depends not on microphysics but on the 
cosmological horizon scale. Finally, both the sensitivity of the trace anomaly 
terms to microphysics and the analogies with many body and condensed 
matter systems suggest that further development of the ideas presented here
may lead to an entirely new basis for the microscopic degrees of freedom of 
quantum gravity and therefore of the constituents of spacetime itself \cite{Maz96}.

\vskip 1cm
\section*{Acknowledgments}

I am very much indebted to all my collaborators, whose work is reviewed in this
article: Paul R. Anderson, Ignatios Antoniadis, Maurizio Giannotti, 
Carmen Molina-P\'aris, Ruslan Vaulin, and especially Pawel O. Mazur, with 
whom the hypothesis of gravitational condensate stars was developed. 
I wish also to thank Michal Praszalowicz and the other organizers of the 
XLIX Krak\'ow School of Theoretical Physics for the invitation to lecture at 
that school. Notes for these lectures eventually grew into this article. 
Finally I wish to thank the I. Antoniadis and the CERN Theoretical Physics Group 
for the Scientific Associateship and hospitality at CERN, Oct., $2009$-Apr., $2010$ 
where and when most of this article was written.

\end{document}